\documentclass[aps,prb,reprint,amsmath,amssymb,nobibnotes]{revtex4-1}

\usepackage{bm}
\usepackage{graphicx}

\begin{document}

\title{Quantum Scattering Theory of Spin Transfer Torque, Spin Pumping and Fluctuations}

\author{Arne Brataas}

\affiliation{Center for Quantum Spintronics, Department of Physics, Norwegian University of Science and Technology, NO-7491 Trondheim, Norway}
\thanks{Arne.Brataas@ntnu.no}

\begin{abstract}
Spin transfer torque and spin pumping are central reciprocal phenomena in spintronics. These phenomena occur in hybrid systems of normal metals and magnets. Spin transfer is the conversion of spin currents in metals to a torque on the magnetization of magnets. Spin pumping is the emission of spin currents from precessing magnets. Here, we demonstrate a general way to understand these effects within a quantum out-of-equilibrium path-integral model. Our results agree with known expressions for spin transfer and spin pumping in terms of transverse (mixing) conductances when there are no fluctuations. However, at a finite temperature, frequency or spin accumulation, the magnet also experiences fluctuating torques. In the classical regime, when the thermal energy is larger than the bias voltage and precession frequency, we reproduce the classical Brownian-Langevin forces associated with spin transfer and spin pumping. At low temperatures, in the quantum regime, we demonstrate that magnetization fluctuations differ in the elastic and inelastic electron transport regimes. Furthermore, we show how additional transport coefficients beyond the mixing conductance govern the fluctuations. Some of these coefficients are related to electron shot noise because of the discrete spin angular momentum of electrons. We estimate the fluctuation coefficients of clean, tunnel, and disordered junctions and in the case of an insulating magnet. Our results open a path for exploring low-temperature magnetization dynamics and spin caloritronics.
\end{abstract}
\date{\today}
\maketitle

\section{Introduction}

Spin transfer and spin pumping are central phenomena in spintronics. Berger proposed the pioneering concept of an interaction between magnetic textures and electric currents in the late 1970s\cite{Berger:JAP79,Freitas:JAP85}. The concept of spin transfer torque\cite{Berger:PRB1996,Slonczewski:JMMM1996} revitalized the fields of spintronics and magnetoelectronics. The seminal experiments in Refs. \onlinecite{Tsoi:PRL1998,Myers:Science1999,Katine:PRL2000,Kiselev:Nature2003,Krivorotov:Science2005} amplified its early recognition. Spin transfer causes a torque on magnets from spin-polarized currents. This phenomenon excites magnets, changes the magnetization orientation, and creates dynamic motion or moving magnetization textures; see the reviews and concepts in Refs. \onlinecite{Silva:JMMM2008,Ralph:JMMM2008,Parkin:Science2008,Brataas:NatMat2012,Manchon:RMP2019}.

In the 1970s, researchers also understood that dynamic magnets emit spin currents\cite{Monod:PRL1972,Silsbee:PRB1979}. Similar to spin transfer torque, this phenomenon has become essential in spintronics, with improved measurements\cite{Mizukami:Jpn2001,Urban:PRL2001} and the development of a quantitative theory and wider applicability of the effect, called spin pumping\cite{Tserkovnyak:PRL2002,Heinrich:PRL2003,Tserkovnyak:RMP2005,Mosendz:PRB2010,Ando:JAP2011,Sandweg:PRL2011,Shikoh:PRL2013,Cheng:PRL2014,Kamra:PRL2017,Johansen:PRB2017,Li:Nature2020,Vaidya:Science2020}.

Most theories describing spin transfer and spin pumping treat magnetization as a classical variable and use conservation of angular momentum to deduce the spin transfer torque or spin pumping\cite{Waintal:PRB2000,Brataas:PRL2000,Brataas:EPB2001,Stiles:PRB2002,Tserkovnyak:PRL2002}. An alternative but also semiclassical approach is to view the spin transfer torque as arising from an exchange potential outside of equilibrium\cite{Nunez:SSC2006}. Finally, a few papers use quantum mechanical language to discuss spin pumping and spin transfer but are often limited to tunnel junctions and no fluctuations\cite{Bender:PRL2012,Bender:PRB2014}.

Foros et al. realized that there are also fluctuating forces related to spin transfer and spin pumping\cite{Foros:PRL2005}. For example, thermal fluctuations of the spin currents in normal metal-ferromagnet junctions give rise to Brownian stochastic forces on the magnet. The strengths of these forces are exactly in agreement with the fluctuation-dissipation theorem for spin pumping-enhanced Gilbert damping, as should be expected. These fluctuations are essential ingredients in the field of ambient temperature spin caloritronics\cite{Xiao:PRB2009,Xiao:PRB2010,Bauer:NMAT2012}. However, when the spin accumulation in the normal metal exceeds the thermal energy, the fluctuation-dissipation theorem no longer holds\cite{Foros:PRL2005}. In this quantum regime, shot noise caused by the discrete nature of the spin angular momentum carried by individual electrons enhances magnetic fluctuations\cite{Foros:PRL2005}. A similar picture emerges in ferromagnet-ferromagnet tunnel junctions\cite{Chudnovskiy:PRL2008,Swiebodzinski:PRB2010}. The latter article introduced path integrals to compute spin transfer, spin pumping, and fluctuations.

Traditionally, spin transfer and spin pumping are explored at ambient temperatures. The thermal energy is larger than the magnetic resonance frequency and the bias charge or spin accumulation. However, new developments in measurements and the prospect of achieving an improved understanding motivate lower-temperature measurements and theory, where quantum fluctuations come into play. In ferromagnet-normal metal-ferromagnet spin valves, Zholud et al. pioneered the observation of quantum effects in the anomalous behavior of magnetization fluctuations at low temperatures\cite{Zholud:PRL2017}. In such spin valves, there are quantum fluctuations that reduce the giant magnetoresistance in the presence of a voltage bias\cite{Zholud:PRL2017,Qaiumzadeh:PRB2018}. 

We find rich and complex quantum fluctuations in normal metal-ferromagnet systems. Our results are relevant to measurements of magnetization fluctuations in such systems. Furthermore, they may be essential in spin transfer-controlled magnonics and Bose-Einstein condensation of magnons\cite{Divinskiy:NatCom2010}. More generally, our results will be central to the development of an understanding of low-temperature spin caloritronics\cite{Bauer:NMAT2012}. Experimental probing of our results for the quantum fluctuations requires low-temperature measurements such as in Ref. \onlinecite{Zholud:PRL2017} carried out at 3.4 K.

This paper describes spin transfer and spin pumping in arbitrary junctions using an out-of-equilibrium quantum picture that considers quantum and classical fluctuations. We describe arbitrary contacts between ferromagnets and normal metals. To this end, we use the scattering theory of electron transport\cite{Buttiker:PRB1992}. We aim to describe the temporal spin dynamics. To do so, we use the out-of-equilibrium path-integral formulation. By integrating out the fermions that may be out of equilibrium with spin and charge accumulations, we find the effective action for the magnons that we can in turn express as an effective Landau-Lifshitz-Gilbert-Slonczewski equation with fluctuating forces.

Our contribution to the field is twofold. We demonstrate that earlier semiclassical approaches to describing spin transfer and spin pumping in arbitrary junctions agree with a complete quantum treatment of magnons and electrons to leading order in the total spin. More importantly, we derive expressions for the quantum and classical fluctuations that the magnet experiences in terms of the reflection and transmission coefficient of the junctions. Each electron carries a discrete amount of spin angular momentum $\hbar/2$. The discreteness of the spin angular momentum causes shot noise, as experienced by the magnet when the bias voltage is larger than the temperature and precession frequency. In our endeavor, we merge earlier semiclassical approaches that describe any junction in terms of the scattering matrix\cite{Brataas:PRL2000,Brataas:EPB2001,Tserkovnyak:PRL2002,Tserkovnyak:RMP2005} with the out-of-equilibrium path-integral formalism introduced in Refs. \onlinecite{Duine:PRB2007,Chudnovskiy:PRL2008,Swiebodzinski:PRB2010}

We consider systems in which the spin angular momentum is conserved for electron transport at the normal metal-ferrromagnet junction and within the ferromagnet. Nevertheless, we include spin-orbit interactions within the normal metals for the electron transport. Additionally,  spin-orbit induced magnetic anisotropy energies can be included in straightforward manners. We believe that generalization of our methods and calculations to spin-orbit torques generated at junctions\cite{Manchon:PRB2008,Manchon:PRB2009,Miron:NatMat2010,Miron:Nature2011,Liu:Science2012,Garello:NatNan2013,Mellnik:Nature2014,MacNeill:NatPhys2017,Zhu:PRL2019,Manchon:RMP2019} and magnonic charge pumping is feasible and deserves our attention.

We organize the manuscript as follows. First, section \ref{sec:system} presents the system and the assumptions. Then, section \ref{accumulations} describes the normal metals with the charge and spin accumulations that may exist therein. We present our main results for the spin dynamics in section \ref{main}. Section \ref{Model} describes the model of the system introduced in section \ref{sec:system}. We use the closed contour path-integral method in section \ref{closed} to compute the main results in section \ref{main} resulting from the model in section \ref{Model}. The derivation of the stochastic Langevin forces is in section \ref{Langevin}. We derive a valuable connection to the scattering formulation of the electron-magnon interaction in section \ref{sec:scattering}. Finally, we conclude the paper in section \ref{sec:conclusion}.

\section{System and Assumptions}
\label{sec:system}

\begin{figure}[htpb]
\includegraphics[width=0.48\textwidth]{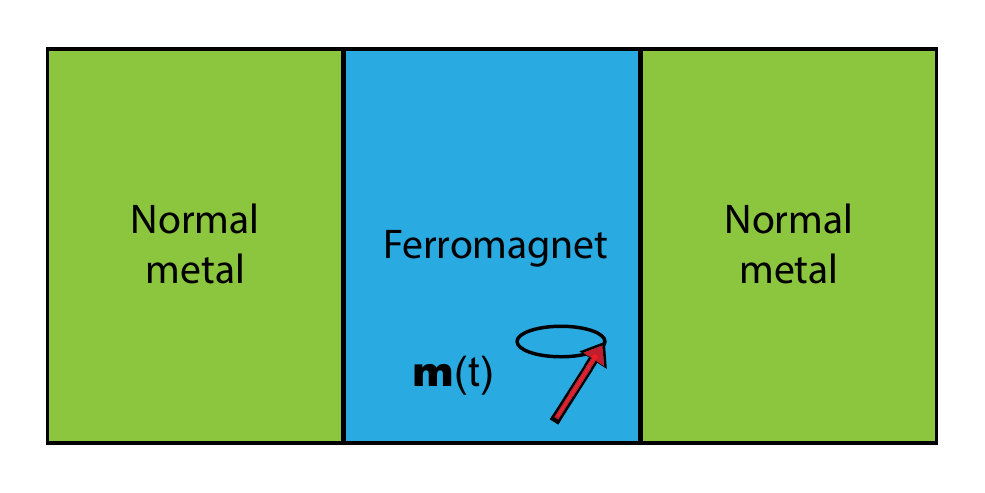}
\caption{Schematic depiction of a normal metal-magnet-normal metal system. Spin accumulations in the normal metals affect the spin dynamics described by precession of the temporal unit vector along the order parameter ${\bf m}(t)$. The figure shows two normal metals. Our results can be straightforwardly generalized to an arbitrary number of normal metals attached to the ferromagnet.}
\label{system}
\end{figure}

We consider the system shown in Fig. \ref{system}. Normal metals surround a ferromagnet. The accumulation of the out-of-equilibrium charge ($\mu_c$) and spin (${\bm \mu}_s$) potentials in normal metals can drive the spin dynamics in the ferromagnet. Furthermore, the spin dynamics can induce spin currents in the metals. The magnet can be insulating or conducting. In either case, the electron transport in the normal metals can influence the spin dynamics.

We will derive how the semiclassical magnetization dynamics respond to spin and charge accumulations. We assume that the total spin of the ferromagnet is large, $S \gg1$. In this sense, our calculation is semiclassical. However, we will also consider fluctuations in the magnetization dynamics. An often-used assumption is that the thermal energy ($k_B T$) is much larger than all the other energy scales related to the magnetization dynamics. This scenario represents the classical regime. While we will cover this classical regime, we will also consider the quantum regime in which the dynamic frequency ($\omega$) and charge and spin accumulations are larger than the thermal energy. This generalization results in shot noise contributions to the fluctuating forces acting on the magnet. Ref. \onlinecite{Foros:PRL2005} pioneered the identification of shot noise contributions to magnetization dynamics and computed some of the terms in some regimes. Quantum fluctuations dominate when the frequency or accumulation is larger than the thermal energy. We will demonstrate that the fluctuations obey a rich and complex behavior that can reveal the nature of the electron transport and how the magnet responds to it.

In all situations, we will assume that the Fermi energy is the largest energy scale in the system. This assumption is valid in metallic systems since the Fermi energy is approximately 1-10 eV, whereas the thermal energy, magnetization frequency, and accumulation are in the meV range. Therefore, consistent with the large Fermi energy assumption, we consider systems much larger than the Fermi wavelength in metals. In other words, the system is much larger than the lattice spacing.

In our explicit calculation, we assume the magnet to be homogeneous. However, we assert that our results are also valid for magnetic textures as long as the variation is larger than the Fermi wavelength, which is always the case except in extreme cases. The underlying reason why the results have broader applicability than only to homogeneous ferromagnets is the significant difference in the length scales of the magnetic variations and the typical wavelength of the electrons. Hence, the electrons only see the magnetization in a range near the Fermi wavelength.

\section{Normal Metal Reservoirs}
\label{accumulations}

This section describes the normal metals. Out-of-equilibrium charge and spin accumulations may exist in the normal metal reservoirs. These may arise from currents from other ferromagnets or normal metals. Alternatively, they may arise from currents in the normal metal reservoirs via the spin Hall effect\cite{Sinova:RMP2015}. We will not repeat such discussions, which are not central to our development. Instead, we will assume that some of these mechanisms within the circuit induce charge and spin accumulations. Furthermore, we will assume that these accumulations are static and set by external forces for clarity. Generalizations to consider the influence of the ferromagnet on the spin and charge accumulations and temporal evolution can be performed via magnetoelectronic circuit theory, as described in Refs. \onlinecite{Brataas:PRL2000,Brataas:PRB2002,Tserkovnyak:RMP2005,Brataas:PhysRep2006,Foros:PRB2007}.

In the reservoirs, the operator $\hat{a}_{\sigma \beta}$ annihilates an electron with spin $\sigma$ in state $\beta$. The statistical occupation of the state is
\begin{equation}
\langle \hat{a}^\dag_{s\alpha} \hat{a}_{\sigma \beta} \rangle = \delta_{\alpha \beta} n_{s\sigma\alpha} \, , 
\label{elocc}
\end{equation}
where $n_{s\sigma\alpha}$ is the $2 \times 2$ out-of-equilibrium occupation of the electrons in state $\alpha$ in spin space. The quantum number $\alpha$ contains lead $\kappa$, transverse waveguide mode $n$, and energy $\epsilon$. In general, in isotropic systems, the $2 \times 2$ out-of-equilibrium occupation of the electrons only depends on the energy $\epsilon$ of the state but can depend on lead $\kappa$:
\begin{align}
n_{s\sigma \kappa }(\epsilon) & =  \left[ f_{\uparrow \kappa}(\epsilon) + f_{\downarrow \kappa}(\epsilon)  \right] 1_{s\sigma}/2 \nonumber \\ & + 
 \left[f_{\uparrow \kappa}(\epsilon) - f_{\downarrow \kappa}(\epsilon)  \right]  {\bm u}_{\kappa} \cdot {\bm \sigma}_{s\sigma}/2 \, , 
\label{ndist}
\end{align}
where $f_\uparrow$ ($f_\downarrow$) is the spin distribution aligned (anti-aligned) with the spin accumulations directed along the unit vector ${\bm u}_{\kappa}$. ${\bm \sigma}$ is a vector of Pauli matrices. 

Electron transport can be inelastic, elastic, or between these regimes. We will demonstrate that the spin fluctuations in the magnet differ in these transport regimes at low temperatures. In other words, measurements of the spin fluctuations will reveal the electron transport regime.

\subsubsection{Inelastic transport}

In the inelastic transport regime, the electron distribution is analogous to that in the equilibrium situation and is determined by Fermi-Dirac functions, but with out-of-equilibrium chemical potentials. There is charge accumulation $\mu_{\kappa c}^{in}$ and spin accumulation ${\bm \mu}_{\kappa s}^{in}$ in reservoir $\kappa$. The spin accumulation ${\bm \mu}_{\kappa s}^{in}$ has a direction along the unit vector ${\bm u}_{\kappa s}$ and an amplitude $\mu_{\kappa s}^{in}$, with ${\bm \mu}_{\kappa s}^{in}= {\bm u}_{\kappa s} \mu_s^{in}$. The inelastic electron spin-up and spin-down distributions are 
\begin{subequations}
\begin{align}
f_{\uparrow \kappa }^{in}(\epsilon) & =f(\epsilon - \mu_{\kappa c}^{in} - \mu_{\kappa s}^{in}/2 ) \, ,  \\ 
f_{\downarrow \kappa}^{in}(\epsilon) & =f(\epsilon - \mu_{\kappa c}^{in} + \mu_{\kappa s}^{in}/2 )\, , 
\end{align}
\label{finelastic}
\end{subequations}
where $f(\epsilon)$ is the Fermi-Dirac distribution function that depends on the temperature $T$.

 \subsubsection{Elastic transport}

In contrast to the inelastic transport regime, in elastic transport, energy is conserved. Let us assume that the elastic transport is a result of transport from two or more large reservoirs with equilibrium electron distributions $f(\epsilon - \mu_1)$, $f(\epsilon - \mu_2)$, etc. Since the energy is conserved, the components of the distributions aligned and anti-aligned with the spin accumulation must be 
\begin{subequations}
\begin{align}
f_{\uparrow \kappa}^{el}(\epsilon) & = \sum_i R_{\uparrow \kappa i} f(\epsilon - \mu_i) \, , \\
f_{\downarrow \kappa}^{el}(\epsilon) & = \sum_i R_{\downarrow \kappa i} f(\epsilon - \mu_i) \, , 
\end{align}
\label{eldist}
\end{subequations}
where $R_{\uparrow \kappa i}$ and $R_{\downarrow \kappa i}$ are transport coefficients that depend on the materials and geometry. We maintain that the bias voltage is much smaller than the Fermi energy, so the transport coefficients $R_{\uparrow \kappa i}$ and $R_{\downarrow \kappa i}$ are constant in the energy range of the bias voltage. At equilibrium, $f_{\uparrow \kappa}^{el}(\epsilon)=f_{\downarrow \kappa}^{el}(\epsilon)=f(\epsilon-\mu_0)$, where $\mu_0$ is the common equilibrium chemical potential. Therefore,
\begin{align}
\sum_i R_{s \kappa i } = 1 \, . 
\end{align}
In the elastic transport regime, for deterministic transport properties such as the spin transfer torque, defining the {\it effective} elastic accumulation as follows is convenient:
\begin{subequations}
\begin{align}
\mu_{\kappa c}^{el} + \mu_{\kappa s}^{el}/2 = \int d\epsilon \left[f^{el}_{\uparrow \kappa}(\epsilon) - f(\epsilon-\mu_0) \right] \nonumber \, , \\
\mu_{\kappa c}^{el} - \mu_{\kappa s}^{el}/2 = \int d\epsilon \left[f^{el}_{\downarrow \kappa}(\epsilon) - f(\epsilon-\mu_0) \right] \, .
\end{align}
\label{muelastic}
\end{subequations}
We will demonstrate that the spin transfer torque is the same in the elastic and inelastic transport regimes provided that the charge and spin accumulations are the same, i.e., $\mu_{\kappa c}^\text{el}=\mu_{\kappa c}^\text{in}$ and ${\bm \mu}_{\kappa s}^\text{el}={\bm \mu}_{\kappa s}^\text{in}$. The reason is that spin transfer and spin pumping are governed by electrons at the Fermi level, which is much larger than the precession frequency and temperature. Then, how the electrons are distributed within a small window close to the Fermi level does not matter. However, we will show that the low-temperature spin fluctuations differ in the two regimes. This dependence on the transport regime occurs because the spin accumulation or precession frequency is much smaller than the Fermi energy and can reveal the low-energy occupation of the electrons.

\section{Main Results - Spin Dynamics}
\label{main}

This section presents our main findings. Our main result is the derivation of a generalized Landau-Lifshitz-Gilbert-Slonczewski (LLGS) equation with stochastic forces that reflect electron transport fluctuations valid from the quantum low-temperature regime to the classical, high-temperature regime. We find that the spin dynamics consist of bulk contributions, spin transfer torque, spin pumping, and electron transport-induced fluctuations:
\begin{equation}
\partial_t {\bm m } = {\bm \tau}_\text{b} + {\bm \tau}_\text{stt} + \tilde{\bm \tau}_\text{stt} + {\bm \tau}_\text{sp} + \tilde{\bm \tau}_\text{sp} + {\bm \tau}_\text{fl} \, .
\end{equation}
The bulk torque is conventional and independent of the electron reservoirs,
\begin{align}
{\bm \tau}_\text{b} =  - \gamma {\bm m} \times \left[ {\bm H}_\text{eff}  + {\bm h}_\text{b}(t) \right] + \alpha_\text{b}  {\bm m} \times \partial_t {\bm m} \, , 
\end{align}
where $\gamma = - g^* \mu_B /\hbar$ is the gyromagnetic ratio, ${\bm H}_\text{eff}=-\partial F / \partial {\bm m} M_s $ is the effective magnetic field governed by the free energy $F$ of the magnet, $\alpha_\text{b}$ is the bulk Gilbert damping coefficient, and ${\bm h}_\text{b}(t)$ is the fluctuating field. The average of the fluctuating field vanishes, i.e., $\langle {\bm h}_\text{b}(t) \rangle=0$, while the fluctuations are related to the Gilbert damping coefficient by the fluctuation-dissipation theorem:
\begin{equation}
\langle \gamma h_{\text{b}i} (t) \gamma h_{\text{b}j}(\tau) \rangle_\omega = \alpha_0  \frac{2 k_B T}{S} \zeta \left( \frac{\omega}{2k_B T} \right) \, , 
\label{hfluct}
\end{equation}
where $S$ is the total spin of the magnet, the Fourier transform on the left-hand side of Eq.\ \eqref{hfluct} is defined in Eq.\ \eqref{Fourier}, and
\begin{equation}
    \zeta(x)=x/\tanh{x} 
    \label{zeta} \, . 
\end{equation} 
In the classical high-temperature limit, $\zeta(x \rightarrow 0)=1$.

With our alternative approach, we reproduce the established result for the spin transfer torques\cite{Brataas:PRL2000,Brataas:EPB2001,Brataas:PRep2006} arising from spin accumulation in the left (${\bm \mu}_s$) and right ($\tilde{\bm \mu}_s$) reservoirs:
\begin{subequations}
\begin{align}
{\bm \tau}_\text{stt} & = 
\alpha^r {\bm m} \times {\bm m}  \times {\bm \mu}_{s}  - \alpha^i   {\bf m}  \times {\bm \mu}_{s}  \, , \\
\tilde{{\bm \tau}}_\text{stt} & =
 \tilde{\alpha}^r {\bm m} \times  {\bm m} \times \tilde{{\bm \mu}}_{s} - \tilde{\alpha}^i  {\bf m} \times \tilde{{\bm \mu}}_{s} \, ,
\end{align}
\label{tauSTT}
\end{subequations}
where $\alpha$ ($\tilde{\alpha}$) is the enhanced Gilbert damping coefficient associated with spin pumping in the left (right) reservoir. The superscript $r$ ($i$) denotes the real (imaginary) part. As previously stated, the expression for the spin transfer torque is similar in the elastic and inelastic transport regimes and only depends on the total amount of out-of-equilibrium spins. We find agreement with previous works\cite{Brataas:PRL2000,Brataas:EPB2001,Brataas:PRep2006,Brataas:arXiv2011} in that the scattering matrix determines the spin transfer (spin pumping) coefficients as follows:
\begin{subequations}
\begin{align}
    \alpha & = \frac{1}{4\pi S} \text{Tr} \left[1 - r^\dag_\uparrow r_\downarrow - t^\dag_\uparrow t_\downarrow \right] \, , \\
    \tilde{\alpha} & = \frac{1}{4\pi S} \text{Tr} \left[1 - \tilde{r}^\dag_\uparrow \tilde{r}_\downarrow - \tilde{t}^\dag_\uparrow \tilde{t}_\downarrow \right] \, , 
\end{align} 
\label{alpha}
\end{subequations}
where $r_s$ ($\tilde{r}_s$) is the reflection amplitude matrix associated with transport from the left (right) lead for electrons with spin $s$ along the magnetization direction and $t_s$ ($\tilde{t}$) is the transmission amplitude matrix associated with transport from the left (right) lead for electrons with spin $s$ along the magnetization direction. 
$r$, $\tilde{r}$, $t$, and $\tilde{t}$ are elements of the scattering matrix defined in Eq. \eqref{Smatrix}.

We also reproduce the established expression for spin pumping in the left and right leads\cite{Tserkovnyak:PRL2002,Brataas:PRB2002,Tserkovnyak:RMP2005}:
\begin{subequations}
\begin{align}
{\bm \tau}_\text{sp} & = 
 \left(\alpha +  \tilde{\alpha} \right) ^r  {\bf m} \times \partial_t {\bm m} \nonumber \\
& -\left( \alpha + \tilde{\alpha} \right)^r {\bm m} \times {\bm m}  \times \partial_t {\bm m}  \, . 
\end{align}
\label{tauSP}
\end{subequations}
The correspondence between the spin transfer torque \eqref{tauSTT} and spin pumping \eqref{tauSP} is a result of Onsager reciprocity relations\cite{Brataas:arXiv2011}. Spin pumping is the reciprocal phenomenon of spin transfer.

Our main result is electron transport-enhanced fluctuations. We find that there is a fluctuating torque
\begin{equation}
{\bm \tau} = - \gamma {\bm m} \times {\bm h}(t) \, , 
\end{equation}
where ${\bm h}(t)$ is the fluctuating field.

At equilibrium, in the absence of bias voltages, we find that the fluctuation-dissipation theorem including the effects of spin pumping holds, so 
\begin{equation}
\langle \gamma h_{i}^\text{e}(t) \gamma h_{j}^\text{e} (\tau) \rangle_\omega =\delta_{ij} \left( \alpha^r + \tilde{\alpha}^r \right) \frac{2 k_B T}{S} \zeta \left( \frac{\omega}{2k_B T} \right) \, ,  
\label{flheq}
\end{equation}
where $\zeta(x)$ was defined in Eq. \eqref{zeta}. Ref. \onlinecite{Foros:PRL2005} first obtained the classical limit, in the regime $k_B T \gg \omega$, of this result. Eq. \eqref{flheq} generalizes this equilibrium result to finite frequencies. As expected, the fluctuations in Eq. \eqref{flheq} are similar to the bulk fluctuations in Eq. \eqref{hfluct}. As established in Ref. \onlinecite{Foros:PRL2005}, this implies that at high temperatures, the LLGS equation holds with fluctuating forces governed by a total Gilbert damping coefficient $\alpha_\text{b} + \alpha^r  + \tilde{\alpha}^r$.

Charge and spin accumulations in the normal metals modify the fluctuations. Out of equilibrium, the fluctuating field ${\bm h}(t)$ has contributions arising from longitudinal and transverse components of the spin accumulation with respect to the magnetization.

In a right-handed coordinate system where the unit vector of the magnetization is along the $z$ direction, we find the general result for the fluctuations:
\begin{widetext}
\begin{subequations}
\begin{align}
\langle \gamma h_x(t) \gamma h_x(\tau)\rangle_\omega & = - \frac{i}{4S} \left[\Sigma^K(\omega) + \Sigma^K(-\omega) \right] + \frac{1}{S} \text{Im} \left[ \tilde{\Sigma}_{\uparrow \downarrow}^K \right] \, , \\
\langle \gamma h_y(t) \gamma h_y(\tau)\rangle_\omega & = - \frac{i}{4S} \left[\Sigma^K(\omega) + \Sigma^K(-\omega) \right] - \frac{1}{S} \text{Im} \left[ \tilde{\Sigma}_{\uparrow \downarrow}^K \right] \, ,\\
\langle \gamma h_x(t) \gamma h_y(\tau) \rangle_\omega & = - \frac{1}{4S} \left[\Sigma^K(\omega) - \Sigma^K(-\omega) \right] + \frac{1}{S} \text{Re} \left[ \tilde{\Sigma}_{\uparrow \downarrow}(\omega) \right] \, , 
\end{align}
\label{fluccorrFourier}
\end{subequations}
\end{widetext}
where $\Sigma^K(\omega)$ and $\tilde{\Sigma}^K_{\uparrow \downarrow}(\omega)$ are Keldysh components of self-energies. $\Sigma^K(\omega)$ is driven by charge accumulation and spin accumulation along the magnetization. In contrast, $\tilde{\Sigma}^K_{\uparrow \downarrow}(\omega)$ is driven by spin accumulation transverse to the magnetization.

The self-energy associated with charge accumulation and spin accumulation along the magnetization is
\begin{align}
    \Sigma^K(\omega) = i \sum_{\kappa \lambda} \sigma_{\kappa \lambda} \pi_{\kappa \lambda}(\omega) \, . 
    \label{Sigmasigma}
\end{align}
The sums over $\kappa$ and $\lambda$ are over the left and right lead indices, and the noise parameters in the left-right lead basis are
\begin{align}
    \sigma = \left(
    \begin{array}{cc}
         \alpha^r - \beta^r &  \beta^r \\
         \tilde{\beta}^r & \tilde{\alpha}^r - \tilde{\beta}^r 
    \end{array}
    \right) \, . 
\label{sigmaKscat}
\end{align}
The spin transfer/spin pumping coefficients $\alpha$ and $\tilde{\alpha}$ are introduced in Eq. \eqref{alpha}. The shot noise parameters relevant when the charge accumulation and longitudinal spin accumulation exceed the thermal energy and resonance frequency are
\begin{subequations}
\begin{align}
    \beta_{\uparrow \downarrow}^r & =  \frac{1}{8\pi S} \text{Tr}\left[1 - \left(r_\uparrow^\dag r_\downarrow +t_\uparrow^\dag t_\downarrow  \right) \left(r_\downarrow^\dag r_\uparrow +t_\downarrow^\dag t_\uparrow  \right)\right] \, , \\
\tilde{\beta}_{\uparrow \downarrow}^r & = \frac{1}{8 \pi S}  \text{Tr} \left[1 - \left(\tilde{r}_\uparrow^\dag \tilde{r}_\downarrow +\tilde{t}_\uparrow^\dag \tilde{t}_\downarrow  \right) \left(\tilde{r}_\downarrow^\dag \tilde{r}_\uparrow +\tilde{t}_\downarrow^\dag \tilde{t}_\uparrow  \right)\right] \, . 
\end{align}
\label{betar}
\end{subequations}

The fluctuations differ in the elastic and inelastic transport regimes. This difference facilitates observation of the electron transport regime by measuring the magnetic fluctuations. At equilibrium, in the inelastic transport regime, and in the elastic transport regime, the contributions are as follows:
\begin{widetext}
\begin{subequations}
\begin{align}
   \pi_{\kappa \lambda}^\text{eq} & = 4 k_B T \zeta\left(\frac{\omega}{2 k_B T} \right) \, ,  \\
   \pi_{\kappa \lambda}^\text{in} & = 4 k_B T \sum_{\sigma s} p_{\uparrow \sigma \kappa} p_{\downarrow s \lambda}  \zeta \left( \frac{\omega + \mu_{\sigma \kappa} - \mu_{s \lambda}}{2 k_B T} \right) \, , \\
   \pi_{\kappa \lambda}^\text{el} & = 4 k_B T \sum_{\sigma s} \sum_{ij} p_{\uparrow \sigma \kappa} p_{\downarrow s \lambda} R_{\sigma \kappa i} R_{s \lambda j} \zeta \left( \frac{\omega + \mu_{i} - \mu_{j}}{2 k_B T} \right) \, , 
\end{align}
\label{pi}
\end{subequations}
\end{widetext}
where the function $\zeta$ was defined in Eq. \eqref{zeta} and we have introduced the projection factor that depends on the longitudinal component of the spin accumulation, $u_{z\kappa}$, in the distribution in Eq. \eqref{ndist}:
\begin{align}
    p_{s\sigma \kappa} & = (1 - u_{z\kappa})/2 + \delta_{s\sigma} u_{z\kappa} \, . 
\end{align}
The coefficient $R$ that governs the linear response distribution in the elastic transport regime is defined in Eq. \eqref{eldist}.

When spin accumulation is transverse to the magnetization, there are contributions from the self-energy:
\begin{align}
    \tilde{\Sigma}_{\uparrow \downarrow}^K(\omega) = - i \sum_{\kappa \lambda} \tilde{\sigma}_{\kappa \lambda} \tilde{\pi}_{\kappa \lambda}(\omega) \, ,
    \label{tildeSigmatildesigma}
\end{align}
where the noise matrix is
\begin{align}
\tilde{\sigma}^K = 
\left(
\begin{array}{cc}
     \alpha - \gamma_{\uparrow \downarrow} &  \delta_{\uparrow \downarrow} \\
     \tilde{\delta_{\uparrow \downarrow}} &  \tilde{\alpha} - \tilde{\gamma}_{\uparrow \downarrow}  
\end{array}
\right) \, , 
\label{tildesigmamatrix}
\end{align}
where the shot noise parameters associated with the transverse spin accumulation are
\begin{align}
\gamma_{\downarrow \uparrow} & = \frac{1}{16 \pi S} \text{Tr} \left[1 - \left( r_\downarrow^\dag r_\uparrow + t_\downarrow^\dag t_\uparrow \right)^2 \right] \, , \\
\delta_{\downarrow \uparrow} & = \frac{1}{8 \pi S} \text{Tr} \left[ \left(r_\downarrow^\dag \tilde{t}_\uparrow + t_\uparrow^\dag \tilde{r}_\uparrow \right) \left( \tilde{t}_\downarrow^\dag r_\uparrow + \tilde{r}_\downarrow^\dag t_\uparrow \right) \right] \, .
\end{align} 
At equilibrium, in the inelastic and elastic transport regimes, we find
\begin{widetext}
\begin{subequations}
\begin{align}
\tilde{\pi}^\text{eq}_{\kappa \lambda} & = 0 \, , \\
\tilde{\pi}^{\text{in}}_{\kappa \lambda} & =  u_{\kappa s-} u_{\lambda s-} \left[  F(\mu_{\lambda \uparrow }  - \mu_{\kappa \uparrow}  - \omega )  + 
F(\mu_{\lambda \downarrow }  - \mu_{\kappa \downarrow}  - \omega ) 
 - F(\mu_{\lambda \uparrow}- \mu_{\kappa \downarrow}  -\omega) - F(\mu_{\lambda \downarrow}  - \mu_{\kappa \uparrow} -\omega  )  \right] \, , \\
\tilde{\pi}_{\kappa \lambda}^\text{el} & =  u_{\kappa s-} u_{\lambda s-}  \sum_{ij}  \left(R_{\uparrow \kappa i } - R_{\downarrow \kappa i} \right) \left(R_{\uparrow \lambda j} - R_{\downarrow \lambda j} \right) F(\mu_j -\mu_i -\omega)  \, , 
 \end{align}
 \label{pitilde}
\end{subequations}
\end{widetext}
where
\begin{align}
F(\mu_2-\mu_1) & = \int_{-\infty}^\infty d\epsilon f(\epsilon - \mu_1) \left[ 1 - f(\epsilon - \mu_2) \right] \nonumber \\
& = \left[ \mu_2 - \mu_1 \right] n(\mu_2 - \mu_1) \, , 
\label{F}
\end{align}
$f(\epsilon)$ is the Fermi-Dirac distribution, and $n(\mu_2-\mu_1)$ is the Bose-Einstein distribution function at temperature $T$. At high temperatures, $F \rightarrow k_B T$, and $\tilde{\pi}^\text{in}=0=\tilde{\psi}^\text{el}$.

We discuss how the spin transfer/spin pumping and shot noise coefficients behave in different junctions in Appendix \ref{sec:trancoeff}. In the following subsection, we discuss three relevant scenarios for the fluctuations.

\subsection{Charge Accumulation-Driven Spin Fluctuations}

Let us consider a scenario in which there is no spin accumulation in the normal metals. In the inelastic transport regime, the out-of-equilibrium chemical potentials are then independent of the spins, $\mu_{\sigma \kappa} \rightarrow \mu_{\kappa}$. Hence,
\begin{subequations}
\begin{align}
    \pi^\text{in}_{\kappa \lambda} &= 4 k_B T \xi \left(\frac{\omega + \mu_\kappa -\mu_\lambda}{2 k_B T} \right) \, , \\
    \pi^\text{el}_{\kappa \lambda} & = 4 k_B T \sum_{ij} R_{\kappa i} R_{\lambda j} \xi \left(\frac{\omega + \mu_i - \mu_j}{2 k_B T} \right) \, , 
\end{align}
\end{subequations}
where $R_{\kappa i}=R_{\sigma \kappa i}$ and is independent of the spin when there is no spin accumulation. Furthermore,
\begin{equation}
    \tilde{\pi}_{\kappa \lambda} = 0 \, .
\end{equation}
Consequently, $\tilde{\Sigma}^K_{\uparrow \downarrow}=0$. In, e.g., the inelastic transport regime,
\begin{widetext}
\begin{align}
   i \Sigma^K = - 4 k_B T \left\{ 
    \left[\left(\alpha^r + \tilde{\alpha}^r\right) - \left(\beta^r + \tilde{\beta} \right) \right] 
    \xi\left(\frac{\omega}{2 k_B T} \right)
 + \beta^r \xi\left(\frac{\omega + \mu_L - \mu_R}{2 k_B T}  \right)
 +  \tilde{\beta}^r \xi \left( \frac{\omega - \mu_L + \mu_R}{2 k_B T} \right)
\right\} \, .
 \label{sigmaKcharge}
\end{align}
\end{widetext}
From Eq. \eqref{sigmaKcharge}, we see that measurements of the magnetic fluctuations reveal the shot noise coefficients $\beta^r$ and $\tilde{\beta}$ at finite bias voltages.

Simple limits of charge bias-driven magnetic fluctuations when there is strong dephasing were discussed in Ref. \onlinecite{Foros:PRB2007}. The more general coefficients $\beta^r$ and $\tilde{\beta}^r$ introduced here captures the results in the earlier work and is more general and valid for any degree of spin dephasing.

\subsection{One Reservoir}

Let us consider a scenario in which there is only one reservoir, the left, and a spin accumulation therein. 

First, we consider a longitudinal spin accumulation, e.g., spin accumulation directed along the magnetization. In this case, $\tilde{\pi}_{\kappa \lambda} = 0$, so $\tilde{\Sigma}^K=0$. Furthermore, $\beta^r=0$, and $\tilde{\beta}^r=0$. We also have that $\tilde{\alpha}=0$. The fluctuations are then determined by
\begin{align}
\Sigma^{K,\text{in}} & = i \alpha 4 k_B T \xi \left(\frac{\omega + \mu_s}{2 k_B T} \right) \, , \\
\Sigma^{K,\text{el}} & = i \alpha 4 k_B T \sum_{ij} R_{\uparrow L i } R_{\downarrow L j} \xi \left( \frac{\omega + \mu_i - \mu_j}{2 k_B T} \right)
\end{align}
and differ in the inelastic and elastic transport regimes. When the spin accumulation or precession frequency exceeds the energy, quantum fluctuations exceed the classical fluctuations.

Second, we consider transverse spin accumulation. Then, fluctuations are governed by the coefficient 

\begin{equation}
    C_{\downarrow \downarrow} = \frac{1}{2} \text{Tr} \left[1 - \left(r_{\downarrow} r_{\downarrow}^\dag \right)^2 \right] \, ,
\end{equation}
and $\delta_{\downarrow \uparrow}=0$. $C_{\downarrow \uparrow}$ resembles the spin transfer/spin pumping coefficient $A_{\uparrow \downarrow}$ of Eq. \eqref{alpha} but is not identical to it. However, when there is strong dephasing, $C_{\downarrow \uparrow}=N/2=A_{\uparrow \downarrow}$, where $N$ is the number of propagating channels (transverse waveguide modes). Furthermore, $\tilde{\pi}_{LR}=\tilde{\pi}_{RL}=\tilde{\pi}_{RR}=0$. In the inelastic and elastic transport regimes, we then find in the case of pure transverse spin accumulation ($u_{zL}=0$)
\begin{widetext}
\begin{subequations}
\begin{align}
-i \tilde{\Sigma}^{K,\text{in}}_{\uparrow \downarrow} & = - \left(\alpha - \gamma_{\uparrow \downarrow} \right)  u_{L-}^2 \left[ 2 F(-\omega) - F(\mu_s - \omega) - F(-\mu_s - \omega) ] \right] \, , 
\label{tildeSigmaperpin}
\\
-i \tilde{\Sigma}^{K,\text{el}}_{\uparrow \downarrow} & = 
- \left(\alpha - \gamma_{\uparrow \downarrow} \right)  u_{L-}^2 \sum_{ij} \left( R_{\uparrow L i} - R_{\downarrow L j} \right) \left(R_{\uparrow L j} - R_{\downarrow L j} \right) F(\mu_j - \mu_i - \omega) \, . 
\label{tildeSigmaperpel}
\end{align}
\end{subequations}
\end{widetext}
At zero temperature and frequency, we find that we can use $F(\omega)=0$ and $F(\mu_s - \mu)+F(-\mu_s-\omega)= - \lvert \mu_s \rvert$ in Eq. \eqref{tildeSigmaperpin} and $F(\mu_j - \mu_i - \omega) = (\mu_i - \mu_j) \theta(\mu_i - \mu_j)$ in Eq. \eqref{tildeSigmaperpel}, where $\Theta(x)$ is the Heaviside function.

\section{Model}
\label{Model}

This section details the model that we use to describe the normal metal-ferromagnet-normal metal system introduced in section \ref{sec:system}. The Hamiltonian is
\begin{equation}
\hat{H} = \int d{\bf r} \hat{\psi}^\dag  \left[ H_0 +u_s \bm{\sigma} \cdot \hat{\bm S} \right] \hat{\psi} + \hat{H}_s(\hat{\bf S})\, , 
\label{H}
\end{equation}
where $\hat{\psi}^\dag=( \hat{\psi}_\uparrow^\dag, \hat{\psi}_\downarrow^\dag )$ is the 2-component itinerant electron field operator, ${\bm \sigma}$ is the vector of Pauli matrices in the $2 \times 2$ spin space, and the spin operator $\hat{\bm S}$ has a total spin angular momentum $S$ such that $\hat{\bm S}^2=S(S+1)$. We set $\hbar=1$. $\hat{H}_s(\hat{\bf S})$ describes the transverse dynamics of the magnetization when it is not interacting with the itinerant electrons. To discuss the influence of the electrons on the magnetization dynamics, we consider a simple example in which the magnetization is subject to an external magnetic field ${\bf H}$,
\begin{equation}
\hat{H}_s = - g^* \mu_B {\bf H} \cdot \hat{\bf S} \, , 
\label{H_s}
\end{equation}
where $g^*$ is the effective g-factor and $\mu_B$ is the Bohr magneton. Our calculations can be straightforwardly generalized to arbitrary forms of $\hat{H}_s$ including magnetic anisotropy energy and demagnetization fields. However, such complications are not the focus of our attention. Our target is the influence of the electrons on the magnet. Therefore, we keep the Hamiltonian $\hat{H}_s$ as simple as possible in our explicit calculations but assert that the results are valid for arbitrary forms of $\hat{H}_s$. Furthermore, we disregard the possible effects of an external magnetic field on the itinerant electrons since these effects are typically small in metallic systems, or we can include them in straightforward generalizations of the spin transport equations\cite{Brataas:EPB2001}. Such complications are irrelevant to our main points.

The spin-independent part of the single-particle electron Hamiltonian is
\begin{equation}
H_0 = - \frac{1}{2m} \nabla^2 + V_c({\bm r}) \, , 
\label{H_0}
\end{equation}
where $V_c({\bm r})$ is the spatially dependent charge potential. In the Hamiltonian \eqref{H}, $u_s (\bf r)$ represents the spatially dependent exchange interaction. It is finite only inside the magnet. In the classical limit of the magnet, the magnitude of the spatially dependent spin potential experienced by the itinerant electrons is $V_s({\bf r})=S u_s (\bf r)$.

We will derive the semiclassical spin dynamics valid for any deviation of the magnetization from the equilibrium orientation. To this end, considering the magnon dynamics near the instantaneous direction of the spin $\hat{{\bm S}}$ and adiabatically adjusting the evolution of the small deviation of the large spin from its instantaneous direction are sufficient\cite{Chudnovskiy:PRL2008,Swiebodzinski:PRB2010}. For simplicity, we consider an instantaneous orientation of the spin along the $z$ direction. We carry out a Holstein-Primakoff transformation of the localized spin operator to the second order in the deviation from the classical ground state:
\begin{subequations}
\begin{align}
\hat{S}_+ & = \hat{S}_x + i \hat{S}_y \approx \sqrt{2S} \hat{b} \, , \\
\hat{S}_- & = \hat{S}_x - i \hat{S}_y \approx \sqrt{2S} \hat{b}^\dag \,  , \\
\hat{S}_z & = \left[ S - \hat{b}^\dag \hat{b} \right] \, , 
\end{align}
\label{HP}
\end{subequations}
where the magnon annihilation and creation operators $\hat{b}$ and $\hat{b}^\dag$ satisfy the boson commutation relation $\left [ \hat{b}, \hat{b}^\dag \right ]=1$. Employing $\bm{\sigma} \cdot \hat{\bm S} = \sigma_- \hat{S}_+ + \sigma_+ \hat{S}_- + \sigma_z \hat{S}_z$, where $\sigma_\pm= (\sigma_x \pm i \sigma_y)/2$, by expanding to the second order in the magnon operators, the Hamiltonian \eqref{H} becomes
\begin{equation}
\hat{H} = \hat{H}_e  + \hat{H}_m + \hat{H}_{em} \, .
\end{equation}
The magnon Hamiltonian $\hat{H}_m$ is directly related to the Hamiltonian $\hat{H}_s$ \eqref{H_s} describing the transverse dynamics of the magnetic moments via the Holstein-Primakoff transformation \eqref{HP}. Disregarding a constant energy contribution, the magnon Hamiltonian up to the second order in the magnon operators is
\begin{equation}
\hat{H}_m =  E_z \hat{b}^\dag \hat{b} + \sqrt{S/2} \left( E_- \hat{b} + E_+ \hat{b}^\dag \right) \,  ,
\label{Hm}
\end{equation}
where the magnon energy is
\begin{align}
    E_z=g^* \mu_B H_z \, .
\end{align}
The magnons are excited via the transverse magnetic field, governed by
\begin{align}
E_\pm = g^* \mu_B H_\pm \, , 
\label{Epm}
\end{align}
where $H_\pm = H_x \pm i H_y$.

The considerable algebra outlined in section \ref{sec:scattering} demonstrates how the electronic Hamiltonian $H_e$ and the electron-magnon Hamiltonian $H_{em}$ become particularly transparent in terms of the scattering states of the itinerant electrons. We find
\begin{equation}
\hat{H}_e = \sum_{s\alpha} \epsilon_\alpha \hat{a}_{s\alpha}^\dag \hat{a}_{s\alpha}  \, , 
\label{H_escat}
\end{equation}  
where $\hat{a}_{s\alpha}$ annihilates an electron with spin $s$ ($s=\uparrow$ or $s=\downarrow$). The quantum number $\alpha=\kappa n\epsilon$ captures the lead ($\kappa=L$ or $\kappa=R$), the transverse waveguide mode $n$ and the electron energy $\epsilon$. The electron energy consists of a transverse contribution $\epsilon_n$ and a longitudinal contribution $\epsilon(k)=k^2/2m$, where $k$ is the longitudinal momentum such that $\epsilon=\epsilon_n + \epsilon(k)$. The eigenenergy is spin degenerate since the leads are paramagnetic. Furthermore, the eigenenergy is independent of the lead index since we consider similar left and right leads. In Eq.\ \eqref{H_escat} and similar expressions to follow, the sum over the states implies that $\sum_\alpha X_{s\alpha}$ = $\sum_{\kappa n} \int_{\epsilon_n}^\infty d\epsilon  X_{s\kappa n}(\epsilon)$. In the scattering basis, the electron-magnon interaction becomes
\begin{equation}
\hat{H}_{em} = \hat{H}_1 + \hat{H}_2 \, . 
\end{equation}
The contribution that is to the first order in the magnon operators is
\begin{equation}
\hat{H}_{1} = \sqrt{\frac{2}{S}} \sum_{\alpha \beta} \left[  \hat{b}  \hat{a}_{\downarrow \alpha}^\dag  W_{\downarrow \uparrow}^{\alpha \beta}\hat{a}_{\uparrow \beta} + \hat{b}^\dag  \hat{a}_{\uparrow \alpha}^\dag  W_{\uparrow \downarrow}^{\alpha \beta}\hat{a}_{\downarrow \beta} \right] \, . 
\label{H1}
\end{equation}
The matrices $W_{\uparrow \downarrow}$ and $W_{\downarrow \uparrow}$ are related. They are governed by the exchange potential $u_s({\bm r})$ and the scattering states. Importantly, we demonstrate in Appendix \ref{sec:eval} that they are proportional to a generalized \emph{inelastic} mode-dependent transverse "mixing" transport property. However, we will demonstrate that to the lowest order in the large spin limit $S \gg 1$, only the conventional elastic "mixing" conductance will contribute to the deterministic part of the spin dynamics. We will derive that new transport coefficients beyond the mixing conductance govern the fluctuations of the spin dynamics. The contribution to the electron-magnon interaction to the second order in the magnon operators is
\begin{equation}
\hat{H}_{2} = - \frac{1}{S} \sum_{\alpha \beta}   \hat{b}^\dag \hat{b}  \left[ \hat{a}_{\uparrow \alpha}^\dag  W_{\uparrow \uparrow}^{\alpha \beta} \hat{a}_{\uparrow \beta} - \hat{a}_{\downarrow \alpha}^\dag W^{\alpha \beta}_{\downarrow \downarrow} \hat{a}_{\downarrow \beta}  \right] \, . 
\label{H2}
\end{equation}
We will show that $\hat{H}_{2}$ does not contribute to the spin dynamics when $S \gg1 $ and hence can be disregarded.

\section{Closed Contour Action}
\label{closed}

This section first presents the out-of-equilibrium path-integral method used to find the influence of electron transport on spin dynamics. Subsequently, we compute the specific form of the effective action from the electrons on the spin dynamics. To this end, we introduce the closed contour action $S$ and partition function $Z$,
\begin{equation}
Z = \int D[\bar{\psi}_\uparrow \psi_\uparrow \bar{\psi}_\downarrow \psi_\downarrow \bar{\phi} \phi ] e^{i S} \, .
\label{Z}
\end{equation}
In action $S$, the integral over time is along the closed contour forward from $t=-\infty$ to $t=+\infty$ and backward from $t=+\infty$ to $t=-\infty$. From the partition function, we can find how the electrons influence the spin dynamics. Action $S$ consists of contributions from the electrons, magnons, and first- and second-order electron-magnon interactions:
\begin{equation}
S=S_e + S_m + S_{1}+S_{2} \, . 
\end{equation}
The electron contribution corresponding to the Hamiltonian \eqref{H_escat} is
\begin{align}
S_e  & = \int_{-\infty}^{\infty} \! \! \! \! dt \left[\bar{\psi}^+_\uparrow \left(i \partial_t - \epsilon \right) \psi^+_\uparrow  - \bar{\psi}^-_\uparrow \left(\partial_t - \epsilon \right) \psi^-_\uparrow \right] + \nonumber \\
& \int_{-\infty}^{\infty} \! \! \! \! dt \left[\bar{\psi}^+_\downarrow \left(i \partial_t - \epsilon \right) \psi^+_\downarrow  - \bar{\psi}^-_\downarrow \left(\partial_t - \epsilon \right) \psi^-_\downarrow \right]  \, , 
\label{S_e}
\end{align}
where $\psi_\sigma$ ($\sigma= \uparrow$ or $\sigma=\downarrow$) is a vector consisting of all Grassmann variables representing the electronic states. The lead, transverse waveguide mode, and electron energy for spin $\sigma$ electrons span the vector. The superscript $+$ ($-$) denotes the forward (backward) path variables. In Eq.\ \eqref{S_e}, $\epsilon$ is a diagonal matrix containing all single-particle energies spanned by the electronic states.

The magnon contribution to the action corresponding to the magnon Hamiltonian \eqref{Hm} is
\begin{align}
&S_m  =  \int_{-\infty}^{\infty} \! \! \! \! dt \left[\bar{\phi}^+ \left(i \partial_t - E_z \right) \phi^+  - \bar{\phi}^- \left(i \partial_t - E_z \right) \phi^- \right] - \nonumber \\
& \sqrt{S/2}  \int_{-\infty}^{\infty} \! \! \! \! dt \left[ E_- \left( \phi^+ - \phi^-  \right)+ E_+ \left(\bar{\phi}^+ - \bar{\phi}^-  \right) \right] \, ,
\label{S_m}
\end{align}
where $\phi$ is a complex number and $\bar{\phi}$ is its conjugate number. The electron-magnon interaction to the first order in the magnon operators corresponding to the Hamiltonian \eqref{H1} is
\begin{align}
S_{1}   &= - \sqrt{\frac{2}{S}} \int_{-\infty}^{\infty} \! \! \! \!  dt  \left[ \bar{\psi}_\downarrow^+ W_{\downarrow \uparrow} \psi^+_\uparrow \phi^+ -\bar{\psi}_\downarrow^- W_{\downarrow \uparrow} \psi^-_\uparrow \phi^- \right] - \nonumber \\
&  \sqrt{\frac{2}{S}} \int_{-\infty}^{\infty} dt  \left[ \bar{\psi}^+_\uparrow W_{\uparrow \downarrow} \psi^+_\downarrow \bar{\phi}^+ -\bar{\psi}_\uparrow^- W_{\uparrow \downarrow} \psi^-_\downarrow \bar{\phi}^- \right] \, . 
\label{S_1}
\end{align}
The electron-magnon interaction to the second order in the magnon operators representing the Hamiltonian \eqref{H2} is
\begin{align}
S_{2}  & =  \frac{1}{S} \int_{-\infty}^{\infty} \! \! \! \! dt  \left[ \bar{\psi}^+_\uparrow W_{\uparrow \uparrow} \psi^+_\uparrow - \bar{\psi}^+_\downarrow W_{\downarrow \downarrow} \psi^+_\downarrow \right] \bar{\phi}^+ \phi^+ \nonumber \\
& -\frac{1}{S} \int_{-\infty}^{\infty} \! \! \! \! dt  \left[ \bar{\psi}^-_\uparrow W_{\uparrow \uparrow} \psi_\uparrow - \bar{\psi}^-_\downarrow W_{\downarrow \downarrow} \psi_\downarrow \right] \bar{\phi}^- \phi^-  \, . 
\label{S_2}
\end{align}
Introducing classical and quantum fields for the bosons and $1-2$ fields for the fermions instead of the $+$ and $-$ fields on the forward and backward paths is convenient and conventional\cite{Kamenev:2011}.

\subsection{Classical Quantum and 1-2 Quantum Fields}

For the bosons, we use the classical and quantum fields
\begin{subequations}
\begin{align}
\phi^{c,q} & = \left(\phi^+ \pm \phi^- \right)/\sqrt{2} \, , \\
\bar{\phi}^{c,q} & = \left(\bar{\phi}^+ \pm \bar{\phi}^- \right)/\sqrt{2} \, .
\end{align}
\label{bosonclqfields}
\end{subequations}
The upper sign is for the classical ($c$) field, and the lower sign is for the quantum field ($q$).
The structure of the boson Green function is therefore
\begin{equation}
D = \left( 
\begin{array}{cc}
D^K & D^R \\
D^A & 0
\end{array}
\right) \, . 
\label{bosonGreen}
\end{equation}
In the representation of Eqs.\ \eqref{bosonclqfields} and \eqref{bosonGreen}, the magnon contribution to the action in Eq.\ \eqref{S_m} becomes
\begin{align}
S_m = \int dt \int d\tau (\bar{\phi}^{c},\bar{\phi}^q) D_0^{-1}
\left( 
\begin{array}{cc}
\phi^{c} \nonumber \\
\phi^q
\end{array}
\right) \\
- \sqrt{S} \int dt \left[ E_- \phi^q + E_+ \bar{\phi}^q \right] \, , 
\label{S_mKeldysh}
\end{align}
where $\left[D_0^{-1}\right]^{R(A)} =\delta(t-\tau) (i\partial_t - E_z \pm i0^+)  $.

For the fermions, we use a different notation of 1-2 fields:
\begin{subequations}
\begin{align}
\psi^{1,2} = \left(\psi^+ \pm \psi^- \right)/\sqrt{2} \, , \\
\bar{\psi}^{1,2} = \left(\bar{\psi}^+ \mp \bar{\psi}^- \right)/\sqrt{2} \, . 
\end{align}
\end{subequations}
The structure of the fermion Green function is then
\begin{equation}
G = \left( 
\begin{array}{cc}
G^R & G^K \\
0 & G^A
\end{array}
\right) \, . 
\end{equation}
The occupations at the initial time $t \rightarrow - \infty$ are included via a standard regularization of the Green functions for bosons and fermions. This procedure results in the Keldysh components for the bosons and fermions. In the Keldysh (classical quantum for bosons and 1-2 for fermions) space, we use the matrices
\begin{equation}
\gamma^{c}=
\left( 
\begin{array}{cc}
1 & 0 \\
0 & 1 
\end{array}
\right) 
\end{equation}
and
\begin{equation}
\gamma^{q}=
\left( 
\begin{array}{cc}
0 & 1 \\
1 & 0 
\end{array}
\right) \, . 
\end{equation}
The contributions to the action involving the electrons, i.e., the electron contribution and the electron-magnon contributions, $S_{e,tot}=S_e + S_{1}+ S_{2}$, are then
\begin{equation}
S_{e,tot} = \int dt \int d \tau \left(\psi^1, \psi^2 \right) G^{-1} 
\left(
\begin{array}{cc}
\psi^1 \\
\psi^2 
\end{array}
\right) \, ,
\end{equation}
where the inverse Green function for the fermions is
\begin{equation}
G^{-1} = G_0^{-1} + \left[ \tilde{W}_1 + \tilde{W}_2 \right]\, , 
\label{Ginvfermions}
\end{equation}
$\left[G_0^{-1}\right]^{R(A)} = \delta(t-\tau) (i\partial_t - \epsilon \pm i0^+) $, $\tilde{W}_1=\delta(t-\tau) W_1$, $\tilde{W}_2=\delta(t-\tau) W_2$, 
\begin{align}
W_1 & = - \frac{1}{\sqrt{S}} \sum_x
\left[
\bar{\psi}_\downarrow W_{\downarrow \uparrow} \gamma^x \psi_\uparrow \phi^x 
+ \bar{\psi}_\uparrow W_{\uparrow \downarrow} \gamma^x \psi_\downarrow \bar{\phi}^x
\right] 
\label{W_1}
\end{align}
and
\begin{align}
 W_2 &= \frac{1}{2S} \sum_x \left[ \bar{\psi}_\uparrow W_{\uparrow \uparrow} \gamma^x \psi_\uparrow - \bar{\psi}_\downarrow W_{\downarrow \downarrow} \gamma^x \psi_\downarrow \right]  \bar{\phi} \gamma^x \phi\, . 
 \label{W_2}
\end{align}
In Eqs.\ \eqref{W_1} and \eqref{W_2}, $x=c,q$. In the partition function \eqref{Z}, we can now integrate over the fermions to find the effective contribution to the boson action
\begin{equation}
S_\text{eff} = -i \text{Tr} \ln \left[1 + G_0 \left( \tilde{W}_1 + \tilde{W}_2 \right) \right] \, .
\label{S_eExact}
\end{equation}
In the following sections, we will expand \eqref{S_eExact} to the second order in the magnon operators, but first, we need to discuss  the unperturbed Green functions of the electrons that contain information about the charge and spin accumulations in the reservoirs discussed in section \ref{accumulations}.

\subsection{Unperturbed Electron Green Function}

The unperturbed electron Green function is the Green function in the absence of the electron-magnon interactions in Eqs.\ \eqref{H1} and \eqref{H2} or, alternatively, the electron-magnon interactions in Eqs.\ \eqref{S_1} and \eqref{S_2}. The retarded and advanced parts of the zeroth-order Green function $G_0$ are diagonal in spin space and all other quantum variables (lead, transverse mode, and energy). In the Fourier representation \eqref{Fourier},
\begin{equation}
G_{0\alpha \beta}^{R/A}(\omega) = \frac{1}{\omega \pm i 0^+ - \epsilon_\alpha} \delta_{\alpha\beta} \, .
\label{G0RA}
\end{equation}
However, the zeroth-order Keldysh component of the Green function $G_0$ has off-diagonal components in spin space since there is a noncollinear spin accumulation in the leads. This out-of-equilibrium noncollinear spin accumulation is what couples to the localized spins.

There are charge and spin accumulations in the reservoirs, as described in section \ref{accumulations}. Consequently, the Keldysh component of the Green function is
\begin{equation}
G^K_{0s\alpha\sigma\beta}(\omega) = - 2\pi i \delta_{\alpha\beta} \delta(\omega - \epsilon_\alpha) \left[\delta_{s\sigma }-  2 n_{s\sigma\alpha} \right] \, ,
\label{G0K}
\end{equation}
where $n_{s\sigma\alpha}$ is the $2 \times 2$ out-of-equilibrium occupation of the electrons introduced in Eq.\ \eqref{ndist}.

At equilibrium, $n_{s\sigma \alpha} = \delta_{s\sigma} f(\epsilon - \mu_0)$, where $\mu_0$ is the equilibrium chemical potential, $\epsilon$ is the energy of state $\alpha$ with spin s, and $f(\epsilon - \mu_0)$ is the Fermi-Dirac distribution function. The fluctuation-dissipation theorem for fermions holds:
\begin{align}
    G_{0s\alpha \sigma \beta}^{K,\text{eq}} = \tanh{\frac{\epsilon - \mu_0}{2 k_B T}} \delta_{s\sigma} \left[G^R_{0\alpha \beta} - G^A_{0\alpha \beta} \right] \, . 
\end{align}
Naturally, we will subsequently find that an analogous relation holds for the (boson) magnon self-energy at equilibrium with a magnon chemical potential of zero.

\subsection{First-Order Corrections to the Magnon Actions}
\label{subsec:first}

By expanding the effective action \eqref{S_eExact} in the electron-magnon coupling, we find the contribution to the action of the fermions to the first order in the magnon operators:
\begin{align}
S_{1}  = &- i \text{Tr} \left[ G_0 \tilde{W}_1 \right] \nonumber \\
            = & -\frac{2}{\sqrt{S}} \sum_\alpha W_{\downarrow \uparrow \alpha\alpha} n_{\uparrow \downarrow\alpha} \int dt  \phi^q  \nonumber \\
& -\frac{2}{\sqrt{S}}  \sum_\alpha W_{\uparrow \downarrow \alpha \alpha} n_{\downarrow \uparrow \alpha}  \int dt \bar{\phi}^q  \, . 
\label{S_1eff}
\end{align}
By comparing the electron-induced action in Eq.\ \eqref{S_1eff} with the contribution to the magnetic action from the transverse magnetic field in Eq.\ \eqref{S_mKeldysh}, we find that the spin accumulation gives rise to an effective transverse magnetic field represented by the effective energies  $E_{+e}= 2 \sum_\alpha W_{\uparrow \downarrow} n_{\downarrow \uparrow \alpha \alpha}/S$ and $E_{-e}= 2\sum_\alpha W_{\downarrow \uparrow} n_{\uparrow \downarrow \alpha \alpha }/S$. In other words, the action \eqref{S_1eff} corresponds to a {\it transverse} magnetic field where the $x$- and $y$-components are $H_{xe}=(E_{+e}+E_{-e})/2g^* \mu_B $ and $H_{ye}=(E_{+e}-E_{-e})/2i g^* \mu_B $. We observe that the effective spin transfer magnetic fields are inversely proportional to the total spin $S$.

We now use Eqs.\ \eqref{Wud_LLRR} and \eqref{WLLRRsum} to find
\begin{equation}
\sum_\alpha W_{\uparrow \downarrow \alpha\alpha} n_{\downarrow \uparrow \alpha} 
= i S \int d\epsilon \left[  \alpha n_{\downarrow \uparrow L}(\epsilon)  + \tilde{\alpha} n_{\downarrow \uparrow R}(\epsilon) \right] \, , 
\end{equation}
where $n_{\downarrow \uparrow L}$ ($n_{\downarrow \uparrow R}$) is a transverse component of the $2 \times 2$ spin distribution in the left (right) reservoir in Eq.\ \eqref{ndist} and we have defined the energy-dependent spin transfer coefficients $\alpha$ and $\tilde{\alpha}$ in Eq. \eqref{alpha}. Furthermore, in metallic systems, the out-of-equilibrium distribution peaks at the Fermi energy, and
\begin{equation}
\sum_\alpha W_{\uparrow \downarrow \alpha\alpha} n_{\uparrow \downarrow \alpha} 
=  i S \left[ \alpha \mu_{\downarrow \uparrow}  + \tilde{\alpha} \tilde{\mu}_{ \downarrow \uparrow} 
\right] \, , 
\label{Wn}
\end{equation}
where the spin transfer coefficients $\alpha$ and $\tilde{\alpha}$ should be evaluated at the Fermi energy and the effective chemical potentials are
\begin{subequations}
\begin{align}
\mu_{\downarrow \uparrow} & = \int d\epsilon_{nk} n_{\downarrow \uparrow L} \, , \\
\tilde{\mu}_{\downarrow \uparrow} &= \int d\epsilon_{nk} n_{\downarrow \uparrow R}
\end{align}
\label{muefftrans}
\end{subequations}
in the left and right reservoirs. We find similar results for $\sum_\alpha W_{\downarrow \uparrow \alpha \alpha} n_{\uparrow \downarrow \alpha}$. With the definition of the effective chemical potential in Eq.\ \eqref{muefftrans}, Eq.\ \eqref{muelastic} and Eq.\ \eqref{finelastic}, we see that $\mu^{in}_{\downarrow \uparrow}=\mu_{\downarrow \uparrow}=\mu^{el}_{\downarrow \uparrow}$ and $\tilde{\mu}^{in}_{\downarrow \uparrow}=\tilde{\mu}_{\downarrow \uparrow}=\tilde{\mu}^{el}_{\downarrow \uparrow}$. Therefore, the spin transfer torques are the same in the elastic and inelastic transport regimes. This is because spin transfer arises from the transport of electrons around the Fermi energy, where the transport coefficients are constant in a metallic system since the out-of-equilibrium accumulations are much smaller than the Fermi energy.

A more transparent representation of the effect of action $S_1$ on the spin dynamics is in terms of the charge accumulation $\mu_c = \text{Tr}\left[ \check{\mu} \right]/2$ and spin accumulation vector ${\bm \mu}_s= \text{Tr} \left[ {\bm \sigma} \check{\mu} \right]/2 $, where $\check{\mu}$ is the $2 \times 2$ distribution function in spin space. Then, $\mu_{\uparrow \downarrow} = \mu_{s,x} - i \mu_{s,y}$, and $\mu_{\downarrow \uparrow} = \mu_{s,x} + i \mu_{s,y}$; thus, the spin transfer effective field is ${\bm H}_\text{stt} = {\bm H}_e + \tilde{{\bm H}}_e$, where
\begin{subequations}
\begin{align}
{\bm H}_e & = \frac{1}{g^* \mu_B}  
\left[ \alpha^r {\bm m} \times {\bm \mu}_{s} + \alpha^i  {\bf m} \times {\bf m} \times {\bm \mu}_{s} \right] \, . 
\label{He} \\
\tilde{{\bm H}}_e & = \frac{1}{g^* \mu_B}  
\left[ \tilde{\alpha}^r {\bm m} \times \tilde{{\bm \mu}}_{s} + \tilde{\alpha}^i  {\bf m} \times {\bf m} \times \tilde{{\bm \mu}}_{s} \right] \, . 
\end{align}
\label{Hetilde}
\end{subequations}
Using the gyromagnetic ratio $\gamma = - g^* \mu_B /\hbar$ while maintaining $\hbar=1$, we can also express the action of the effective fields in Eq.\ \eqref{He} on the spin dynamics as the spin transfer torques in Eq.\ \eqref{tauSTT}.
 
In Eqs.\ \eqref{He} and \eqref{Hetilde} and what follows, the superscript $r$ denotes the real part, and the superscript $i$ denotes the imaginary part. The spin transfer and spin pumping efficiencies associated with the left and right reservoirs introduced in Eq. \eqref{alpha} are related to the "mixing" conductances $A_{\uparrow \downarrow}$ and $\tilde{A}_{\uparrow \downarrow}$ defined in Eqs. \eqref{Aupdo} and \eqref{Atildeupdo} as follows:
\begin{subequations}
\begin{align}
\alpha & = \frac{A_{\uparrow \downarrow}}{4 \pi S} \, ,  \\
\tilde{\alpha} & = \frac{\tilde{A}_{\uparrow \downarrow}}{4 \pi S} \, .
\end{align}
\label{alpharelatedtoA}
\end{subequations}
Since the mixing conductances $A_{\uparrow \downarrow}$ and $\tilde{A}_{\uparrow \downarrow}$ in Eqs. \eqref{Aupdo} and \eqref{Atildeupdo} are normal metal-ferromagnet interface properties proportional to the cross section and the total spin $S$ is proportional to the volume of the ferromagnet, the spin transfer efficiencies \eqref{alpha} are inversely proportional to the ferromagnet thickness.

Additionally, we will soon see the well-known fact that spin pumping gives rise to enhanced damping that acquires contributions from the left and right interfaces governed by the spin transfer/spin pumping efficiencies $\alpha$ and $\tilde{\alpha}$ in Eq.\ \eqref{alpha}.

\subsection{Second-Order Corrections to the Magnon Action}

The contribution to the action of the fermions to the second order in the magnon operators is $S_2  = S_{2a} + S_{2b}$, where
\begin{subequations}
\begin{align}
S_{2a} & = - i \text{Tr} \left[ G_0 W_2 \right] \, , \label{S2a} \\
S_{2b} & =  \frac{i}{2} \text{Tr} \left[ G_0 W_1 G_0 W_1\right] \label{S2b} \, . 
\end{align}
\end{subequations}

We will compute these contributions separately. First, we compute
\begin{widetext}
\begin{equation}
S_{2a} = - \frac{1}{2S} \sum_\alpha \left[ (1 - 2 n_{\uparrow \uparrow\alpha} )W_{\uparrow \uparrow \alpha} - (1- 2 n_{\downarrow \downarrow\alpha}) W_{\downarrow \downarrow \alpha} \right] \int dt \bar{\phi} \gamma^q \phi \, . 
\label{S2aeff}
\end{equation}
\end{widetext}
Action $S_{2a}$ therefore renormalizes the magnon energy $E_z$, as it appears in the magnon action \eqref{S_mKeldysh} and corresponds to a {\it longitudinal} magnetic field driven by the {\it longitudinal} spin accumulation represented by $n_{\uparrow \uparrow \alpha}$ and $n_{\downarrow \downarrow \alpha}$. $E_z \rightarrow E_z + E_{ze}$. $E_{ze}$ is finite even at equilibrium due to the exchange coupling between the itinerant electrons and the localized spins. Furthermore, the renormalized magnon energy can change due to the out-of-equilibrium longitudinal spin accumulation, $E_{ez}=E_{ez,eq}+E_{ez,ne}$. However, longitudinal magnetic fields have no consequences for the spin dynamics since, in the instantaneous reference frame, such contributions correspond to a total free energy of the form $a \hat{\bm S}^2$, where $a$ is a constant\cite{Chudnovskiy:PRL2008,Swiebodzinski:PRB2010}. Hence, we will not discuss action $S_{2a}$ in Eq.\ \eqref{S2aeff} further.

Next, we consider the second term $S_{2b}$ in Eq.\ \eqref{S2b}. There are four contributions to this term:
\begin{widetext}
\begin{align}
S_{2b} =& \frac{i}{2S}  \int dt \int d\tau \sum_{xy} \text{Tr} \left[ 
G_{0\uparrow \downarrow}(\tau,t) W_{\downarrow \uparrow} \gamma^x 
G_{0\uparrow \downarrow}(t,\tau) W_{\downarrow \uparrow} \gamma^y \right] \phi^x(t) \phi^y(\tau) \nonumber \\
+ &\frac{i}{2S}  \int dt \int d\tau \sum_{xy} \text{Tr} \left[ 
G_{0\uparrow \uparrow}(\tau,t) W_{\uparrow \downarrow} \gamma^x 
G_{0\downarrow \downarrow}(t,\tau) W_{\downarrow \uparrow} \gamma^y \right] \bar{\phi}^x(t) \phi^y(\tau) \nonumber \\
+&\frac{i}{2S}  \int dt \int d\tau \sum_{xy} \text{Tr} \left[ 
G_{0\downarrow \downarrow}(\tau,t) W_{\downarrow \uparrow} \gamma^x 
G_{0\uparrow \uparrow}(t,\tau) W_{\uparrow \downarrow} \gamma^y \right] \phi^x(t) \bar{\phi}^y(\tau) \nonumber \\
+&\frac{i}{2S}  \int dt \int d\tau \sum_{xy} \text{Tr} \left[ 
G_{o\downarrow \uparrow}(t,\tau) W_{\uparrow \downarrow} \gamma^x 
G_{o\downarrow \uparrow}(t,\tau) W_{\uparrow \downarrow} \gamma^y \right] \bar{\phi}^x(t) \bar{\phi}^y(\tau) \, . 
\label{S2b_four}
\end{align}
\end{widetext}
We will now discuss the contributions to $S_{2b}$ line by line. The first line of Eq. \eqref{S2b_four} becomes
\begin{equation}
S_{2b}^{qq} =  \int dt \int d\tau \phi^q(t) \tilde{\Sigma}_{\uparrow \downarrow }^K(t,\tau) \phi^q(\tau) \, ,
\label{S2qq}
\end{equation}
where we have introduced the out-of-equilibrium transverse spin accumulation-induced Keldysh component
\begin{equation}
\tilde{\Sigma}_{\uparrow \downarrow}^{K}(t,\tau) = \frac{i}{2S} \sum_{\alpha \beta} G_{0\uparrow \downarrow \alpha \alpha}^K(\tau,t) W_{\downarrow \uparrow \alpha \beta} G_{0\uparrow \downarrow \beta \beta}^K(t,\tau) W_{\downarrow \uparrow \beta \alpha}  \, .
\end{equation}
Inserting the electron Green functions given by Eqs. \eqref{G0RA} and \eqref{G0K}, we compute
\begin{widetext}
\begin{equation}
\tilde{\Sigma}_{\uparrow \downarrow}^K(t-\tau) = - \frac{2i}{S} \sum_{\alpha \beta} n_{\uparrow \downarrow \alpha} W_{\downarrow \uparrow \alpha \beta} n_{\uparrow \downarrow \beta} W_{\downarrow \uparrow \beta \alpha} e^{-i (\epsilon_\alpha - \epsilon_\beta) (\tau-t)} \, . 
\label{tildeD_updown}
\end{equation}
\end{widetext}
Importantly, inspecting the self-energy $ \tilde{\Sigma}_{\uparrow \downarrow}^K(t-\tau)$ in Eq. \eqref{tildeD_updown} demonstrates that it is an even function of the relative time $t-\tau$:
\begin{align}
\tilde{\Sigma}_{\uparrow \downarrow}^K(t-\tau)=\tilde{\Sigma}_{\uparrow \downarrow}^K(\tau-t)
\label{tildeSigmaKeven} \, . 
\end{align}
This symmetry is essential in its contribution to the fluctuating field experienced by the ferromagnet.

Similarly, the fourth line of Eq. \eqref{S2b_four} becomes
\begin{equation}
S_{2b}^{\bar{q}\bar{q}} =  \int dt \int d\tau \bar{\phi}^q(t) \tilde{\Sigma}_{\downarrow \uparrow }^K(t,\tau) \bar{\phi}^q(\tau) \, ,
\end{equation}
where $\tilde{\Sigma}_{\downarrow \uparrow}^K$ is similar to $\tilde{\Sigma}_{\uparrow \downarrow}^K$ in Eq. \eqref{tildeD_updown} by flipping both spin indices, i.e.,
\begin{widetext}
\begin{align}
    \tilde{\Sigma}^K_{\downarrow \uparrow}(t-\tau) = - \frac{2i}{S} \sum_{\alpha \beta} n_{\downarrow \uparrow \alpha} W_{\uparrow \downarrow \alpha} n_{\downarrow \uparrow \beta} W_{\downarrow \uparrow \beta \alpha} e^{- i (\epsilon_\alpha - \epsilon_\beta) (t- \tau)} \, .
    \label{tildeSigmadownup}
\end{align}
\end{widetext}
By using the properties of the complex conjugate of the out-of-equilibrium distribution in Eq. \eqref{ndist}, $n_{\downarrow \uparrow}^*=n_{\uparrow \downarrow}$, we can relate $\tilde{\Sigma}^K_{\uparrow \downarrow}$ in Eq. \eqref{tildeSigmadownup} to $\tilde{\Sigma}^K_{\downarrow \uparrow}$ in Eq. \eqref{tildeD_updown}:
\begin{align}
    \tilde{\Sigma}_{\downarrow \uparrow}^K(t-\tau) = - \left[ \tilde{\Sigma}^{K}_{\uparrow \downarrow}(t-\tau) \right]^* \, , 
    \label{Stilrel}
\end{align}
where the superscript $*$ implies the complex conjugate. We will demonstrate that this relation \eqref{Stilrel} is essential in ensuring that the fluctuations of the effective transverse fields are real numbers.

From Eq. \eqref{tildeD_updown}, we see that $\tilde{\Sigma}^K_{\uparrow \downarrow}$ vanishes at equilibrium. $\tilde{\Sigma}^K_{\uparrow \downarrow}$ is quadratic in the out-of-equilibrium transverse spin accumulation in the classical limit when the thermal energy is much larger than the spin accumulation, $k_B T \gg \mu_s$. Therefore, actions $S_{2b}^{qq}$ and $S_{2b}^{\bar{q}\bar{q}}$ can be disregarded in the classical limit. 

Interestingly, there are contributions from $\tilde{\Sigma}^K_{\uparrow \downarrow}$ related to quantum shot noise in the quantum limit at low temperatures when $k_B T \le \mu_s$. We find the general expression for $\tilde{\Sigma}_{\uparrow \downarrow}^K$ valid for arbitrary ratios between the spin accumulation and thermal energy by Fourier transforming Eq. \eqref{tildeD_updown}:
\begin{widetext}
\begin{equation}
\tilde{\Sigma}_{\uparrow \downarrow}^K(\omega) = - \frac{4\pi i }{S} \sum_{\alpha \beta} n_{\uparrow \downarrow \alpha} W_{\downarrow \uparrow \alpha \beta} n_{\uparrow \downarrow \beta} W_{\downarrow \uparrow \beta \alpha} \delta(\omega + (\epsilon_\alpha - \epsilon_\beta))
\label{tildeD_updownFourier} \, . 
\end{equation}
\end{widetext}
Subsequently, we disregard small terms of the order $\mu_s/\epsilon_F$ and $k_B T/\epsilon_F$ and define the transverse spin accumulation-driven parameters
\begin{equation}
\tilde{\pi}_{\kappa \lambda}(\omega) =  - 4 \int_{-\infty}^{\infty} d\epsilon n_{\uparrow \downarrow \kappa} (\epsilon) n_{\uparrow \downarrow \lambda}(\epsilon + \omega)
\label{Y}
\end{equation}
so that the Fourier transform of $\tilde{\Sigma}^K_{\uparrow \downarrow}$ becomes Eq. \eqref{tildeSigmatildesigma}, where
\begin{equation}
\tilde{\sigma}_{\uparrow \downarrow \kappa \lambda} = - \frac{\pi}{S} \sum_{nm} W_{\downarrow \uparrow \kappa n \lambda m}W_{\downarrow \uparrow \lambda m \kappa n} 
\label{tildesigmadef}
\end{equation}
and all energy arguments of the matrix elements are at the Fermi energy. To evaluate $\tilde{\pi}_{\kappa \lambda}(\omega)$ in Eq.\ \eqref{Y}, we use distribution \eqref{ndist} and ${\bm u}_{\kappa} \cdot {\bm \sigma}_{\uparrow \downarrow}/2=(u_{\kappa x}- i u_{\kappa y})/2 \equiv u_{\kappa -}/2$ to find
\begin{align}
& \tilde{\pi}_{\kappa \lambda}(\omega) = - u_{\kappa s-} u_{\lambda s-} \nonumber \\
&\int_{-\infty}^\infty d\epsilon \left[f_{\uparrow \kappa}(\epsilon) - f_{\downarrow \kappa}(\epsilon)\right] \left[ f_{\uparrow \lambda}(\epsilon + \omega) - f_{\downarrow \lambda}(\epsilon + \omega)\right] \, . 
\end{align}
We compute $\tilde{\pi}^K$ at equilibrium and in the inelastic and elastic transport regimes and find the result in Eq. \eqref{pitilde}.
When the system is at equilibrium, we find $\tilde{\pi}_{\kappa \lambda}(\omega)=0$. Furthermore, we observe that $\tilde{\Sigma}_{\uparrow \downarrow}^K=0$ in the classical limit when $k_B T \gg \mu_s$, as anticipated above.

Next, we can see that the second and third lines in Eq.\ \eqref{S2b_four} are identical by interchanging $x \leftrightarrow y$ and $ t \leftrightarrow \tau$ and using the cyclic properties of the trace. Furthermore, the $cc$ components vanish because they are combinations of products of retarded Green functions with positive and negative time and products of advanced Green functions with positive and negative time. The second and third lines in Eq.\ \eqref{S2b_four} give a contribution to the $qc$ component of
\begin{equation}
S_{2b}^{qc} =  \int dt \int d\tau \bar{\phi}^q(t) \Sigma^R(t,\tau) \phi^{c}(\tau) \, , 
\end{equation}
where the retarded self-energy is
\begin{align}
\Sigma^R(t,\tau) =   & \frac{i}{S}  \text{Tr} \left[G_{0\uparrow \uparrow}^K(\tau,t) W_{\uparrow \downarrow} G_{0\downarrow \downarrow}^R(t,\tau) W_{\downarrow \uparrow} \right] \nonumber \\
+& \frac{i}{S} \text{Tr} \left[G_{0\uparrow \uparrow}^A (\tau,t) W_{\uparrow \downarrow} G_{0\downarrow \downarrow}^K(t,\tau) W_{\uparrow \downarrow}
 \right]  \, . 
\end{align}
We then compute that the retarded self-energy is
\begin{widetext}
\begin{equation}
\Sigma^R(t-\tau) =   i\theta(t-\tau) \frac{2}{S} \sum_{\alpha \beta} \left( n_{\uparrow \uparrow \alpha} - n_{\downarrow \downarrow \beta} \right)  \lvert W_{\uparrow \downarrow \alpha \beta} \rvert^2 e^{i (\epsilon_\alpha - \epsilon_\beta)(t-\tau)} \,  ,
\end{equation}
\end{widetext}
where $W_{\uparrow \downarrow \alpha \beta}^* = W_{\downarrow \uparrow \beta \alpha}$. Similarly, we find that the contribution to the $cq$ component from the second and third terms of Eq.\ \eqref{S2b_four} is
\begin{equation}
S_{2b}^{cq} = \int dt \int d\tau \phi^{c} (t) D^A(t,\tau) \phi^{q}(\tau),
\end{equation}
where the advanced self-energy is
\begin{widetext}
\begin{equation}
\Sigma^A(t-\tau) =  - i\theta(\tau-t) \frac{2}{S} \sum_{\alpha \beta} \left( n_{\uparrow \uparrow \alpha} - n_{\downarrow \downarrow \beta} \right)  \lvert W_{\uparrow \downarrow \alpha \beta} \rvert^2 e^{i (\epsilon_\alpha - \epsilon_\beta)(t-\tau)} \, . 
\end{equation}
Finally, the contribution to the $qq$ component from the second and third terms of Eq.\ \eqref{S2b_four} is
\begin{equation}
S_{2b}^{qq} =  \int dt \int d\tau \bar{\phi}^q(t) \Sigma^K(t,\tau) \phi^q (\tau) \, ,
\end{equation}
where the Keldysh self-energy is
\begin{equation} 
\Sigma^K(t,\tau) = \frac{2i}{S} \sum_{\alpha \beta} \left(- 2n_{\uparrow \uparrow \alpha} n_{\downarrow \downarrow \beta} + n_{\uparrow \uparrow \alpha} + n_{\downarrow \downarrow \beta} \right) \lvert W_{\uparrow \downarrow \alpha \beta} \rvert^2 e^{i (\epsilon_\alpha - \epsilon_\beta) (t-\tau)} \, . 
\label{SigmaKtime}
\end{equation}
\end{widetext}
We note that the sum $i \Sigma^K(t-\tau)+ i \Sigma^K(\tau-t)$ and the difference $\Sigma^K(t-\tau) - \Sigma^K(\tau-t)$ are real numbers. These properties are crucial in ensuring that the fluctuating transverse fields acting on the magnetization are real numbers.

Collecting the $qc$, $cq$ and $qq$ components of $S_{2b}$, we see that the contributions from the second and third terms of Eq.\ \eqref{S2b_four} can be expressed as
\begin{equation}
S_{2b}^{\bar{c}q} + S_{2b}^{\bar{q}c} + S_{2b}^{\bar{q}q} = \int dt \int d\tau \left(\bar{\phi}^{c} \bar{\phi}^q \right) \Sigma
\left( 
\begin{array}{cc}
\phi^{c} \\
\phi^q
\end{array}
\right) \, , 
\label{S2223}
\end{equation}
where the self-energy is
\begin{equation}
\Sigma = 
\left( 
\begin{array}{cc}
0 & \Sigma^A \\
\Sigma^R & \Sigma^K
\end{array}
\right) \, . 
\label{Sigma}
\end{equation}

Using the Fourier transform of Eq.\ \eqref{Fourier}, we find that the Fourier transforms of the retarded and advanced self-energies are
\begin{align}
\Sigma^{R/A} =- \frac{2}{S} \sum_{\alpha \beta} \frac{n_{\uparrow \uparrow \alpha}-n_{\downarrow \downarrow \beta}}{\omega + \epsilon_\alpha - \epsilon_\beta \pm i 0^+} \lvert W_{\uparrow \downarrow \alpha \beta}\rvert^2 \, .
\label{sigmaRAomega}
\end{align}
The Fourier transform of the Keldysh component of the self-energy is
\begin{widetext}
\begin{equation} 
\Sigma^K = i \frac{4\pi}{S} \sum_{\alpha \beta} \delta(\omega + \epsilon_\alpha - \epsilon_\beta) \left(- 2n_{\uparrow \uparrow \alpha } n_{\downarrow \downarrow \beta} + n_{\uparrow \uparrow \alpha} + n_{\downarrow \downarrow \beta} \right) \lvert W_{\uparrow \downarrow \alpha \beta} \rvert^2  \, . 
\label{sigmaKomega}
\end{equation}
\end{widetext}
As we see, $\Sigma^K$ is purely imaginary:
\begin{align}
    \Sigma^K(\omega) = i \text{Im} \left[ \Sigma^K(\omega) \right] \, ,
    \label{SigmaKimaginary}
\end{align}
which is important because $ i \Sigma^K(\omega)$ controls magnetization fluctuations.
Furthermore, the imaginary part is positive definite. This feature is important because the self-energy appears as $\exp\left[- \text{Im}\Sigma(\omega) \lvert \phi^q(\omega)\rvert^2\right]$ in the partition function of Eq. \eqref{Z}. Consequently, the partition function is well defined.

Disregarding small terms of the order $\omega/\epsilon_F$, we compute from Eq. \eqref{sigmaRAomega} that
\begin{subequations}
\begin{align}
    \Sigma^R - \Sigma^A & = i \omega \frac{1}{4\pi S} 2 \left[A_{\uparrow \downarrow}^r + \tilde{A}_{\uparrow \downarrow}^r \right] \\
    & = i \omega 2 \left[\alpha^r + \tilde\alpha^r \right] \, ,
\end{align}
\end{subequations}
where $A_{\uparrow \downarrow}$ and $\tilde{A}_{\uparrow \downarrow}$ are defined in Eq. \eqref{WLLRRsum} and the spin transfer coefficients $\alpha$ and $\tilde{\alpha}$ are defined in Eq. \eqref{alpha}. Additionally, we have the symmetry
\begin{align}
    \Sigma^K(\omega) = \left[ \Sigma^K( - \omega)  \right]_{\uparrow \uparrow \leftrightarrow \downarrow \downarrow} \, . 
    \label{SigmaKomegasymmetry}
\end{align}
In other words, Eq. \eqref{SigmaKomegasymmetry} describes that flipping the spin accumulation along the magnetization and reversing the frequency have the same effect on the self-energy $\Sigma^K(\omega)$.

In general, we disregard small terms of the order $k_B T/\epsilon_F$ and $(\mu_\uparrow - \mu_\downarrow)/\epsilon_F$ and find from Eq. \eqref{sigmaKomega} that the self-energy can be expressed as Eq. \eqref{Sigmasigma}, where the noise matrix is defined as
\begin{equation}
    \sigma_{\kappa \lambda}= \frac{2\pi}{S} \sum_{nm} \lvert W_{\uparrow \downarrow \kappa n \lambda m} \rvert^2 \, , 
\end{equation}
and the matrix element $W_{\uparrow \downarrow \kappa n \lambda m}$ should be evaluated at the Fermi energy. We can relate $\sigma_{\kappa \lambda}$ to the scattering matrix, as shown in Appendix \ref{sec:eval}. On the basis of the lead (left or right), we find from Eq. \eqref{Wprobabilities} that $\sigma$ can be expressed as in Eq. \eqref{sigmaKscat}, where the spin transfer/spin pumping coefficients $\alpha$ and $\tilde{\alpha}$ are defined in Eq. \eqref{alpha} and the shot noise parameters $\beta^r$ and $\tilde{\beta}^r$ are defined in Eq. \eqref{betar}.

In the expression for the Keldysh component of the self-energy in Eq. \eqref{Sigmasigma}, the matrix elements of $\pi$ depend on the temperature, spin accumulation, and frequency as follows:
\begin{widetext}
\begin{align}
\pi_{\kappa \lambda}(\omega) = - 2 \int d\epsilon   \left[ 2 n_{\uparrow \uparrow \kappa} (\epsilon) n_{\downarrow \downarrow \lambda}(\epsilon + \omega) - n_{\uparrow \uparrow \kappa} (\epsilon)- n_{\downarrow \downarrow \lambda} (\epsilon + \omega) \right] \, , 
\end{align}
\end{widetext}
where the distributions $n_{\uparrow \uparrow}$ and $n_{\downarrow \downarrow}$ are defined in Eq. \eqref{ndist}. We compute $\pi^K$ at equilibrium, in the inelastic transport regime, and in the elastic transport regime and find Eq. \eqref{pi}.

At high temperatures, $k_B T \gg \lvert \mu_\uparrow - \mu_\downarrow \rvert$ and $k_B T \gg \omega$, $\pi_{\kappa \lambda}^\text{eq}=\pi_{\kappa \lambda}^\text{in}=\pi_{\kappa \lambda}^\text{el}=4 k_B T$, and we see from Eqs. \eqref{Sigmasigma} and \eqref{sigmaKscat} that the self-energy approaches its classical value, $\Sigma^K \rightarrow \Sigma^{K \text{cl}}$,
\begin{equation}
\Sigma^{K \text{cl}} = 4 k_B T  i \left( \alpha_{\uparrow \downarrow}^r + \tilde{\alpha}_{\uparrow \downarrow}^r \right) \, . 
\label{sigmaKcl}
\end{equation}

The advanced and retarded components of the self-energy \eqref{Sigma} can be expanded in terms of frequency $\omega$. To the zeroth order in frequency, the retarded and advanced components of the self-energy in \eqref{Sigma} appearing in action \eqref{S2223} only renormalize the magnon energy $E_z$, as discussed above for action $S_{2a}$. We disregard such terms since they have no consequences for the spin dynamics\cite{Chudnovskiy:PRL2008,Swiebodzinski:PRB2010}. Expanding the retarded and advanced self-energies in Eq. \eqref{sigmaRAomega} to the first order in frequency and maintaining the limit $\mu_\uparrow - \mu_\downarrow \ll \epsilon_F$, we find
\begin{align}
& \Sigma^{R/A}(\omega) - \Sigma^{R/A}(\omega=0) \nonumber \\
& \approx  \pm i\pi \omega \frac{2}{S} \sum_{\alpha \beta} \left( - \frac{\partial f}{\partial \epsilon} \right)_{\epsilon  =\epsilon_\alpha} \delta(\epsilon_\alpha-\epsilon_\beta)\lvert W_{\uparrow \downarrow \alpha \beta} \rvert^2 \nonumber \\
& = \pm i \omega \left[\alpha^r + \tilde{\alpha}^r \right],
\label{SigmaRA}
\end{align}
where we used the relation \eqref{optheorem} and the definition of the spin transfer/spin pumping efficiency \eqref{alpha}.

Summarizing this section, we find that the effective action that determines the spin dynamics is
\begin{equation}
\tilde{S}_m = S_m + S_{1}  + S_{2b}^d + S_{2b}^f \, , 
\end{equation}
where $S_1$ governs the spin transfer torque, 
\begin{equation}
S_{2b}^d= S_{2b}^{\bar{q}c}  + S_{2b}^{\bar{c}q}
\end{equation}
governs the spin pumping-enhanced dissipation (Gilbert damping), and
\begin{equation}
    S_{2b}^f=S_{2b}^{\bar{q}\bar{q}} + S_{2b}^{qq} + S_{2b}^{\bar{q} q}
\end{equation}
governs the fluctuations.

In the low-frequency limit, we use Eq. \eqref{SigmaRA} and can collect the dissipative contributions to the spin dynamics as follows:\footnote{We generalize to a complex spin pumping coefficient that is consistent with our finding for the spin transfer torque.}
\begin{align}
S_{2b}^{\bar{c}q} + S_{2b}^{c\bar{q}} & = \left( \alpha + \tilde{\alpha } \right)  \int dt \phi^q \partial_t \bar{\phi}^c \nonumber \\ 
+ & \left( \alpha^* + \tilde{\alpha}^* \right) \int dt  \bar{\phi}^q \partial_t \phi^c  \, . 
\end{align}
In analogy with the magnon action $S_m$ in Eq. \eqref{S_mKeldysh}, we see that actions $S_{2b}^{\bar{c}q} + S_{2b}^{c\bar{q}}$ contribute to the spin dynamics similar to an effective dissipative transverse field of the form
\begin{subequations}
\begin{align}
E_{d}^+  & = - \left( \alpha + \tilde{\alpha} \right) \partial_t S_+ / S \, , \\
E_{d}^- & = - \left( \alpha + \tilde{\alpha} \right)  \partial_t S^- / S \, , 
\end{align}
\end{subequations}
where we have used $\phi^c=S_+/\sqrt{S}$ in the semiclassical limit. Consequently, we find the spin pumping torques in Eq. \eqref{tauSP}.

\section{Stochastic Langevin Forces}
\label{Langevin}

This section derives the stochastic Langevin Forces from the effective action derived in the previous section \ref{closed}. We will find that fluctuating magnetic fields describe the Langevin forces. 

To see the effects of the actions that are quadratic in the quantum components of the boson fields, $\phi^q$ and $\bar{\phi}^q$, we use the Gaussian integral over the real variables ${h_i}$:
\begin{widetext}
\begin{equation}
\exp{\left[ - \frac{1}{2} \sum_{ij} \phi_i \left( a \right)_{ij} \phi_j \right]}  = 
\frac{1}{\sqrt{\text{det}(a)}} \int \Pi_{j}  \left( \frac{dh_j}{\sqrt{2\pi}} \right) \exp{\left[  \sum_i i h_i \phi_i - \frac{1}{2} \sum_{ij } h_i \left( a^{-1} \right)_{ij} h_j \right]} \, ,  
\label{realGaussian}
\end{equation}
assuming that the integral converges. Additionally, we use the Gaussian integral over the complex variable $z=z^r + i z^i$:
\begin{equation}
\exp{ \left[ - \sum_{ij} \phi_i b _{ij} \phi_j \right]} 
= \frac{1}{\text{det}(b)}
\int \Pi_{j}  D[\bar{z}_j, z_j] \exp{ \left[ \sum_i i \bar{z}_i \phi_i + z_i \bar{\phi}_i- \sum_{ij } \bar{z}_i \left( b^{-1} \right)_{ij} z_j \right]}  \, , 
\label{complexGaussian}
\end{equation}
\end{widetext}
assuming that the integral converges and where $D[\bar{z},z]= d z^R d zI/\pi$.

We will also use the results for the expectation value and fluctuations of the real variables $h$ with the distribution determined by matrix $a$ as in the right-hand side of Eq. \eqref{realGaussian} with $\phi=0$, for example,
\begin{subequations}
\begin{align}
\langle h_l \rangle & = 0 \, , \\
\langle h_l h_m \rangle & = a_{ml} \, . 
\end{align}
\label{hexpfluc}
\end{subequations}
We note that although the variables $h_i$ appearing in Eq.\ \eqref{realGaussian} are real, the variance in Eq.\ \eqref{hexpfluc} governed by complex distribution $a$ in Eq.\ \eqref{realGaussian} can be complex.

Similarly, the expectation value and fluctuations of the complex variables $z$ with the distribution as in the right-hand side of Eq.\ \eqref{complexGaussian} with $\phi=0$ are
\begin{subequations}
\begin{align}
\langle \bar{z}_l \rangle & = 0 \, , \\
\langle z_l z_m \rangle & = 0 \, , \\
\langle \bar{z}_l z_m \rangle & = b_{ml} \, . 
\end{align}
\end{subequations}

By using the Gaussian integral over the real variables \eqref{realGaussian}, which is a Hubbard-Stratonovich decoupling scheme, we separately remove the quadratic terms in actions $S_{sb}^{\bar{q}\bar{q}}$ and $S_{2b}^{qq}$  with two independent and real auxiliary fields $h^{\bar{q}\bar{q}}(t)$ and $h^{qq}(t)$. For example, we use
\begin{widetext}
\begin{align}
    \exp{i S_{2b}^{qq}} & = \exp{i \int dt \int d\tau \phi^q(t) \tilde{\Sigma}^K_{\uparrow \downarrow}(t-\tau) \phi^q(\tau)} \\
    & = \frac{1}{\sqrt{\text{det}\left[-2 i \tilde{\Sigma}_{\uparrow \downarrow}^K \right]}} \int \Pi_j \frac{d h_{qq}(t_j)}{\sqrt{2\pi}} 
    \exp{\sum_i i h_{qq}(t_i) \phi^q(t_i) - \frac{1}{2} \sum_{ij} h_{qq}(t_i) (-2 i \tilde{\Sigma}_{\uparrow \downarrow}^K)^{-1}(t_i,t_j) h_{qq}(t_j) } \, ,
\end{align}
\end{widetext}
and we treat the effect of the contribution $\exp{i S_{2b}^{\bar{q} \bar{q}}}$ to the partition function in a similar manner. Then, the expectation values and correlations of the auxiliary fields in our action satisfy
\begin{subequations}
\begin{align}
\langle h^{qq}(t) \rangle & =0 \, , \\
\langle h^{qq}(t) h^{qq}(\tau) \rangle & = - 2 i  \tilde{\Sigma}^K_{\uparrow \downarrow}(t-\tau)
\end{align}
\end{subequations}
and
\begin{subequations}
\begin{align}
\langle h^{\bar{q}\bar{q}}(t) \rangle & = 0
 \, , \\
\langle h^{\bar{q}\bar{q}}(t) h^{\bar{q}\bar{q}}(\tau) \rangle & = - 2 i \tilde{\Sigma}^K_{\downarrow \uparrow} (t-\tau) \, .
\end{align}
\label{realhcorr}
\end{subequations}
Similarly, by using the Gaussian integral over complex variables \eqref{complexGaussian}, we remove the quadratic terms in action $S_{2b}^{\bar{q}q}$: 
\begin{widetext}
\begin{align}
    \exp{i S_{2b}^{\bar{q}q}} & = \exp{i \int dt \int d\tau \bar{\phi}^q(t) \Sigma^K_{\uparrow \downarrow}(t-\tau) \phi^q(\tau)} \\
    & = \frac{1}{\text{det}\left[- i \Sigma^K \right]} \int \Pi_j \frac{d h_{qq}(t_j)}{\sqrt{2\pi}} 
    \exp{\sum_i i \bar{h}_{\bar{q}q}(t_i) \phi^q(t_i) + i h_{\bar{q}q}(t_i) \bar{\phi}^q(t_i) - \sum_{ij} \bar{h}_{\bar{q}q}(t_i) (- i \Sigma^K)^{-1}(t_i,t_j) h_{\bar{q}q}(t_j) } \, .
\end{align}
\end{widetext}
The complex auxiliary field $h^{\bar{q}q}(t)$ obeys the statistics
\begin{subequations}
\begin{align}
\langle h^{\bar{q}q}(t) \rangle & = 0  \, , \\
\langle h^{\bar{q}q} ( t) h^{\bar{q} q} (\tau) \rangle  & = 0 \, , \\
\langle \bar{h}^{\bar{q}q} ( t) h^{\bar{q} q} (\tau) \rangle  & =  - i \Sigma^K (t-\tau) \, .
\end{align} 
\label{complexhcorr}
\end{subequations}
Note the essential difference of a factor of 2 in the correlations between Eq. \eqref{realhcorr} and Eq. \eqref{complexhcorr}, which is essential in fulfilling the fluctuation-dissipation theorem for the magnetic fluctuations at equilibrium.

As a result of the Hubbard-Stratonovich decoupling scheme, we obtain an effective action that is linear in the boson fields $\phi^q$ and $\bar{\phi}^q$ and depends on the auxiliary and stochastic temporal fields $h^{\bar{q}\bar{q}}$, $h^{qq}$, and $h^{\bar{q}q}$.

The fluctuating fields then effectively contribute to the action as follows:
\begin{align}
    S_{2b}^f & = \int dt \left[ \bar{h}^{\bar{q}q}(t) + h^{qq}(t) \right] \phi^q(t) \nonumber \\
    & + \int dt \left[ h^{\bar{q}q}(t) + h^{\bar{q}\bar{q}}(t) \right] \bar{\phi}^q(t) \, .
\end{align}
From Eq.\ \eqref{S_mKeldysh}, we see that the fluctuating fields correspond to transverse magnetic fields represented by effective energies
\begin{subequations}
\begin{align}
    E_-^{f} & = -\frac{1}{\sqrt{S}} \left[ \bar{h}^{\bar{q}q}(t) + h^{qq}(t) \right] \, , \\
    E_+^{f} & = -\frac{1}{\sqrt{S}} \left[ h^{\bar{q}q}(t) + h^{\bar{q}\bar{q}}(t) \right] \, . 
\end{align}
\end{subequations}
Using the relation between effective energies and effective transverse magnetic fields, as we introduced in Eq. \eqref{Epm}, the effective transverse fluctuating fields are
\begin{subequations}
\begin{align}
 H_x & = \frac{1}{2 \sqrt{S} g^* \mu_B} \left[ h^{\bar{q}q} + \bar{h}^{\bar{q}q} + h^{qq}+ h^{\bar{q}\bar{q}} \right] \, , \\
 H_y & = \frac{1}{i 2 \sqrt{S} g^* \mu_B} \left[ h^{\bar{q}q} - \bar{h}^{\bar{q}q} -  h^{qq}+ h^{\bar{q}\bar{q}} \right] \, . 
\end{align}
\end{subequations}
By using the correlations of the auxiliary fields in Eqs.\ \eqref{realhcorr} and \eqref{complexhcorr} and the symmetries of the self-energies in Eqs. \eqref{tildeSigmaKeven} and \eqref{SigmaKimaginary}, we find that the correlations between the fluctuating fields are
\begin{widetext}
\begin{subequations}
\begin{align}
\langle \gamma H_x(t) \gamma H_x(\tau) \rangle &= - \frac{i}{ 4 S}  \left[  \Sigma^K(t-\tau) + \Sigma^K(\tau - t) \right] + \frac{1}{S} \text{Im} \tilde{\Sigma}^K_{\uparrow \downarrow} (t-\tau)  \, , \\
\langle \gamma H_y(t) \gamma H_y(\tau) \rangle &= - \frac{i}{ 4 S}  \left[  \Sigma^K(t-\tau) + \Sigma^K(\tau - t) \right]  - \frac{1}{S} \text{Im} \tilde{\Sigma}^K_{\uparrow \downarrow} (t-\tau)  \, , \\
\langle \gamma H_x(t) \gamma H_y(\tau) \rangle &= - \frac{1}{ 4 S}  \left[  \Sigma^K(t-\tau) - \Sigma^K(\tau - t)  \right] + \frac{1}{S} \text{Re} \tilde{\Sigma}^K_{\uparrow \downarrow} (t-\tau). 
\end{align}
\label{fluccorr}
\end{subequations}
\end{widetext}
From the earlier derived relations in this section that $i ( \Sigma^K(t-\tau) + \Sigma^K(\tau - t) ) $ and $( \Sigma^K(t-\tau) - \Sigma^K(\tau - t) ) $ are real numbers and $\tilde{\Sigma}^K_{\uparrow \downarrow} (t-\tau) $ is an even function of time, we see that all correlations between the fluctuating fields are real numbers and that the correlations \eqref{fluccorr} are even functions of the time difference $t-\tau$, as should be expected. Furthermore, we see from Eq. \eqref{SigmaKtime} that $ - \Sigma^K(t-\tau)  \ge 0 $, as expected for the equal time correlations.

By Fourier transforming Eq. \eqref{fluccorr}, we find the correlations introduced in Eq. \eqref{fluccorrFourier}.

%
%

\section{Scattering Formulation of the Electron-Magnon Interaction}
\label{sec:scattering}

This section presents how the electron-magnon interaction can be expressed in terms of scattering matrices of electron transport. In the classical ground state of the magnet, the itinerant electron contribution to the Hamiltonian \eqref{H} is
\begin{equation}
\hat{H}_e = \int d{\bf r} \hat{\psi}^\dag \left[ H_0 + V_s \sigma_z \right] \hat{\psi}  \, . 
\label{H_e}
\end{equation}
From the Hamiltonian \eqref{H}, by using the Holstein-Primakoff transformation \eqref{HP}, the electron-magnon coupling is
\begin{align}
H_{em} & =  b^\dag \sqrt{\frac{2}{S}} \int d{\bf r} \hat{\psi}_\uparrow V_s \hat{\psi}_\downarrow +  b \sqrt{\frac{2}{S}} \int d{\bf r} \hat{\psi}_\downarrow V_s \hat{\psi}_\uparrow  \nonumber \\
& - \hat{b}^\dag \hat{b} \frac{1}{S} \left[ \int d{\bf r} \hat{\psi}_\uparrow V_s \hat{\psi}_\uparrow -  \int d{\bf r} \hat{\psi}_\downarrow V_s \hat{\psi}_\downarrow \right] \, .
\label{H_em}
\end{align}
We use scattering states to express the itinerant electron field operators. As a first representation, the field operators can be expressed as
\begin{equation}
\hat{\psi}_s  = \int_0^\infty \frac{dk}{\sqrt{2\pi}} \sum_{\kappa n} \psi_{s \kappa nk}(x{\bm \rho}) \hat{a}_{s\kappa nk} \, , 
\label{efield}
\end{equation}
where $\hat{l}_{s\kappa nk}$ annihilates an electron with spin $s$ in reservoir $\kappa$ ($\kappa=L$ or $\kappa=R$) with transverse waveguide mode $n$ and longitudinal momentum $k$. The associated spatially dependent wave function is $\psi_{s\kappa nk}(x {\bm \rho})$, where $x$ is the longitudinal coordinate and ${\bm \rho}$ denotes the transverse coordinates. The field operators satisfy the anti-commutation relations
\begin{align}
\left\{ \hat{a}_{s \kappa nk},\hat{a}^\dag_{t \lambda mp} \right\} & = \delta_{st} \delta_{\kappa \lambda} \delta_{nm} \delta(k-p) \, .
\end{align}
From Eq.\ \eqref{H_e}, we identify the spin-dependent single-particle electron Hamiltonian $H_s$, where $H_\uparrow=H_0 + V_s$ and $H_\downarrow = H_0 - V_s$. In general, in the lead and the scattering region, the Schr\"{o}dinger equation associated with an incoming electron from the left or right is $H_s \psi_{s \kappa nk}(x {\bm \rho}) = \epsilon_{nk} \psi_{s \kappa nk}(x {\bm \rho})$. In the following, it is essential that the single-particle energy $\epsilon_{nk}$ is spin degenerate because the leads are paramagnetic, i.e., the potential $V_s$ vanishes in the leads.

Expressing the field operator \eqref{efield} and the scattering states in terms of the total energy of the states $\epsilon=\epsilon_n + \epsilon_k$ by substituting $dk \rightarrow d\epsilon (\partial \epsilon/dk)^{-1}$, $\psi_{s \kappa nk} \rightarrow \psi_{s \kappa n}(\epsilon) (\partial \epsilon/dk)^{1/2}$, and $\hat{a}_{s \kappa nk} \rightarrow \hat{a}_{s \kappa n}(\epsilon) (\partial \epsilon/dk)^{1/2}$ is more convenient. A second representation of the field operator is therefore
\begin{equation}
\hat{\psi}_s = \frac{1}{\sqrt{2\pi}} \sum_{\kappa n} \int_{\epsilon_n}^\infty d\epsilon \psi_{s\kappa n}(\epsilon x {\bm \rho}) \hat{a}_{s \kappa n}(\epsilon) \, . 
\label{fieldscat}
\end{equation}
On the basis of the scattering states in the total energy representation \eqref{fieldscat}, using the orthogonality of the single-particle states and assuming that the system size is much larger than the Fermi wavelength, the electronic Hamilton operator becomes Eq.\ \eqref{H_escat}.

In terms of scattering states \eqref{fieldscat}, the electron-magnon interaction contains first- and second-order terms in the magnon operators, $\hat{H}_{em}= \hat{H}_{1}+\hat{H}_{2}$. Using Eq.\ \eqref{H_em}, we see that the first-order terms in the magnon operators are expressed as in Eq.\ \eqref{H1} and the second-order terms in the magnon operators are given by Eq.\ \eqref{H2}, where the matrix elements are
\begin{equation}
W_{\uparrow \downarrow \alpha \beta} = \frac{1}{2\pi}\int d{\bf r} \psi_{\uparrow \alpha}^* V_s \psi_{\downarrow \beta} \, , 
\label{Wupdown}
\end{equation}
$W_{\downarrow \uparrow \beta \alpha}= (W_{\uparrow \downarrow \alpha \beta})^*$, and
\begin{equation}
W_{\sigma\sigma \alpha \beta} = \frac{1}{2\pi} \int d{\bf r} \psi_{\sigma \alpha}^* V_s \psi_{\sigma \beta} \, . 
\label{Wsigmasigma}
\end{equation}
We will demonstrate in Appendix \ref{sec:eval} that the matrix elements $W_{\uparrow \downarrow \alpha \beta}$ and $W_{\downarrow \uparrow \beta \alpha}$ are related to a generalized {\it inelastic} (since the energy is not conserved) $2 \times 2$ transverse (mixing) conductance. However, the final results for the spin dynamics demonstrate that they nevertheless only relate to the elastic $2 \times 2$ transverse (mixing) conductance, as expected since the magnon frequency is tiny compared to the Fermi energy. Furthermore, we have demonstrated in the previous sections that the matrix elements $W_{\sigma\sigma \alpha \beta}$ diagonal in the spin indices are irrelevant to the spin dynamics. Only spin-flip events are relevant in the semiclassical limit $S \gg 1$.

\section{Conclusion}
\label{sec:conclusion}

We have presented a quantum scattering theory for spin transport and magnetization dynamics in normal metal-ferromagnet systems. The path-integral formalism generalizes earlier semiclassical results to the quantum low-temperature regime and provides a valuable connection to the scattering formalism of electron transport. At low temperatures, the chemical potential in the normal metals controls quantum contributions to the magnetic noise. Some of these contributions are related to shot noise because of the discrete spin angular momentum of electrons. These quantum contributions dominate when the temperature is lower than the spin or charge chemical potentials.

\appendix

\section{Evaluation of Spin-Flip Matrix Elements}
\label{sec:eval}

In this section, we demonstrate how the matrix elements $W_{\uparrow \downarrow \alpha \beta} $ \eqref{Wupdown} are governed by the scattering matrix, i.e., the reflection and transmission amplitudes. This implies that the matrix elements are proportional to the surface area between the ferromagnet and the normal metal instead of the volume of the ferromagnet, as one might naively expect from the definition \eqref{Wupdown}.

In the lead, $V_s$ vanishes, and $H_0$ decomposes into longitudinal and transverse parts as follows:
\begin{equation}
H_0 = - \frac{1}{2m} \frac{\partial^2}{\partial x^2} + H_{\perp}({\bm \rho}) \, ,
\end{equation}
so the transverse wave function $\phi_{n}$ obeys the Schr\"{o}dinger equation:
\begin{equation}
H_{\perp}({\bm \rho}) \phi_{n}({\bm \rho})= \epsilon_{n} \phi_{n}({\bm \rho}) \, ,
\end{equation}
where $\epsilon_{n}$ is the transverse component of the single-particle energy 
 $\epsilon_{nk}=\epsilon_k + \epsilon_{n}$, where $\epsilon_k= k^2/2m$.

We introduce the energy $\epsilon$-dependent reflection and transmission matrices $r$, $t$, $\tilde{r}$, and $\tilde{t}$ that constitute the S-matrix:
\begin{equation}
S =\left( 
\begin{array}{cc}
r & \tilde{t} \\
t & \tilde{r} 
\end{array}
\right) \, ,
\label{Smatrix}
\end{equation}
which should not be confused with the total spin of the magnet $S$. The S-matrix obeys the unitarity relations
\begin{subequations}
\begin{align}
S S^\dag & = 1 \, , \\
S^\dag S & = 1 \, . 
\end{align}
\label{Sunitarity}
\end{subequations}

In the left lead, when $x<x_L$, the scattering states at energy $\epsilon$ are
\begin{subequations}
\begin{align}
\psi_{sLn} & = 
\sum_m \frac{\phi_m({\bm \rho}) }{\sqrt{v_m}}
\left[  \delta^{mn} e^{ i k_n \tilde{x}_L} +  r_{s}^{mn} e^{- i k_m \tilde{x}_L} \right] \,  , \\
\psi_{sRn} & = 
\sum_m \frac{\phi_m({\bm \rho})}{\sqrt{k_m}} 
\tilde{t}_{s}^{mn} e^{- i k \tilde{x}_L}  \, , 
\end{align}
\label{asymptleft}
\end{subequations}
where $v_m= \hbar k_m/m$, the longitudinal wavevector $k_m$ is determined by $\hbar^2 k_n^2/2m+\epsilon_n=\epsilon$, and $\tilde{x}_L=x-x_L$ is a local coordinate relative to the scattering region containing the ferromagnet that originates at $x=x_L$ and ends at $x=x_R$. Similarly, in the right lead, when $x>x_R$, the scattering states are
\begin{subequations}
\begin{align}
\psi_{sLn} & = \sum_m \frac{\phi_m({\bm \rho}) }{\sqrt{k_m}} t_{s}^{mn} e^{i k \tilde{x}_R} \, , \\
\psi_{sRn} & = 
\sum_m \frac{\phi_m({\bm \rho})}{\sqrt{k_m}} \left[  \delta^{mn} e^{ - i k \tilde{x}_R} +  \tilde{r}_{s}^{mn} e^{ i k \tilde{x}_R}\right] \, , 
\end{align}
\label{asymptright}
\end{subequations}
where $\tilde{x}_R=x-x_R$ is the local coordinate in the right lead.

Using $V_s=(H_\uparrow - H_\downarrow)/2$, partial integration of the kinetic energy along the transport direction, and the fact that the single-particle scattering state energies are spin degenerate, we find that
\begin{align}
W_{\uparrow \downarrow \alpha \beta} & =  \frac{1}{4\pi}   \int_{x_L}^{x_R} \! \! \! \! \! \! dx \int \! \! d{\bm \rho}\left( \psi_{\uparrow \alpha}^* H_\uparrow \psi_{\downarrow \beta} -  \psi_{\uparrow \alpha}^* H_\downarrow \psi_{\downarrow \beta} \right)  \nonumber \\
& =  \frac{1}{4\pi i}  \left[ j_{\uparrow \downarrow \alpha \beta} (x=x_R)-j_{\uparrow \downarrow \alpha \beta} (x=x_L) \right] \, .
\label{Wudcurrent}
\end{align}
We have introduced the off-diagonal component of the spin current in the transport direction:
\begin{equation}
j_{\uparrow \downarrow \alpha \beta} (x)= \frac{1}{2mi} \int {d {\bm \rho}} \left[ \psi_{\uparrow \alpha \beta}^* \frac{\partial \psi_{\downarrow \beta}}{\partial x}   -   \frac{\partial \psi_{\uparrow \alpha}^*}{\partial x}  \psi_{\downarrow \beta}\right] \, . 
\label{Icurperp}
\end{equation}
The transverse component of the spin current with respect to the equilibrium orientation of the magnetization flowing in the transport direction \eqref{Icurperp} can be computed from the asymptotic behavior of the wave functions in Eqs.\ \eqref{asymptleft} and \eqref{asymptright}.

We can now evaluate the spin current in terms of the scattering matrix. In the final results for the effective magnon action, only matrix elements at the Fermi energy are needed. For $RR$ components, we find
\begin{subequations}
\begin{align}
j_{\uparrow \downarrow RR}^{nnm}(x_R) 
& =-  \left( 1-\tilde{r}_{\uparrow}^{\dag} \tilde{r}_{\downarrow } 
 \right)^{nm} \, , \\
j_{\uparrow \downarrow RR}^{nnm}(x_L) & = - 
 \left( \tilde{t}_{\uparrow k}^{\dag} \tilde{t}_{\downarrow } \right)^{nm} \,  ,
 \end{align}
 \label{jRR}
 \end{subequations}
 the $LL$ components are 
 \begin{subequations}
 \begin{align}
 j_{\uparrow\downarrow LL}^{nm}(x_R) 
&
 =  \left( 
 t_{\uparrow k}^{\dag} t_{\downarrow k} \right)^{nm} \, , \\
 j_{\uparrow\downarrow LL}^{nm}(x_L) & = 
 \left(1-r_{\uparrow}^{\dag} r_{\downarrow} 
 \right)^{nm}  \, , 
 \end{align}
 \label{jLL}
 \end{subequations}
the $LR$ components are 
\begin{subequations}
\begin{align}
j_{\uparrow \downarrow LR}^{nm}(x_R) & = \left( t_\uparrow^\dag \tilde{r}_\downarrow \right)^{nm} \, ,  \\
j_{\uparrow \downarrow LR}^{nm}(x_L) & =  - \left(r_\uparrow^\dag \tilde{t}_\downarrow \right)^{nm} \, , 
\end{align}
\label{jLR}
\end{subequations}
and the $RL$ components are 
\begin{subequations}
\begin{align}
j_{\uparrow \downarrow RL}^{nn}(x_R) & = \left(\tilde{r}_\uparrow^\dag t_\downarrow \right)^{nm} \, , \\
j_{\uparrow \downarrow RL}^{nn}(x_L) & = - \left(\tilde{t}_\uparrow^\dag r_\downarrow \right)^{nm}
\, . 
\end{align}
\label{jRL}
\end{subequations}
In the expressions \eqref{jRR}, \eqref{jLL}, \eqref{jLR}, and\eqref{jRL}, the superscript $nm$ denotes the matrix element, and matrix multiplication of the reflection and transmission coefficients is implied.

Consequently, we find from Eqs.\ \eqref{Wudcurrent}, \eqref{jRR}, \eqref{jLL}, \eqref{jLR}, and \eqref{jRL}
\begin{subequations}
\begin{align}
W_{\uparrow \downarrow LL}^{nm} & = i \frac{1}{4\pi} A_{\uparrow \downarrow}^{nm} \, ,
\label{Wud_LL} \\
W_{\uparrow \downarrow RR}^{nm} & = i \frac{1}{4\pi} \tilde{A}_{\uparrow \downarrow}^{nm} \, ,
\label{Wud_RR} 
\end{align}
\label{Wud_LLRR}
\end{subequations}
where
\begin{align}
A_{\uparrow \downarrow}^{nm} &= g_{r}^{nm} - g_t^{nm}  \, , \\
\tilde{A}_{\uparrow \downarrow}^{nm} & = g_{\tilde{r}}^{nm} - g_{\tilde{t}}^{nm} \, , 
\label{Aupdown}
\end{align}
and we have introduced the dimensionless partial conductance matrix elements
\begin{subequations}
\begin{align}
g_{r}^{nm} & = \left( 1 - r_{\uparrow }^\dag r_\downarrow \right)^{nm} \, ,  \\
g_{\tilde{r}}^{nm} & = \left( 1 - \tilde{r}_{\uparrow }^{\dag} \tilde{r}_\downarrow \right)^{nm} \, ,  \\
g_{t}^{nm} & =  \left( t_{\uparrow }^{\dag} t_\downarrow \right)^{nm} \, , \\
g_{\tilde{t}}^{nm} & = \left( \tilde{t}_{\uparrow }^{\dag} \tilde{t}_\downarrow \right)^{nm} \, .
\end{align}
\end{subequations}

We also define\cite{Tserkovnyak:PRL2002}
\begin{subequations}
\begin{align}
A_{\uparrow \downarrow} & =  \sum_{nn} W_{\uparrow \downarrow LL}^{nn} = g_r - g_t \, ,\label{Aupdo} \\
\ \tilde{A}_{\uparrow \downarrow} & = \sum_{nn} W_{\uparrow \downarrow RR}^{nn} = g_{\tilde{r}} - g_{\tilde{t}}  \, , \label{Atildeupdo} 
\end{align}
\label{WLLRRsum}
\end{subequations}
where the dimensionless transverse (mixing) conductances are
\begin{subequations}
\begin{align}
g_{r} & = \text{Tr} \left( 1 - r_{\uparrow }^\dag r_\downarrow \right) \, ,  \\
g_{\tilde{r}} & = \text{Tr} \left( 1 - \tilde{r}_{\uparrow }^{\dag} \tilde{r}_\downarrow \right) \, ,  \\
g_{t} & =  \text{Tr} \left( t_{\uparrow }^{\dag} t_\downarrow \right) \, , \\
g_{\tilde{t}} & = \text{Tr} \left( \tilde{t}_{\uparrow }^{\dag} \tilde{t}_\downarrow \right) \, .
\end{align}
\end{subequations}

The spin transfer/spin pumping coefficients are defined as
\begin{subequations}
\begin{align}
    \alpha & = \frac{A_{\uparrow \downarrow}}{4 \pi S} \, , \\
    \tilde{\alpha} & = \frac{\tilde{A}_{\uparrow \downarrow}}{4 \pi S} \, ,
\end{align}
\end{subequations}
which can be expressed as in Eq. \eqref{alpha}.

Additionally, we find that
\begin{subequations}
\begin{align}
W_{\uparrow \downarrow LR}^{nm} & = - i \frac{1}{4\pi} \left(r_\uparrow^\dag \tilde{t}_\downarrow + t_{\uparrow}^\dag \tilde{r}_\downarrow \right)^{nm} \, , \\
W_{\uparrow \downarrow RL}^{nm} & = - i \frac{1}{4\pi} \left(\tilde{t}_\uparrow^\dag r_\downarrow + \tilde{r}_{\uparrow}^\dag t_\downarrow \right)^{nm} \, .
\end{align}
\label{WLRRL}
\end{subequations}

By using Eq.\ \eqref{Wud_LLRR}, Eq.\ \eqref{WLRRL}, and the unitarity of the S-matrix \eqref{Sunitarity}, we also find that
\begin{subequations}
\begin{align}
\sum_{nm} \lvert W_{\uparrow \downarrow LL}^{nm} \rvert^2  & = \left( \frac{1}{4\pi} \right)^2 2 \left( A_{\uparrow \downarrow}^r - B_{\uparrow \downarrow}^r \right) \, , \\
\sum_{nm} \lvert W_{\uparrow \downarrow RR}^{nm} \rvert^2  & = \left( \frac{1}{4\pi} \right)^2 2 \left( \tilde{A}_{\uparrow \downarrow}^r - \tilde{B}_{\uparrow \downarrow}^r \right) \, , \\
\sum_{nm} \lvert W_{\uparrow \downarrow LR}^{nm} \rvert^2  & = \left( \frac{1}{4\pi} \right)^2 2 B_{\uparrow \downarrow}^r \, , \\
\sum_{nm} \lvert W_{\uparrow \downarrow RL}^{nm} \rvert^2  & = \left( \frac{1}{4\pi} \right)^2 2 \tilde{B}_{\uparrow \downarrow}^r \, ,
\end{align}
\label{Wprobabilities}
\end{subequations}
where the shot noise parameters are
\begin{subequations}
\begin{align}
B_{\uparrow \downarrow}^r & = \frac{1}{2}  \text{Tr} \left[1 - \left(r_\uparrow^\dag r_\downarrow +t_\uparrow^\dag t_\downarrow  \right) \left(r_\downarrow^\dag r_\uparrow +t_\downarrow^\dag t_\uparrow  \right)\right] \, , \\
\tilde{B}_{\uparrow \downarrow}^r & =  \frac{1}{2}  \text{Tr}\left[1 - \left(\tilde{r}_\uparrow^\dag \tilde{r}_\downarrow +\tilde{t}_\uparrow^\dag \tilde{t}_\downarrow  \right) \left(\tilde{r}_\downarrow^\dag \tilde{r}_\uparrow +\tilde{t}_\downarrow^\dag \tilde{t}_\uparrow  \right)\right]
\end{align}
\label{Bupdo}
\end{subequations}
and $A_{\uparrow \downarrow}$ was introduced in Eq.\ \eqref{Aupdown}.

The noise matrix parameters are
\begin{align}
    \sigma = \frac{1}{4 \pi S}
    \left( 
\begin{array}{cc}
A_{\uparrow \downarrow}^r - B_{\uparrow \downarrow}^r & B_{\uparrow \downarrow}^r \\
\tilde{B}_{\uparrow \downarrow}^r & \tilde{A}_{\uparrow \downarrow}^r - \tilde{B}_{\uparrow \downarrow}^r 
\end{array}
\right) \, ,
\end{align}
which can be expressed as in Eq. \eqref{sigmaKscat}.

From the definition of the absolute squares of the matrix elements \eqref{Wprobabilities}, we see that the following conditions are fulfilled:
\begin{subequations}
\begin{align}
A_{\uparrow \downarrow}^r & \ge 0 \text{ , }  \tilde{A}_{\uparrow \downarrow}^r  \ge 0 \, , \\
B_{\uparrow \downarrow}^r & \ge 0 \text{ , } \tilde{B}_{\uparrow \downarrow} ^r \ge 0 \, , \\
A_{\uparrow \downarrow}^r & \ge B_{\uparrow \downarrow}^r, \text{ , } \tilde{A}_{\uparrow \downarrow}^r  \ge  \tilde{B}_{\uparrow \downarrow}^r \, .
\end{align}
\end{subequations}
Hence, all the matrix elements in Eq. \eqref{betar} are positive definite, as expected for fluctuations.

We can also find the sum rule
\begin{subequations}
\begin{align}
\sum_{nm} \left[
\lvert W_{\uparrow \downarrow LL}^{nm} \rvert^2 + \lvert W_{\uparrow \downarrow LR}^{nm} \rvert^2  \right] &= \left( \frac{1}{4\pi} \right)^2 2 A^r_{\uparrow \downarrow} \, , \\
\sum_{nm} \left[ 
\lvert W_{\uparrow \downarrow RR}^{nm} \rvert^2 + \lvert W_{\uparrow \downarrow RL}^{nm} \rvert^2  \right] & = \left( \frac{1}{4\pi} \right)^2 2 \tilde{A}^R_{\uparrow \downarrow} \, , 
\end{align} 
\label{optheorem}
\end{subequations}
which we classify as an optical theorem for the transport of spins transverse to the magnetization direction. The optical theorem \eqref{optheorem} relates sums of reflection and transmission probabilities to the real part of the reflection and transmission amplitudes.

We also evaluate
\begin{align}
\sum_{nm}  W_{\downarrow \uparrow LL}^{nm}  W_{\downarrow \uparrow LL}^{mn}     &= -  \left( \frac{1}{4\pi} \right)^2 2  \left( A_{\downarrow \uparrow} - C_{\downarrow \uparrow} \right) \, , 
\end{align}
where
\begin{equation}
A_{\downarrow \uparrow} =  A_{\uparrow \downarrow}^* \, , 
\end{equation}
$A_{\uparrow \downarrow}$ was defined in Eq.\ \eqref{Aupdo}, and
\begin{equation}
C_{\downarrow \uparrow} = \frac{1}{2} \text{Tr} \left[1 - \left( r_\downarrow^\dag r_\uparrow + t_\downarrow^\dag t_\uparrow \right)^2 \right] \, . 
\end{equation} 
We note that in general, $C_{\downarrow \uparrow} \ne B_{\downarrow \uparrow}$. Furthermore, we find
\begin{equation}
\sum_{nm} W_{\downarrow \uparrow LR }^{nm} W_{\downarrow \uparrow RL}^{mn} = - \left( \frac{1}{4\pi} \right)^2 2 D_{\downarrow \uparrow} \, , 
\end{equation}
where
\begin{equation}
D_{\downarrow \uparrow} = \text{Tr} \left[ \left(r_\downarrow^\dag \tilde{t}_\uparrow + t_\uparrow^\dag \tilde{r}_\uparrow \right) \left( \tilde{t}_\downarrow^\dag r_\uparrow + \tilde{r}_\downarrow^\dag t_\uparrow \right) \right] \, .
\end{equation}
We also find
\begin{equation}
\sum_{nm} W_{\downarrow \uparrow RL }^{nm} W_{\downarrow \uparrow LR}^{mn} = - \left( \frac{1}{4\pi} \right)^2 2 D_{\downarrow \uparrow} \, , 
\end{equation}
and
\begin{equation}
\sum_{nm} W_{\downarrow \uparrow RR }^{nm} W_{\downarrow \uparrow RR}^{mn} = - \left( \frac{1}{4\pi} \right)^2 2 \left(\tilde{A}_{\downarrow \uparrow} - \tilde{C}_{\downarrow \uparrow} \right)\, .
\end{equation}
We use these results to compute the transverse spin accumulation noise matrix elements as defined in Eq. \eqref{tildesigmadef}:
\begin{align}
    \tilde{\sigma} = \frac{1}{8\pi S} \left(
    \begin{array}{cc}
    A_{\downarrow \uparrow} - C_{\downarrow \uparrow} & D_{\downarrow \uparrow} \\
    D_{\downarrow \uparrow} & \tilde{A}_{\downarrow \uparrow} - \tilde{C}_{\downarrow \uparrow} 
    \end{array}
    \right) \, , 
\end{align}
which can be expressed as in Eq. \eqref{tildesigmamatrix}.

\section{Transport Coefficients}
\label{sec:trancoeff}

The transport coefficients can be computed by ab initio band structure calculations. This has been carried out for the spin transfer and spin pumping efficiencies reflected in the mixing conductances in Eq.\ \eqref{alpha} in Ref. \onlinecite{Zwierzycki:PRB2005}.

Here, we consider some scenarios for electron transport.

\subsection{Magnetic Insulator}

In the case of a magnetic insulator, $t=0$, and $\tilde{t}=0$; thus, that we find from Eq. \ \eqref{WLLRRsum} that the spin transfer and spin pumping parameters are determined by the reflection coefficients:
\begin{subequations}
\begin{align}
A_{\uparrow \downarrow} & = g_r \, , \\
\tilde{A}_{\uparrow \downarrow} & = \tilde{g}_r \, .  
\end{align}
\end{subequations}
The spin transfer/spin pumping efficiencies are determined by the mixing conductance associated with the reflection of the incoming waveguide modes.

We also find that
\begin{align}
C_{\downarrow \uparrow} & = \frac{1}{2} \text{Tr} \left[1 - \left(r_\uparrow^\dag r_\uparrow \right)^2 \right]\, , \\
\tilde{C}_{\downarrow \uparrow} & = \frac{1}{2} \text{Tr} \left[1 - \left(\tilde{r}_\uparrow^\dag \tilde{r}_\uparrow \right)^2 \right]\, , \\
D_{\downarrow \uparrow} &=0 \, , \\
\tilde{D}_{\downarrow \uparrow} & = 0 \, .
\end{align}
%

In the simplest case of complete dephasing, $g_r=N=\tilde{g}_r$, where $N$ is the number of transverse waveguide modes. Furthermore, from Eq. \eqref{Bupdo}, we find that
\begin{subequations}
\begin{align}
B_{\uparrow \downarrow} & =0 \, , \\
\tilde{B}_{\uparrow \downarrow} & = 0 \, ,   
\end{align}
\end{subequations}
and $C_{\downarrow \uparrow}=N/2=C_{\uparrow \downarrow}$. The latter implies that there is noise associated with transverse spin accumulation transport. This is due to the discrete angular momentum carried by each electron that may be flipped inside the magnet.

In contrast, there are no shot noise contributions arising from $B_{\uparrow \downarrow}$ due to the fluctuations of magnetic insulators. This is because $B_{\uparrow \downarrow}$ governs the transport of longitudinal spins that is absent in magnetic insulators.

\subsection{Clean Junctions}

In a simple model of a clean junction, the reflection and transmission coefficients are diagonal in the transverse mode basis. To illustrate the main physics, let us furthermore assume that only the phases of the reflection and transmission coefficients differ. Then, for each mode, the spin-dependent scattering coefficients are
\begin{subequations}
\begin{align}
r_s & =  \sqrt{1-T} \exp{ i \phi_{rs}} \, ,  \\
t_s & =  \sqrt{T} \exp{ i \phi_{rt}} \, ,  \\
\tilde{r}_s & =  \sqrt{1-T} \exp{ i \phi_{\tilde{r}s}} \, ,  \\
\tilde{t}_s & =  \sqrt{T} \exp{ i \phi_{\tilde{t}s}}  \, .
\end{align}
\end{subequations}
Then, we find for a single mode
\begin{subequations}
\begin{align}
A_{\uparrow \downarrow}(1) & = \frac{1}{4 \pi S} \left(1 - (1-T) e^{i (\phi_{r\uparrow} - \phi_{r\downarrow})}  - T ^{i (\phi_{t\uparrow} - \phi_{t\downarrow})} \right) \, , \\
\tilde{A}_{\uparrow \downarrow}(1) & = \frac{1}{4 \pi S} \left(1 - (1-T) e^{i (\phi_{r\uparrow} - \phi_{r\downarrow})}  - T ^{i (\phi_{t\uparrow} - \phi_{t\downarrow})} \right) \, ,
\end{align}
\end{subequations}
\begin{subequations}
\begin{align}
B_{\uparrow \downarrow}(1) & = T \left(1 - T \right) \left[1 - e^{-i (\phi_{r\uparrow} - \phi_{r\downarrow}) } e^{-i(\phi_{t\uparrow} - \phi_{t\downarrow})} \right] \, , \nonumber \\
\tilde{B}_{\uparrow \downarrow}(1) & = T \left(1 - T \right) \left[1 - e^{-i (\phi_{\tilde{r}\uparrow} - \phi_{\tilde{r}\downarrow}) } e^{-i(\phi_{\tilde{t}\uparrow} - \phi_{\tilde{t}\downarrow})} \right] \, , 
\end{align}
\end{subequations}
\begin{equation}
    C_{\downarrow \uparrow} = \frac{1}{2} \left[ 1 - R e^{-i(\phi_{r\uparrow} - \phi_{r\downarrow})} + T e^{-i (\phi_{t\uparrow} - \phi_{t\downarrow})} \right] \, ,
\end{equation}
\begin{equation}
    D_{\downarrow \uparrow} = 0 \, .
\end{equation}
As is common in the shot noise characteristics for fermions, the shot noise parameter $B_{\uparrow \downarrow}$ is proportional to the factor $T(1-T)$ due to the Fermi exclusion principle ensuring that only one particle can propagate at the same time. Furthermore, when we have many channels and average over all of them with different scattering phases, dephasing results in
\begin{subequations}
\begin{align}
A_{\uparrow \downarrow}(N) & = N \, , \\
\tilde{A}_{\uparrow \downarrow} (N)& = N \, ,
\end{align}
\label{alphaN}
\end{subequations}
\begin{subequations}
\begin{align}
B_{\uparrow \downarrow} (N)& =  N T \left(1 - T \right)  \, , \nonumber \\
\tilde{B}_{\uparrow \downarrow}(N) & = N T \left(1 - T \right) \, , 
\end{align}
\end{subequations}
\begin{equation}
    C_{\downarrow \uparrow} = N/2 \, , 
\end{equation}
and
\begin{equation}
    D_{\downarrow \uparrow} =0 \, ,
\end{equation}
where $N$ is the number of transverse waveguide modes. Eq.\ \eqref{alphaN} is the well-known result that the mixing conductance roughly equals the Sharvin conductance.

\subsection{Disordered Junctions}

Let us consider a random potential scatter in the magnet and normal metals. Furthermore, we assume that the random impurity scattering differs for spin-up and spin-down electrons and that these scatterings are uncorrelated. The statistical averages of the spin-up and spin-down electrons are then independent. When the number of transverse modes $N$ is large, we use the averages
\begin{align}
\langle \lvert t_{nm} \rvert^2 \rangle & = g/N^2 \, , \\
\langle \lvert r_{nm} \rvert^2 \rangle & = \frac{1-g/N}{N} \, ,
\end{align}
where $g$ is the dimensionless conductance. We then find that $B_{\uparrow \downarrow} = \frac{g_\uparrow + g_\downarrow}{N^2} \rightarrow^{N \rightarrow \infty} 0 $, so 
\begin{subequations}
\begin{align}
A_{\uparrow \downarrow}^{dis} & = N \, , \\
\tilde{A}_{\uparrow \downarrow}^{dis} & = N \, ,  \\
\end{align}
\end{subequations}
\begin{subequations}
\begin{align}
B_{\uparrow \downarrow}^{dis}& = 0 \, , \nonumber \\
\tilde{B}_{\uparrow \downarrow}^{dis} & = 0 \, , 
\end{align}
\end{subequations}
\begin{align}
    C_{\downarrow \uparrow} & = N/2 \, ,  \nonumber \\
    \tilde{C}_{\downarrow \uparrow} & = N/2 \, ,
\end{align}
\begin{align}
    D_{\downarrow \uparrow } & =0 \, , \nonumber \\
    \tilde{D}_{\downarrow \uparrow } & = 0 \, .
\end{align}

\section{Fourier Transform}

We define the Fourier transform of the Green functions as
\begin{subequations}
\begin{align}
G_0(\omega) & = \int_{-\infty}^{\infty} d (t-\tau) G(t-\tau) e^{ i \omega (t-\tau)} \, , \\
G_0(t-\tau) & = \frac{1}{2\pi} \int_{-\infty}^\infty d  \omega G_0(\omega) e^{-i \omega (t-\tau)} \, . 
\end{align}
\label{Fourier}
\end{subequations}

\begin{acknowledgements}
This work was supported by the Research Council of Norway through its Centres of Excellence funding scheme, project number 262633, "QuSpin". 
\end{acknowledgements}


\begin{thebibliography}{66}%
\makeatletter
\providecommand \@ifxundefined [1]{%
 \@ifx{#1\undefined}
}%
\providecommand \@ifnum [1]{%
 \ifnum #1\expandafter \@firstoftwo
 \else \expandafter \@secondoftwo
 \fi
}%
\providecommand \@ifx [1]{%
 \ifx #1\expandafter \@firstoftwo
 \else \expandafter \@secondoftwo
 \fi
}%
\providecommand \natexlab [1]{#1}%
\providecommand \enquote  [1]{``#1''}%
\providecommand \bibnamefont  [1]{#1}%
\providecommand \bibfnamefont [1]{#1}%
\providecommand \citenamefont [1]{#1}%
\providecommand \href@noop [0]{\@secondoftwo}%
\providecommand \href [0]{\begingroup \@sanitize@url \@href}%
\providecommand \@href[1]{\@@startlink{#1}\@@href}%
\providecommand \@@href[1]{\endgroup#1\@@endlink}%
\providecommand \@sanitize@url [0]{\catcode `\\12\catcode `\$12\catcode
  `\&12\catcode `\#12\catcode `\^12\catcode `\_12\catcode `\%12\relax}%
\providecommand \@@startlink[1]{}%
\providecommand \@@endlink[0]{}%
\providecommand \url  [0]{\begingroup\@sanitize@url \@url }%
\providecommand \@url [1]{\endgroup\@href {#1}{\urlprefix }}%
\providecommand \urlprefix  [0]{URL }%
\providecommand \Eprint [0]{\href }%
\providecommand \doibase [0]{http://dx.doi.org/}%
\providecommand \selectlanguage [0]{\@gobble}%
\providecommand \bibinfo  [0]{\@secondoftwo}%
\providecommand \bibfield  [0]{\@secondoftwo}%
\providecommand \translation [1]{[#1]}%
\providecommand \BibitemOpen [0]{}%
\providecommand \bibitemStop [0]{}%
\providecommand \bibitemNoStop [0]{.\EOS\space}%
\providecommand \EOS [0]{\spacefactor3000\relax}%
\providecommand \BibitemShut  [1]{\csname bibitem#1\endcsname}%
\let\auto@bib@innerbib\@empty
\bibitem [{\citenamefont {Berger}(1979)}]{Berger:JAP79}%
  \BibitemOpen
  \bibfield  {author} {\bibinfo {author} {\bibfnamefont {L.}~\bibnamefont
  {Berger}},\ }\href@noop {} {\bibfield  {journal} {\bibinfo  {journal} {J.
  Appl. Phys.}\ }\textbf {\bibinfo {volume} {50}},\ \bibinfo {pages} {7102}
  (\bibinfo {year} {1979})}\BibitemShut {NoStop}%
\bibitem [{\citenamefont {Freitas}\ and\ \citenamefont
  {Berger}(1985)}]{Freitas:JAP85}%
  \BibitemOpen
  \bibfield  {author} {\bibinfo {author} {\bibfnamefont {P.~P.}\ \bibnamefont
  {Freitas}}\ and\ \bibinfo {author} {\bibfnamefont {L.}~\bibnamefont
  {Berger}},\ }\href {http://dx.doi.org/doi/10.1063/1.334524} {\bibfield
  {journal} {\bibinfo  {journal} {J. Appl. Phys.}\ }\textbf {\bibinfo {volume}
  {57}},\ \bibinfo {pages} {1266} (\bibinfo {year} {1985})}\BibitemShut
  {NoStop}%
\bibitem [{\citenamefont {Berger}(1996)}]{Berger:PRB1996}%
  \BibitemOpen
  \bibfield  {author} {\bibinfo {author} {\bibfnamefont {L.}~\bibnamefont
  {Berger}},\ }\href {http://link.aps.org/abstract/PRB/v54/p9353} {\bibfield
  {journal} {\bibinfo  {journal} {Physical Review B}\ }\textbf {\bibinfo
  {volume} {54}},\ \bibinfo {pages} {9353} (\bibinfo {year}
  {1996})}\BibitemShut {NoStop}%
\bibitem [{\citenamefont {Slonczewski}(1996)}]{Slonczewski:JMMM1996}%
  \BibitemOpen
  \bibfield  {author} {\bibinfo {author} {\bibfnamefont {J.~C.}\ \bibnamefont
  {Slonczewski}},\ }\href {<Go to ISI>://A1996VH55200001} {\bibfield  {journal}
  {\bibinfo  {journal} {Journal of Magnetism and Magnetic Materials}\ }\textbf
  {\bibinfo {volume} {159}},\ \bibinfo {pages} {L1} (\bibinfo {year}
  {1996})}\BibitemShut {NoStop}%
\bibitem [{\citenamefont {Tsoi}\ \emph {et~al.}(1998)\citenamefont {Tsoi},
  \citenamefont {Jansen}, \citenamefont {Bass}, \citenamefont {Chiang},
  \citenamefont {Seck}, \citenamefont {Tsoi},\ and\ \citenamefont
  {Wyder}}]{Tsoi:PRL1998}%
  \BibitemOpen
  \bibfield  {author} {\bibinfo {author} {\bibfnamefont {M.}~\bibnamefont
  {Tsoi}}, \bibinfo {author} {\bibfnamefont {A.~G.~M.}\ \bibnamefont {Jansen}},
  \bibinfo {author} {\bibfnamefont {J.}~\bibnamefont {Bass}}, \bibinfo {author}
  {\bibfnamefont {W.~C.}\ \bibnamefont {Chiang}}, \bibinfo {author}
  {\bibfnamefont {M.}~\bibnamefont {Seck}}, \bibinfo {author} {\bibfnamefont
  {V.}~\bibnamefont {Tsoi}}, \ and\ \bibinfo {author} {\bibfnamefont
  {P.}~\bibnamefont {Wyder}},\ }\href
  {http://link.aps.org/doi/10.1103/PhysRevLett.80.4281} {\bibfield  {journal}
  {\bibinfo  {journal} {Physical Review Letters}\ }\textbf {\bibinfo {volume}
  {80}},\ \bibinfo {pages} {4281} (\bibinfo {year} {1998})}\BibitemShut
  {NoStop}%
\bibitem [{\citenamefont {Myers}\ \emph {et~al.}(1999)\citenamefont {Myers},
  \citenamefont {Ralph}, \citenamefont {Katine}, \citenamefont {Louie},\ and\
  \citenamefont {Buhrman}}]{Myers:Science1999}%
  \BibitemOpen
  \bibfield  {author} {\bibinfo {author} {\bibfnamefont {E.~B.}\ \bibnamefont
  {Myers}}, \bibinfo {author} {\bibfnamefont {D.~C.}\ \bibnamefont {Ralph}},
  \bibinfo {author} {\bibfnamefont {J.~A.}\ \bibnamefont {Katine}}, \bibinfo
  {author} {\bibfnamefont {R.~N.}\ \bibnamefont {Louie}}, \ and\ \bibinfo
  {author} {\bibfnamefont {R.~A.}\ \bibnamefont {Buhrman}},\ }\href {\doibase
  10.1126/science.285.5429.867} {\bibfield  {journal} {\bibinfo  {journal}
  {Science}\ }\textbf {\bibinfo {volume} {285}},\ \bibinfo {pages} {867}
  (\bibinfo {year} {1999})}\BibitemShut {NoStop}%
\bibitem [{\citenamefont {Katine}\ \emph {et~al.}(2000)\citenamefont {Katine},
  \citenamefont {Albert}, \citenamefont {Buhrman}, \citenamefont {Myers},\ and\
  \citenamefont {Ralph}}]{Katine:PRL2000}%
  \BibitemOpen
  \bibfield  {author} {\bibinfo {author} {\bibfnamefont {J.~A.}\ \bibnamefont
  {Katine}}, \bibinfo {author} {\bibfnamefont {F.~J.}\ \bibnamefont {Albert}},
  \bibinfo {author} {\bibfnamefont {R.~A.}\ \bibnamefont {Buhrman}}, \bibinfo
  {author} {\bibfnamefont {E.~B.}\ \bibnamefont {Myers}}, \ and\ \bibinfo
  {author} {\bibfnamefont {D.~C.}\ \bibnamefont {Ralph}},\ }\href {<Go to
  ISI>://000086243800032} {\bibfield  {journal} {\bibinfo  {journal} {Physical
  Review Letters}\ }\textbf {\bibinfo {volume} {84}},\ \bibinfo {pages} {3149}
  (\bibinfo {year} {2000})}\BibitemShut {NoStop}%
\bibitem [{\citenamefont {Kiselev}\ \emph {et~al.}(2003)\citenamefont
  {Kiselev}, \citenamefont {Sankey}, \citenamefont {Krivorotov}, \citenamefont
  {Emley}, \citenamefont {Schoelkopf}, \citenamefont {Buhrman},\ and\
  \citenamefont {Ralph}}]{Kiselev:Nature2003}%
  \BibitemOpen
  \bibfield  {author} {\bibinfo {author} {\bibfnamefont {S.~I.}\ \bibnamefont
  {Kiselev}}, \bibinfo {author} {\bibfnamefont {J.~C.}\ \bibnamefont {Sankey}},
  \bibinfo {author} {\bibfnamefont {I.~N.}\ \bibnamefont {Krivorotov}},
  \bibinfo {author} {\bibfnamefont {N.~C.}\ \bibnamefont {Emley}}, \bibinfo
  {author} {\bibfnamefont {R.~J.}\ \bibnamefont {Schoelkopf}}, \bibinfo
  {author} {\bibfnamefont {R.~A.}\ \bibnamefont {Buhrman}}, \ and\ \bibinfo
  {author} {\bibfnamefont {D.~C.}\ \bibnamefont {Ralph}},\ }\href
  {http://dx.doi.org/10.1038/nature01967} {\bibfield  {journal} {\bibinfo
  {journal} {Nature}\ }\textbf {\bibinfo {volume} {425}},\ \bibinfo {pages}
  {380} (\bibinfo {year} {2003})}\BibitemShut {NoStop}%
\bibitem [{\citenamefont {Krivorotov}\ \emph {et~al.}(2005)\citenamefont
  {Krivorotov}, \citenamefont {Emley}, \citenamefont {Sankey}, \citenamefont
  {Kiselev}, \citenamefont {Ralph},\ and\ \citenamefont
  {Buhrman}}]{Krivorotov:Science2005}%
  \BibitemOpen
  \bibfield  {author} {\bibinfo {author} {\bibfnamefont {I.~N.}\ \bibnamefont
  {Krivorotov}}, \bibinfo {author} {\bibfnamefont {N.~C.}\ \bibnamefont
  {Emley}}, \bibinfo {author} {\bibfnamefont {J.~C.}\ \bibnamefont {Sankey}},
  \bibinfo {author} {\bibfnamefont {S.~I.}\ \bibnamefont {Kiselev}}, \bibinfo
  {author} {\bibfnamefont {D.~C.}\ \bibnamefont {Ralph}}, \ and\ \bibinfo
  {author} {\bibfnamefont {R.~A.}\ \bibnamefont {Buhrman}},\ }\href {<Go to
  ISI>://000226361900034} {\bibfield  {journal} {\bibinfo  {journal} {Science}\
  }\textbf {\bibinfo {volume} {307}},\ \bibinfo {pages} {228} (\bibinfo {year}
  {2005})}\BibitemShut {NoStop}%
\bibitem [{\citenamefont {Silva}\ and\ \citenamefont
  {Rippard}(2008)}]{Silva:JMMM2008}%
  \BibitemOpen
  \bibfield  {author} {\bibinfo {author} {\bibfnamefont {T.~J.}\ \bibnamefont
  {Silva}}\ and\ \bibinfo {author} {\bibfnamefont {W.~H.}\ \bibnamefont
  {Rippard}},\ }\href {\doibase 10.1016/j.jmmm.2007.12.022} {\bibfield
  {journal} {\bibinfo  {journal} {Journal of Magnetism and Magnetic Materials}\
  }\textbf {\bibinfo {volume} {320}},\ \bibinfo {pages} {1260} (\bibinfo {year}
  {2008})}\BibitemShut {NoStop}%
\bibitem [{\citenamefont {Ralph}\ and\ \citenamefont
  {Stiles}(2008)}]{Ralph:JMMM2008}%
  \BibitemOpen
  \bibfield  {author} {\bibinfo {author} {\bibfnamefont {D.~C.}\ \bibnamefont
  {Ralph}}\ and\ \bibinfo {author} {\bibfnamefont {M.~D.}\ \bibnamefont
  {Stiles}},\ }\href
  {http://www.sciencedirect.com/science/article/B6TJJ-4RFSD1M-2/2/f35a2bc5e9c53f19f6883d74c20dbb69}
  {\bibfield  {journal} {\bibinfo  {journal} {Journal of Magnetism and Magnetic
  Materials}\ }\textbf {\bibinfo {volume} {320}},\ \bibinfo {pages} {1190}
  (\bibinfo {year} {2008})}\BibitemShut {NoStop}%
\bibitem [{\citenamefont {Parkin}\ \emph {et~al.}(2008)\citenamefont {Parkin},
  \citenamefont {Hayashi},\ and\ \citenamefont {Thomas}}]{Parkin:Science2008}%
  \BibitemOpen
  \bibfield  {author} {\bibinfo {author} {\bibfnamefont {S.~S.~P.}\
  \bibnamefont {Parkin}}, \bibinfo {author} {\bibfnamefont {M.}~\bibnamefont
  {Hayashi}}, \ and\ \bibinfo {author} {\bibfnamefont {L.}~\bibnamefont
  {Thomas}},\ }\href {\doibase 10.1126/science.1145799} {\bibfield  {journal}
  {\bibinfo  {journal} {Science}\ }\textbf {\bibinfo {volume} {320}},\ \bibinfo
  {pages} {190} (\bibinfo {year} {2008})}\BibitemShut {NoStop}%
\bibitem [{\citenamefont {Brataas}\ \emph {et~al.}(2012)\citenamefont
  {Brataas}, \citenamefont {Kent},\ and\ \citenamefont
  {Ohno}}]{Brataas:NatMat2012}%
  \BibitemOpen
  \bibfield  {author} {\bibinfo {author} {\bibfnamefont {A.}~\bibnamefont
  {Brataas}}, \bibinfo {author} {\bibfnamefont {A.~D.}\ \bibnamefont {Kent}}, \
  and\ \bibinfo {author} {\bibfnamefont {H.}~\bibnamefont {Ohno}},\ }\href
  {\doibase 10.1038/nmat3311} {\bibfield  {journal} {\bibinfo  {journal}
  {Nature Materials}\ }\textbf {\bibinfo {volume} {11}},\ \bibinfo {pages}
  {372} (\bibinfo {year} {2012})}\BibitemShut {NoStop}%
\bibitem [{\citenamefont {Manchon}\ \emph {et~al.}(2019)\citenamefont
  {Manchon}, \citenamefont {Zelesny}, \citenamefont {Miron}, \citenamefont
  {Jungwirth}, \citenamefont {Sinova}, \citenamefont {Thiaville}, \citenamefont
  {Garello},\ and\ \citenamefont {Gambardella}}]{Manchon:RMP2019}%
  \BibitemOpen
  \bibfield  {author} {\bibinfo {author} {\bibfnamefont {A.}~\bibnamefont
  {Manchon}}, \bibinfo {author} {\bibfnamefont {J.}~\bibnamefont {Zelesny}},
  \bibinfo {author} {\bibfnamefont {I.~M.}\ \bibnamefont {Miron}}, \bibinfo
  {author} {\bibfnamefont {T.}~\bibnamefont {Jungwirth}}, \bibinfo {author}
  {\bibfnamefont {J.}~\bibnamefont {Sinova}}, \bibinfo {author} {\bibfnamefont
  {A.}~\bibnamefont {Thiaville}}, \bibinfo {author} {\bibfnamefont
  {K.}~\bibnamefont {Garello}}, \ and\ \bibinfo {author} {\bibfnamefont
  {P.}~\bibnamefont {Gambardella}},\ }\href {\doibase
  10.1103/RevModPhys.91.035004} {\bibfield  {journal} {\bibinfo  {journal}
  {Reviews of Modern Physics}\ }\textbf {\bibinfo {volume} {91}},\ \bibinfo
  {pages} {035004} (\bibinfo {year} {2019})}\BibitemShut {NoStop}%
\bibitem [{\citenamefont {Monod}\ \emph {et~al.}(1972)\citenamefont {Monod},
  \citenamefont {Hurdequint}, \citenamefont {Janossy}, \citenamefont {Obert},\
  and\ \citenamefont {Chaumont}}]{Monod:PRL1972}%
  \BibitemOpen
  \bibfield  {author} {\bibinfo {author} {\bibfnamefont {P.}~\bibnamefont
  {Monod}}, \bibinfo {author} {\bibfnamefont {H.}~\bibnamefont {Hurdequint}},
  \bibinfo {author} {\bibfnamefont {A.}~\bibnamefont {Janossy}}, \bibinfo
  {author} {\bibfnamefont {J.}~\bibnamefont {Obert}}, \ and\ \bibinfo {author}
  {\bibfnamefont {J.}~\bibnamefont {Chaumont}},\ }\href
  {http://link.aps.org/doi/10.1103/PhysRevLett.29.1327} {\bibfield  {journal}
  {\bibinfo  {journal} {Physical Review Letters}\ }\textbf {\bibinfo {volume}
  {29}},\ \bibinfo {pages} {1327} (\bibinfo {year} {1972})}\BibitemShut
  {NoStop}%
\bibitem [{\citenamefont {Silsbee}\ \emph {et~al.}(1979)\citenamefont
  {Silsbee}, \citenamefont {Janossy},\ and\ \citenamefont
  {Monod}}]{Silsbee:PRB1979}%
  \BibitemOpen
  \bibfield  {author} {\bibinfo {author} {\bibfnamefont {R.~H.}\ \bibnamefont
  {Silsbee}}, \bibinfo {author} {\bibfnamefont {A.}~\bibnamefont {Janossy}}, \
  and\ \bibinfo {author} {\bibfnamefont {P.}~\bibnamefont {Monod}},\ }\href
  {http://link.aps.org/doi/10.1103/PhysRevB.19.4382} {\bibfield  {journal}
  {\bibinfo  {journal} {Physical Review B}\ }\textbf {\bibinfo {volume} {19}},\
  \bibinfo {pages} {4382} (\bibinfo {year} {1979})}\BibitemShut {NoStop}%
\bibitem [{\citenamefont {Mizukami}\ \emph {et~al.}(2001)\citenamefont
  {Mizukami}, \citenamefont {Ando},\ and\ \citenamefont
  {Miazaki}}]{Mizukami:Jpn2001}%
  \BibitemOpen
  \bibfield  {author} {\bibinfo {author} {\bibfnamefont {S.}~\bibnamefont
  {Mizukami}}, \bibinfo {author} {\bibfnamefont {Y.}~\bibnamefont {Ando}}, \
  and\ \bibinfo {author} {\bibfnamefont {T.}~\bibnamefont {Miazaki}},\
  }\href@noop {} {\bibfield  {journal} {\bibinfo  {journal} {Jpn. J. Appl.
  Phys.}\ }\textbf {\bibinfo {volume} {40}},\ \bibinfo {pages} {580} (\bibinfo
  {year} {2001})}\BibitemShut {NoStop}%
\bibitem [{\citenamefont {Urban}\ \emph {et~al.}(2001)\citenamefont {Urban},
  \citenamefont {Woltersdorf},\ and\ \citenamefont {Heinrich}}]{Urban:PRL2001}%
  \BibitemOpen
  \bibfield  {author} {\bibinfo {author} {\bibfnamefont {R.}~\bibnamefont
  {Urban}}, \bibinfo {author} {\bibfnamefont {G.}~\bibnamefont {Woltersdorf}},
  \ and\ \bibinfo {author} {\bibfnamefont {B.}~\bibnamefont {Heinrich}},\
  }\href {http://link.aps.org/doi/10.1103/PhysRevLett.87.217204} {\bibfield
  {journal} {\bibinfo  {journal} {Physical Review Letters}\ }\textbf {\bibinfo
  {volume} {87}},\ \bibinfo {pages} {217204} (\bibinfo {year}
  {2001})}\BibitemShut {NoStop}%
\bibitem [{\citenamefont {Tserkovnyak}\ \emph {et~al.}(2002)\citenamefont
  {Tserkovnyak}, \citenamefont {Brataas},\ and\ \citenamefont
  {Bauer}}]{Tserkovnyak:PRL2002}%
  \BibitemOpen
  \bibfield  {author} {\bibinfo {author} {\bibfnamefont {Y.}~\bibnamefont
  {Tserkovnyak}}, \bibinfo {author} {\bibfnamefont {A.}~\bibnamefont
  {Brataas}}, \ and\ \bibinfo {author} {\bibfnamefont {G.~E.~W.}\ \bibnamefont
  {Bauer}},\ }\href@noop {} {\bibfield  {journal} {\bibinfo  {journal}
  {Physical Review Letters}\ }\textbf {\bibinfo {volume} {88}},\ \bibinfo
  {pages} {117601} (\bibinfo {year} {2002})}\BibitemShut {NoStop}%
\bibitem [{\citenamefont {Heinrich}\ \emph {et~al.}(2003)\citenamefont
  {Heinrich}, \citenamefont {Tserkovnyak}, \citenamefont {Woltersdorf},
  \citenamefont {Brataas}, \citenamefont {Urban},\ and\ \citenamefont
  {Bauer}}]{Heinrich:PRL2003}%
  \BibitemOpen
  \bibfield  {author} {\bibinfo {author} {\bibfnamefont {B.}~\bibnamefont
  {Heinrich}}, \bibinfo {author} {\bibfnamefont {Y.}~\bibnamefont
  {Tserkovnyak}}, \bibinfo {author} {\bibfnamefont {G.}~\bibnamefont
  {Woltersdorf}}, \bibinfo {author} {\bibfnamefont {A.}~\bibnamefont
  {Brataas}}, \bibinfo {author} {\bibfnamefont {R.}~\bibnamefont {Urban}}, \
  and\ \bibinfo {author} {\bibfnamefont {G.~E.~W.}\ \bibnamefont {Bauer}},\
  }\href@noop {} {\bibfield  {journal} {\bibinfo  {journal} {Physical Review
  Letters}\ }\textbf {\bibinfo {volume} {90}},\ \bibinfo {pages} {187601}
  (\bibinfo {year} {2003})}\BibitemShut {NoStop}%
\bibitem [{\citenamefont {Tserkovnyak}\ \emph {et~al.}(2005)\citenamefont
  {Tserkovnyak}, \citenamefont {Brataas}, \citenamefont {Bauer},\ and\
  \citenamefont {Halperin}}]{Tserkovnyak:RMP2005}%
  \BibitemOpen
  \bibfield  {author} {\bibinfo {author} {\bibfnamefont {Y.}~\bibnamefont
  {Tserkovnyak}}, \bibinfo {author} {\bibfnamefont {A.}~\bibnamefont
  {Brataas}}, \bibinfo {author} {\bibfnamefont {G.~E.~W.}\ \bibnamefont
  {Bauer}}, \ and\ \bibinfo {author} {\bibfnamefont {B.~I.}\ \bibnamefont
  {Halperin}},\ }\href {http://link.aps.org/abstract/RMP/v77/p1375} {\bibfield
  {journal} {\bibinfo  {journal} {Reviews of Modern Physics}\ }\textbf
  {\bibinfo {volume} {77}},\ \bibinfo {pages} {1375} (\bibinfo {year}
  {2005})}\BibitemShut {NoStop}%
\bibitem [{\citenamefont {Mosendz}\ \emph {et~al.}(2010)\citenamefont
  {Mosendz}, \citenamefont {Vlaminck}, \citenamefont {Pearson}, \citenamefont
  {Fradin}, \citenamefont {Bauer}, \citenamefont {Bader},\ and\ \citenamefont
  {Hoffmann}}]{Mosendz:PRB2010}%
  \BibitemOpen
  \bibfield  {author} {\bibinfo {author} {\bibfnamefont {O.}~\bibnamefont
  {Mosendz}}, \bibinfo {author} {\bibfnamefont {V.}~\bibnamefont {Vlaminck}},
  \bibinfo {author} {\bibfnamefont {J.~E.}\ \bibnamefont {Pearson}}, \bibinfo
  {author} {\bibfnamefont {F.~Y.}\ \bibnamefont {Fradin}}, \bibinfo {author}
  {\bibfnamefont {G.~E.~W.}\ \bibnamefont {Bauer}}, \bibinfo {author}
  {\bibfnamefont {S.~D.}\ \bibnamefont {Bader}}, \ and\ \bibinfo {author}
  {\bibfnamefont {A.}~\bibnamefont {Hoffmann}},\ }\href
  {http://link.aps.org/doi/10.1103/PhysRevB.82.214403} {\bibfield  {journal}
  {\bibinfo  {journal} {Physical Review B}\ }\textbf {\bibinfo {volume} {82}},\
  \bibinfo {pages} {214403} (\bibinfo {year} {2010})}\BibitemShut {NoStop}%
\bibitem [{\citenamefont {Ando}\ \emph {et~al.}(2011)\citenamefont {Ando},
  \citenamefont {Takahashi}, \citenamefont {Ieda}, \citenamefont {Kajiwara},
  \citenamefont {Nakayama}, \citenamefont {Yoshino}, \citenamefont {Harii},
  \citenamefont {Fujikawa}, \citenamefont {Matsuo}, \citenamefont {Maekawa},\
  and\ \citenamefont {Saitoh}}]{Ando:JAP2011}%
  \BibitemOpen
  \bibfield  {author} {\bibinfo {author} {\bibfnamefont {K.}~\bibnamefont
  {Ando}}, \bibinfo {author} {\bibfnamefont {S.}~\bibnamefont {Takahashi}},
  \bibinfo {author} {\bibfnamefont {J.}~\bibnamefont {Ieda}}, \bibinfo {author}
  {\bibfnamefont {Y.}~\bibnamefont {Kajiwara}}, \bibinfo {author}
  {\bibfnamefont {H.}~\bibnamefont {Nakayama}}, \bibinfo {author}
  {\bibfnamefont {T.}~\bibnamefont {Yoshino}}, \bibinfo {author} {\bibfnamefont
  {K.}~\bibnamefont {Harii}}, \bibinfo {author} {\bibfnamefont
  {Y.}~\bibnamefont {Fujikawa}}, \bibinfo {author} {\bibfnamefont
  {M.}~\bibnamefont {Matsuo}}, \bibinfo {author} {\bibfnamefont
  {S.}~\bibnamefont {Maekawa}}, \ and\ \bibinfo {author} {\bibfnamefont
  {E.}~\bibnamefont {Saitoh}},\ }\href {\doibase 10.1063/1.3587173} {\bibfield
  {journal} {\bibinfo  {journal} {Journal of Applied Physics}\ }\textbf
  {\bibinfo {volume} {109}},\ \bibinfo {pages} {103913} (\bibinfo {year}
  {2011})}\BibitemShut {NoStop}%
\bibitem [{\citenamefont {Sandweg}\ \emph {et~al.}(2011)\citenamefont
  {Sandweg}, \citenamefont {Kajiwara}, \citenamefont {Chumak}, \citenamefont
  {Serga}, \citenamefont {Vasyuchka}, \citenamefont {Jungfleisch},
  \citenamefont {Saitoh},\ and\ \citenamefont {Hillebrands}}]{Sandweg:PRL2011}%
  \BibitemOpen
  \bibfield  {author} {\bibinfo {author} {\bibfnamefont {C.~W.}\ \bibnamefont
  {Sandweg}}, \bibinfo {author} {\bibfnamefont {Y.}~\bibnamefont {Kajiwara}},
  \bibinfo {author} {\bibfnamefont {A.~V.}\ \bibnamefont {Chumak}}, \bibinfo
  {author} {\bibfnamefont {A.~A.}\ \bibnamefont {Serga}}, \bibinfo {author}
  {\bibfnamefont {V.~I.}\ \bibnamefont {Vasyuchka}}, \bibinfo {author}
  {\bibfnamefont {M.~B.}\ \bibnamefont {Jungfleisch}}, \bibinfo {author}
  {\bibfnamefont {E.}~\bibnamefont {Saitoh}}, \ and\ \bibinfo {author}
  {\bibfnamefont {B.}~\bibnamefont {Hillebrands}},\ }\href
  {http://link.aps.org/doi/10.1103/PhysRevLett.106.216601} {\bibfield
  {journal} {\bibinfo  {journal} {Physical Review Letters}\ }\textbf {\bibinfo
  {volume} {106}},\ \bibinfo {pages} {216601} (\bibinfo {year}
  {2011})}\BibitemShut {NoStop}%
\bibitem [{\citenamefont {Shikoh}\ \emph {et~al.}(2013)\citenamefont {Shikoh},
  \citenamefont {Ando}, \citenamefont {Kubo}, \citenamefont {Saitoh},
  \citenamefont {Shinjo},\ and\ \citenamefont {Shiraishi}}]{Shikoh:PRL2013}%
  \BibitemOpen
  \bibfield  {author} {\bibinfo {author} {\bibfnamefont {E.}~\bibnamefont
  {Shikoh}}, \bibinfo {author} {\bibfnamefont {K.}~\bibnamefont {Ando}},
  \bibinfo {author} {\bibfnamefont {K.}~\bibnamefont {Kubo}}, \bibinfo {author}
  {\bibfnamefont {E.}~\bibnamefont {Saitoh}}, \bibinfo {author} {\bibfnamefont
  {T.}~\bibnamefont {Shinjo}}, \ and\ \bibinfo {author} {\bibfnamefont
  {M.}~\bibnamefont {Shiraishi}},\ }\href
  {http://link.aps.org/doi/10.1103/PhysRevLett.110.127201} {\bibfield
  {journal} {\bibinfo  {journal} {Physical Review Letters}\ }\textbf {\bibinfo
  {volume} {110}},\ \bibinfo {pages} {127201} (\bibinfo {year}
  {2013})}\BibitemShut {NoStop}%
\bibitem [{\citenamefont {Cheng}\ \emph {et~al.}(2014)\citenamefont {Cheng},
  \citenamefont {Xiao}, \citenamefont {Niu},\ and\ \citenamefont
  {Brataas}}]{Cheng:PRL2014}%
  \BibitemOpen
  \bibfield  {author} {\bibinfo {author} {\bibfnamefont {R.}~\bibnamefont
  {Cheng}}, \bibinfo {author} {\bibfnamefont {J.}~\bibnamefont {Xiao}},
  \bibinfo {author} {\bibfnamefont {Q.}~\bibnamefont {Niu}}, \ and\ \bibinfo
  {author} {\bibfnamefont {A.}~\bibnamefont {Brataas}},\ }\href
  {http://link.aps.org/doi/10.1103/PhysRevLett.113.057601} {\bibfield
  {journal} {\bibinfo  {journal} {Physical Review Letters}\ }\textbf {\bibinfo
  {volume} {113}},\ \bibinfo {pages} {057601} (\bibinfo {year}
  {2014})}\BibitemShut {NoStop}%
\bibitem [{\citenamefont {Kamra}\ and\ \citenamefont
  {Belzig}(2017)}]{Kamra:PRL2017}%
  \BibitemOpen
  \bibfield  {author} {\bibinfo {author} {\bibfnamefont {A.}~\bibnamefont
  {Kamra}}\ and\ \bibinfo {author} {\bibfnamefont {W.}~\bibnamefont {Belzig}},\
  }\href {https://link.aps.org/doi/10.1103/PhysRevLett.119.197201} {\bibfield
  {journal} {\bibinfo  {journal} {Physical Review Letters}\ }\textbf {\bibinfo
  {volume} {119}},\ \bibinfo {pages} {197201} (\bibinfo {year}
  {2017})}\BibitemShut {NoStop}%
\bibitem [{\citenamefont {Johansen}\ and\ \citenamefont
  {Brataas}(2017)}]{Johansen:PRB2017}%
  \BibitemOpen
  \bibfield  {author} {\bibinfo {author} {\bibfnamefont {O.}~\bibnamefont
  {Johansen}}\ and\ \bibinfo {author} {\bibfnamefont {A.}~\bibnamefont
  {Brataas}},\ }\href {https://link.aps.org/doi/10.1103/PhysRevB.95.220408}
  {\bibfield  {journal} {\bibinfo  {journal} {Physical Review B}\ }\textbf
  {\bibinfo {volume} {95}},\ \bibinfo {pages} {220408} (\bibinfo {year}
  {2017})}\BibitemShut {NoStop}%
\bibitem [{\citenamefont {Li}\ \emph {et~al.}(2020)\citenamefont {Li},
  \citenamefont {Wilson}, \citenamefont {Cheng}, \citenamefont {Lohmann},
  \citenamefont {Kavand}, \citenamefont {Yuan}, \citenamefont {Aldosary},
  \citenamefont {Agladze}, \citenamefont {Wei}, \citenamefont {Sherwin},\ and\
  \citenamefont {Shi}}]{Li:Nature2020}%
  \BibitemOpen
  \bibfield  {author} {\bibinfo {author} {\bibfnamefont {J.}~\bibnamefont
  {Li}}, \bibinfo {author} {\bibfnamefont {C.~B.}\ \bibnamefont {Wilson}},
  \bibinfo {author} {\bibfnamefont {R.}~\bibnamefont {Cheng}}, \bibinfo
  {author} {\bibfnamefont {M.}~\bibnamefont {Lohmann}}, \bibinfo {author}
  {\bibfnamefont {M.}~\bibnamefont {Kavand}}, \bibinfo {author} {\bibfnamefont
  {W.}~\bibnamefont {Yuan}}, \bibinfo {author} {\bibfnamefont {M.}~\bibnamefont
  {Aldosary}}, \bibinfo {author} {\bibfnamefont {N.}~\bibnamefont {Agladze}},
  \bibinfo {author} {\bibfnamefont {P.}~\bibnamefont {Wei}}, \bibinfo {author}
  {\bibfnamefont {M.~S.}\ \bibnamefont {Sherwin}}, \ and\ \bibinfo {author}
  {\bibfnamefont {J.}~\bibnamefont {Shi}},\ }\href {\doibase
  10.1038/s41586-020-1950-4} {\bibfield  {journal} {\bibinfo  {journal}
  {Nature}\ }\textbf {\bibinfo {volume} {578}},\ \bibinfo {pages} {70}
  (\bibinfo {year} {2020})}\BibitemShut {NoStop}%
\bibitem [{\citenamefont {Vaidya}\ \emph {et~al.}(2020)\citenamefont {Vaidya},
  \citenamefont {Morley}, \citenamefont {van Tol}, \citenamefont {Liu},
  \citenamefont {Cheng}, \citenamefont {Brataas}, \citenamefont {Lederman},\
  and\ \citenamefont {del Barco}}]{Vaidya:Science2020}%
  \BibitemOpen
  \bibfield  {author} {\bibinfo {author} {\bibfnamefont {P.}~\bibnamefont
  {Vaidya}}, \bibinfo {author} {\bibfnamefont {S.~A.}\ \bibnamefont {Morley}},
  \bibinfo {author} {\bibfnamefont {J.}~\bibnamefont {van Tol}}, \bibinfo
  {author} {\bibfnamefont {Y.}~\bibnamefont {Liu}}, \bibinfo {author}
  {\bibfnamefont {R.}~\bibnamefont {Cheng}}, \bibinfo {author} {\bibfnamefont
  {A.}~\bibnamefont {Brataas}}, \bibinfo {author} {\bibfnamefont
  {D.}~\bibnamefont {Lederman}}, \ and\ \bibinfo {author} {\bibfnamefont
  {E.}~\bibnamefont {del Barco}},\ }\href {\doibase 10.1126/science.aaz4247}
  {\bibfield  {journal} {\bibinfo  {journal} {Science}\ }\textbf {\bibinfo
  {volume} {368}},\ \bibinfo {pages} {160} (\bibinfo {year}
  {2020})}\BibitemShut {NoStop}%
\bibitem [{\citenamefont {Waintal}\ \emph {et~al.}(2000)\citenamefont
  {Waintal}, \citenamefont {Myers}, \citenamefont {Brouwer},\ and\
  \citenamefont {Ralph}}]{Waintal:PRB2000}%
  \BibitemOpen
  \bibfield  {author} {\bibinfo {author} {\bibfnamefont {X.}~\bibnamefont
  {Waintal}}, \bibinfo {author} {\bibfnamefont {E.~B.}\ \bibnamefont {Myers}},
  \bibinfo {author} {\bibfnamefont {P.~W.}\ \bibnamefont {Brouwer}}, \ and\
  \bibinfo {author} {\bibfnamefont {D.~C.}\ \bibnamefont {Ralph}},\ }\href
  {http://link.aps.org/abstract/PRB/v62/p12317} {\bibfield  {journal} {\bibinfo
   {journal} {Physical Review B}\ }\textbf {\bibinfo {volume} {62}},\ \bibinfo
  {pages} {12317} (\bibinfo {year} {2000})}\BibitemShut {NoStop}%
\bibitem [{\citenamefont {Brataas}\ \emph {et~al.}(2000)\citenamefont
  {Brataas}, \citenamefont {Nazarov},\ and\ \citenamefont
  {Bauer}}]{Brataas:PRL2000}%
  \BibitemOpen
  \bibfield  {author} {\bibinfo {author} {\bibfnamefont {A.}~\bibnamefont
  {Brataas}}, \bibinfo {author} {\bibfnamefont {Y.~V.}\ \bibnamefont
  {Nazarov}}, \ and\ \bibinfo {author} {\bibfnamefont {G.~E.~W.}\ \bibnamefont
  {Bauer}},\ }\href {http://link.aps.org/doi/10.1103/PhysRevLett.84.2481}
  {\bibfield  {journal} {\bibinfo  {journal} {Physical Review Letters}\
  }\textbf {\bibinfo {volume} {84}},\ \bibinfo {pages} {2481} (\bibinfo {year}
  {2000})}\BibitemShut {NoStop}%
\bibitem [{\citenamefont {Brataas}\ \emph {et~al.}(2001)\citenamefont
  {Brataas}, \citenamefont {Nazarov},\ and\ \citenamefont
  {Bauer}}]{Brataas:EPB2001}%
  \BibitemOpen
  \bibfield  {author} {\bibinfo {author} {\bibfnamefont {A.}~\bibnamefont
  {Brataas}}, \bibinfo {author} {\bibfnamefont {Y.~V.}\ \bibnamefont
  {Nazarov}}, \ and\ \bibinfo {author} {\bibfnamefont {G.~E.~W.}\ \bibnamefont
  {Bauer}},\ }\href {<Go to ISI>://000170373200012} {\bibfield  {journal}
  {\bibinfo  {journal} {European Physical Journal B}\ }\textbf {\bibinfo
  {volume} {22}},\ \bibinfo {pages} {99} (\bibinfo {year} {2001})}\BibitemShut
  {NoStop}%
\bibitem [{\citenamefont {Stiles}\ and\ \citenamefont
  {Zangwill}(2002)}]{Stiles:PRB2002}%
  \BibitemOpen
  \bibfield  {author} {\bibinfo {author} {\bibfnamefont {M.~D.}\ \bibnamefont
  {Stiles}}\ and\ \bibinfo {author} {\bibfnamefont {A.}~\bibnamefont
  {Zangwill}},\ }\href {http://link.aps.org/doi/10.1103/PhysRevB.66.014407}
  {\bibfield  {journal} {\bibinfo  {journal} {Physical Review B}\ }\textbf
  {\bibinfo {volume} {66}},\ \bibinfo {pages} {014407} (\bibinfo {year}
  {2002})}\BibitemShut {NoStop}%
\bibitem [{\citenamefont {Nunez}\ and\ \citenamefont
  {MacDonald}(2006)}]{Nunez:SSC2006}%
  \BibitemOpen
  \bibfield  {author} {\bibinfo {author} {\bibfnamefont {A.~S.}\ \bibnamefont
  {Nunez}}\ and\ \bibinfo {author} {\bibfnamefont {A.~H.}\ \bibnamefont
  {MacDonald}},\ }\href {\doibase http://dx.doi.org/10.1016/j.ssc.2006.05.004}
  {\bibfield  {journal} {\bibinfo  {journal} {Solid State Communications}\
  }\textbf {\bibinfo {volume} {139}},\ \bibinfo {pages} {31} (\bibinfo {year}
  {2006})}\BibitemShut {NoStop}%
\bibitem [{\citenamefont {Bender}\ \emph {et~al.}(2012)\citenamefont {Bender},
  \citenamefont {Duine},\ and\ \citenamefont {Tserkovnyak}}]{Bender:PRL2012}%
  \BibitemOpen
  \bibfield  {author} {\bibinfo {author} {\bibfnamefont {S.~A.}\ \bibnamefont
  {Bender}}, \bibinfo {author} {\bibfnamefont {R.~A.}\ \bibnamefont {Duine}}, \
  and\ \bibinfo {author} {\bibfnamefont {Y.}~\bibnamefont {Tserkovnyak}},\
  }\href {http://link.aps.org/doi/10.1103/PhysRevLett.108.246601} {\bibfield
  {journal} {\bibinfo  {journal} {Physical Review Letters}\ }\textbf {\bibinfo
  {volume} {108}},\ \bibinfo {pages} {246601} (\bibinfo {year}
  {2012})}\BibitemShut {NoStop}%
\bibitem [{\citenamefont {Bender}\ \emph {et~al.}(2014)\citenamefont {Bender},
  \citenamefont {Duine}, \citenamefont {Brataas},\ and\ \citenamefont
  {Tserkovnyak}}]{Bender:PRB2014}%
  \BibitemOpen
  \bibfield  {author} {\bibinfo {author} {\bibfnamefont {S.~A.}\ \bibnamefont
  {Bender}}, \bibinfo {author} {\bibfnamefont {R.~A.}\ \bibnamefont {Duine}},
  \bibinfo {author} {\bibfnamefont {A.}~\bibnamefont {Brataas}}, \ and\
  \bibinfo {author} {\bibfnamefont {Y.}~\bibnamefont {Tserkovnyak}},\ }\href
  {http://link.aps.org/doi/10.1103/PhysRevB.90.094409} {\bibfield  {journal}
  {\bibinfo  {journal} {Physical Review B}\ }\textbf {\bibinfo {volume} {90}},\
  \bibinfo {pages} {094409} (\bibinfo {year} {2014})}\BibitemShut {NoStop}%
\bibitem [{\citenamefont {Foros}\ \emph {et~al.}(2005)\citenamefont {Foros},
  \citenamefont {Brataas}, \citenamefont {Tserkovnyak},\ and\ \citenamefont
  {Bauer}}]{Foros:PRL2005}%
  \BibitemOpen
  \bibfield  {author} {\bibinfo {author} {\bibfnamefont {J.}~\bibnamefont
  {Foros}}, \bibinfo {author} {\bibfnamefont {A.}~\bibnamefont {Brataas}},
  \bibinfo {author} {\bibfnamefont {Y.}~\bibnamefont {Tserkovnyak}}, \ and\
  \bibinfo {author} {\bibfnamefont {G.}~\bibnamefont {Bauer}},\ }\href
  {\doibase 10.1103/PhysRevLett.95.016601} {\bibfield  {journal} {\bibinfo
  {journal} {Physical Review Letters}\ }\textbf {\bibinfo {volume} {95}}
  (\bibinfo {year} {2005}),\ 10.1103/PhysRevLett.95.016601}\BibitemShut
  {NoStop}%
\bibitem [{\citenamefont {Xiao}\ \emph {et~al.}(2009)\citenamefont {Xiao},
  \citenamefont {Bauer}, \citenamefont {Maekawa},\ and\ \citenamefont
  {Brataas}}]{Xiao:PRB2009}%
  \BibitemOpen
  \bibfield  {author} {\bibinfo {author} {\bibfnamefont {J.}~\bibnamefont
  {Xiao}}, \bibinfo {author} {\bibfnamefont {G.~E.~W.}\ \bibnamefont {Bauer}},
  \bibinfo {author} {\bibfnamefont {S.}~\bibnamefont {Maekawa}}, \ and\
  \bibinfo {author} {\bibfnamefont {A.}~\bibnamefont {Brataas}},\ }\href
  {http://link.aps.org/abstract/PRB/v79/e174415} {\bibfield  {journal}
  {\bibinfo  {journal} {Physical Review B (Condensed Matter and Materials
  Physics)}\ }\textbf {\bibinfo {volume} {79}},\ \bibinfo {pages} {174415}
  (\bibinfo {year} {2009})}\BibitemShut {NoStop}%
\bibitem [{\citenamefont {Xiao}\ \emph {et~al.}(2010)\citenamefont {Xiao},
  \citenamefont {Bauer}, \citenamefont {Uchida}, \citenamefont {Saitoh},\ and\
  \citenamefont {Maekawa}}]{Xiao:PRB2010}%
  \BibitemOpen
  \bibfield  {author} {\bibinfo {author} {\bibfnamefont {J.}~\bibnamefont
  {Xiao}}, \bibinfo {author} {\bibfnamefont {G.~E.~W.}\ \bibnamefont {Bauer}},
  \bibinfo {author} {\bibfnamefont {K.-c.}\ \bibnamefont {Uchida}}, \bibinfo
  {author} {\bibfnamefont {E.}~\bibnamefont {Saitoh}}, \ and\ \bibinfo {author}
  {\bibfnamefont {S.}~\bibnamefont {Maekawa}},\ }\href
  {http://link.aps.org/doi/10.1103/PhysRevB.81.214418} {\bibfield  {journal}
  {\bibinfo  {journal} {Physical Review B}\ }\textbf {\bibinfo {volume} {81}},\
  \bibinfo {pages} {214418} (\bibinfo {year} {2010})}\BibitemShut {NoStop}%
\bibitem [{\citenamefont {Bauer}\ \emph {et~al.}(2012)\citenamefont {Bauer},
  \citenamefont {Saitoh},\ and\ \citenamefont {van Wees}}]{Bauer:NMAT2012}%
  \BibitemOpen
  \bibfield  {author} {\bibinfo {author} {\bibfnamefont {G.~E.~W.}\
  \bibnamefont {Bauer}}, \bibinfo {author} {\bibfnamefont {E.}~\bibnamefont
  {Saitoh}}, \ and\ \bibinfo {author} {\bibfnamefont {B.~J.}\ \bibnamefont {van
  Wees}},\ }\href {http://dx.doi.org/10.1038/nmat3301} {\bibfield  {journal}
  {\bibinfo  {journal} {Nat Mater}\ }\textbf {\bibinfo {volume} {11}},\
  \bibinfo {pages} {391} (\bibinfo {year} {2012})}\BibitemShut {NoStop}%
\bibitem [{\citenamefont {Chudnovskiy}\ \emph {et~al.}(2008)\citenamefont
  {Chudnovskiy}, \citenamefont {Swiebodzinski},\ and\ \citenamefont
  {Kamenev}}]{Chudnovskiy:PRL2008}%
  \BibitemOpen
  \bibfield  {author} {\bibinfo {author} {\bibfnamefont {A.~L.}\ \bibnamefont
  {Chudnovskiy}}, \bibinfo {author} {\bibfnamefont {J.}~\bibnamefont
  {Swiebodzinski}}, \ and\ \bibinfo {author} {\bibfnamefont {A.}~\bibnamefont
  {Kamenev}},\ }\href {\doibase 10.1103/PhysRevLett.101.066601} {\bibfield
  {journal} {\bibinfo  {journal} {Physical Review Letters}\ }\textbf {\bibinfo
  {volume} {101}},\ \bibinfo {pages} {066601} (\bibinfo {year}
  {2008})}\BibitemShut {NoStop}%
\bibitem [{\citenamefont {Swiebodzinski}\ \emph {et~al.}(2010)\citenamefont
  {Swiebodzinski}, \citenamefont {Chudnovskiy}, \citenamefont {Dunn},\ and\
  \citenamefont {Kamenev}}]{Swiebodzinski:PRB2010}%
  \BibitemOpen
  \bibfield  {author} {\bibinfo {author} {\bibfnamefont {J.}~\bibnamefont
  {Swiebodzinski}}, \bibinfo {author} {\bibfnamefont {A.}~\bibnamefont
  {Chudnovskiy}}, \bibinfo {author} {\bibfnamefont {T.}~\bibnamefont {Dunn}}, \
  and\ \bibinfo {author} {\bibfnamefont {A.}~\bibnamefont {Kamenev}},\ }\href
  {\doibase 10.1103/PhysRevB.82.144404} {\bibfield  {journal} {\bibinfo
  {journal} {Physical Review B}\ }\textbf {\bibinfo {volume} {82}},\ \bibinfo
  {pages} {144404} (\bibinfo {year} {2010})}\BibitemShut {NoStop}%
\bibitem [{\citenamefont {Zholud}\ \emph {et~al.}(2017)\citenamefont {Zholud},
  \citenamefont {Freeman}, \citenamefont {Cao}, \citenamefont {Srivastava},\
  and\ \citenamefont {Urazhdin}}]{Zholud:PRL2017}%
  \BibitemOpen
  \bibfield  {author} {\bibinfo {author} {\bibfnamefont {A.}~\bibnamefont
  {Zholud}}, \bibinfo {author} {\bibfnamefont {R.}~\bibnamefont {Freeman}},
  \bibinfo {author} {\bibfnamefont {R.}~\bibnamefont {Cao}}, \bibinfo {author}
  {\bibfnamefont {A.}~\bibnamefont {Srivastava}}, \ and\ \bibinfo {author}
  {\bibfnamefont {S.}~\bibnamefont {Urazhdin}},\ }\href
  {https://link.aps.org/doi/10.1103/PhysRevLett.119.257201} {\bibfield
  {journal} {\bibinfo  {journal} {Physical Review Letters}\ }\textbf {\bibinfo
  {volume} {119}},\ \bibinfo {pages} {257201} (\bibinfo {year}
  {2017})}\BibitemShut {NoStop}%
\bibitem [{\citenamefont {Qaiumzadeh}\ and\ \citenamefont
  {Brataas}(2018)}]{Qaiumzadeh:PRB2018}%
  \BibitemOpen
  \bibfield  {author} {\bibinfo {author} {\bibfnamefont {A.}~\bibnamefont
  {Qaiumzadeh}}\ and\ \bibinfo {author} {\bibfnamefont {A.}~\bibnamefont
  {Brataas}},\ }\href {\doibase 10.1103/PhysRevB.98.220408} {\bibfield
  {journal} {\bibinfo  {journal} {Physical Review B}\ }\textbf {\bibinfo
  {volume} {98}},\ \bibinfo {pages} {220408} (\bibinfo {year}
  {2018})}\BibitemShut {NoStop}%
\bibitem [{\citenamefont {Divinskiy}\ \emph {et~al.}(2021)\citenamefont
  {Divinskiy}, \citenamefont {Merbouche}, \citenamefont {Demidov},
  \citenamefont {Nikolaev}, \citenamefont {Soumah}, \citenamefont {Gouéré},
  \citenamefont {Lebrun}, \citenamefont {Cros}, \citenamefont {Youssef},
  \citenamefont {Bortolotti}, \citenamefont {Anane},\ and\ \citenamefont
  {Demokritov}}]{Divinskiy:NatCom2010}%
  \BibitemOpen
  \bibfield  {author} {\bibinfo {author} {\bibfnamefont {B.}~\bibnamefont
  {Divinskiy}}, \bibinfo {author} {\bibfnamefont {H.}~\bibnamefont
  {Merbouche}}, \bibinfo {author} {\bibfnamefont {V.~E.}\ \bibnamefont
  {Demidov}}, \bibinfo {author} {\bibfnamefont {K.~O.}\ \bibnamefont
  {Nikolaev}}, \bibinfo {author} {\bibfnamefont {L.}~\bibnamefont {Soumah}},
  \bibinfo {author} {\bibfnamefont {D.}~\bibnamefont {Gouéré}}, \bibinfo
  {author} {\bibfnamefont {R.}~\bibnamefont {Lebrun}}, \bibinfo {author}
  {\bibfnamefont {V.}~\bibnamefont {Cros}}, \bibinfo {author} {\bibfnamefont
  {J.~B.}\ \bibnamefont {Youssef}}, \bibinfo {author} {\bibfnamefont
  {P.}~\bibnamefont {Bortolotti}}, \bibinfo {author} {\bibfnamefont
  {A.}~\bibnamefont {Anane}}, \ and\ \bibinfo {author} {\bibfnamefont {S.~O.}\
  \bibnamefont {Demokritov}},\ }\href {\doibase 10.1038/s41467-021-26790-y}
  {\bibfield  {journal} {\bibinfo  {journal} {Nature Communications}\ }\textbf
  {\bibinfo {volume} {12}},\ \bibinfo {pages} {6541} (\bibinfo {year}
  {2021})}\BibitemShut {NoStop}%
\bibitem [{\citenamefont {Büttiker}(1992)}]{Buttiker:PRB1992}%
  \BibitemOpen
  \bibfield  {author} {\bibinfo {author} {\bibfnamefont {M.}~\bibnamefont
  {Büttiker}},\ }\href {http://link.aps.org/abstract/PRB/v46/p12485}
  {\bibfield  {journal} {\bibinfo  {journal} {Physical Review B}\ }\textbf
  {\bibinfo {volume} {46}},\ \bibinfo {pages} {12485} (\bibinfo {year}
  {1992})}\BibitemShut {NoStop}%
\bibitem [{\citenamefont {Duine}\ \emph {et~al.}(2007)\citenamefont {Duine},
  \citenamefont {Nunez}, \citenamefont {Sinova}, ,\ and\ \citenamefont
  {MacDonald}}]{Duine:PRB2007}%
  \BibitemOpen
  \bibfield  {author} {\bibinfo {author} {\bibfnamefont {R.}~\bibnamefont
  {Duine}}, \bibinfo {author} {\bibfnamefont {A.~S.}\ \bibnamefont {Nunez}},
  \bibinfo {author} {\bibfnamefont {J.}~\bibnamefont {Sinova}}, , \ and\
  \bibinfo {author} {\bibfnamefont {A.~H.}\ \bibnamefont {MacDonald}},\
  }\href@noop {} {\bibfield  {journal} {\bibinfo  {journal} {Physical Review
  B}\ }\textbf {\bibinfo {volume} {75}},\ \bibinfo {pages} {214420} (\bibinfo
  {year} {2007})}\BibitemShut {NoStop}%
\bibitem [{\citenamefont {Manchon}\ and\ \citenamefont
  {Zhang}(2008)}]{Manchon:PRB2008}%
  \BibitemOpen
  \bibfield  {author} {\bibinfo {author} {\bibfnamefont {A.}~\bibnamefont
  {Manchon}}\ and\ \bibinfo {author} {\bibfnamefont {S.}~\bibnamefont
  {Zhang}},\ }\href {http://link.aps.org/abstract/PRB/v78/e212405} {\bibfield
  {journal} {\bibinfo  {journal} {Physical Review B (Condensed Matter and
  Materials Physics)}\ }\textbf {\bibinfo {volume} {78}},\ \bibinfo {pages}
  {212405} (\bibinfo {year} {2008})}\BibitemShut {NoStop}%
\bibitem [{\citenamefont {Manchon}\ and\ \citenamefont
  {Zhang}(2009)}]{Manchon:PRB2009}%
  \BibitemOpen
  \bibfield  {author} {\bibinfo {author} {\bibfnamefont {A.}~\bibnamefont
  {Manchon}}\ and\ \bibinfo {author} {\bibfnamefont {S.}~\bibnamefont
  {Zhang}},\ }\href {http://link.aps.org/abstract/PRB/v79/e094422} {\bibfield
  {journal} {\bibinfo  {journal} {Physical Review B (Condensed Matter and
  Materials Physics)}\ }\textbf {\bibinfo {volume} {79}},\ \bibinfo {pages}
  {094422} (\bibinfo {year} {2009})}\BibitemShut {NoStop}%
\bibitem [{\citenamefont {Miron}\ \emph {et~al.}(2010)\citenamefont {Miron},
  \citenamefont {Gaudin}, \citenamefont {Auffret}, \citenamefont {Rodmacq},
  \citenamefont {Schuhl}, \citenamefont {Pizzini}, \citenamefont {Vogel},\ and\
  \citenamefont {Gambardella}}]{Miron:NatMat2010}%
  \BibitemOpen
  \bibfield  {author} {\bibinfo {author} {\bibfnamefont {I.~M.}\ \bibnamefont
  {Miron}}, \bibinfo {author} {\bibfnamefont {G.}~\bibnamefont {Gaudin}},
  \bibinfo {author} {\bibfnamefont {S.}~\bibnamefont {Auffret}}, \bibinfo
  {author} {\bibfnamefont {B.}~\bibnamefont {Rodmacq}}, \bibinfo {author}
  {\bibfnamefont {A.}~\bibnamefont {Schuhl}}, \bibinfo {author} {\bibfnamefont
  {S.}~\bibnamefont {Pizzini}}, \bibinfo {author} {\bibfnamefont
  {J.}~\bibnamefont {Vogel}}, \ and\ \bibinfo {author} {\bibfnamefont
  {P.}~\bibnamefont {Gambardella}},\ }\href
  {http://dx.doi.org/10.1038/nmat2613} {\bibfield  {journal} {\bibinfo
  {journal} {Nature Materials}\ }\textbf {\bibinfo {volume} {9}},\ \bibinfo
  {pages} {230} (\bibinfo {year} {2010})}\BibitemShut {NoStop}%
\bibitem [{\citenamefont {Miron}\ \emph {et~al.}(2011)\citenamefont {Miron},
  \citenamefont {Garello}, \citenamefont {Gaudin}, \citenamefont {Zermatten},
  \citenamefont {Costache}, \citenamefont {Auffret}, \citenamefont {Bandiera},
  \citenamefont {Rodmacq}, \citenamefont {Schuhl},\ and\ \citenamefont
  {Gambardella}}]{Miron:Nature2011}%
  \BibitemOpen
  \bibfield  {author} {\bibinfo {author} {\bibfnamefont {I.~M.}\ \bibnamefont
  {Miron}}, \bibinfo {author} {\bibfnamefont {K.}~\bibnamefont {Garello}},
  \bibinfo {author} {\bibfnamefont {G.}~\bibnamefont {Gaudin}}, \bibinfo
  {author} {\bibfnamefont {P.-J.}\ \bibnamefont {Zermatten}}, \bibinfo {author}
  {\bibfnamefont {M.~V.}\ \bibnamefont {Costache}}, \bibinfo {author}
  {\bibfnamefont {S.}~\bibnamefont {Auffret}}, \bibinfo {author} {\bibfnamefont
  {S.}~\bibnamefont {Bandiera}}, \bibinfo {author} {\bibfnamefont
  {B.}~\bibnamefont {Rodmacq}}, \bibinfo {author} {\bibfnamefont
  {A.}~\bibnamefont {Schuhl}}, \ and\ \bibinfo {author} {\bibfnamefont
  {P.}~\bibnamefont {Gambardella}},\ }\href {\doibase
  http://www.nature.com/nature/journal/v476/n7359/abs/nature10309.html#supplementary-information}
  {\bibfield  {journal} {\bibinfo  {journal} {Nature}\ }\textbf {\bibinfo
  {volume} {476}},\ \bibinfo {pages} {189} (\bibinfo {year}
  {2011})}\BibitemShut {NoStop}%
\bibitem [{\citenamefont {Liu}\ \emph {et~al.}(2012)\citenamefont {Liu},
  \citenamefont {Pai}, \citenamefont {Li}, \citenamefont {Tseng}, \citenamefont
  {Ralph},\ and\ \citenamefont {Buhrman}}]{Liu:Science2012}%
  \BibitemOpen
  \bibfield  {author} {\bibinfo {author} {\bibfnamefont {L.}~\bibnamefont
  {Liu}}, \bibinfo {author} {\bibfnamefont {C.-F.}\ \bibnamefont {Pai}},
  \bibinfo {author} {\bibfnamefont {Y.}~\bibnamefont {Li}}, \bibinfo {author}
  {\bibfnamefont {H.~W.}\ \bibnamefont {Tseng}}, \bibinfo {author}
  {\bibfnamefont {D.~C.}\ \bibnamefont {Ralph}}, \ and\ \bibinfo {author}
  {\bibfnamefont {R.~A.}\ \bibnamefont {Buhrman}},\ }\href {\doibase
  10.1126/science.1218197} {\bibfield  {journal} {\bibinfo  {journal}
  {Science}\ }\textbf {\bibinfo {volume} {336}},\ \bibinfo {pages} {555}
  (\bibinfo {year} {2012})}\BibitemShut {NoStop}%
\bibitem [{\citenamefont {Garello}\ \emph {et~al.}(2013)\citenamefont
  {Garello}, \citenamefont {Miron}, \citenamefont {Avci}, \citenamefont
  {Freimuth}, \citenamefont {Mokrousov}, \citenamefont {Blugel}, \citenamefont
  {Auffret}, \citenamefont {Boulle}, \citenamefont {Gaudin},\ and\
  \citenamefont {Gambardella}}]{Garello:NatNan2013}%
  \BibitemOpen
  \bibfield  {author} {\bibinfo {author} {\bibfnamefont {K.}~\bibnamefont
  {Garello}}, \bibinfo {author} {\bibfnamefont {I.~M.}\ \bibnamefont {Miron}},
  \bibinfo {author} {\bibfnamefont {C.~O.}\ \bibnamefont {Avci}}, \bibinfo
  {author} {\bibfnamefont {F.}~\bibnamefont {Freimuth}}, \bibinfo {author}
  {\bibfnamefont {Y.}~\bibnamefont {Mokrousov}}, \bibinfo {author}
  {\bibfnamefont {S.}~\bibnamefont {Blugel}}, \bibinfo {author} {\bibfnamefont
  {S.}~\bibnamefont {Auffret}}, \bibinfo {author} {\bibfnamefont
  {O.}~\bibnamefont {Boulle}}, \bibinfo {author} {\bibfnamefont
  {G.}~\bibnamefont {Gaudin}}, \ and\ \bibinfo {author} {\bibfnamefont
  {P.}~\bibnamefont {Gambardella}},\ }\href@noop {} {\bibfield  {journal}
  {\bibinfo  {journal} {Nature Nanotechnology}\ }\textbf {\bibinfo {volume}
  {8}},\ \bibinfo {pages} {587} (\bibinfo {year} {2013})}\BibitemShut {NoStop}%
\bibitem [{\citenamefont {Mellnik}\ \emph {et~al.}(2014)\citenamefont
  {Mellnik}, \citenamefont {Lee}, \citenamefont {Richardella}, \citenamefont
  {Grab}, \citenamefont {Mintun}, \citenamefont {Fischer}, \citenamefont
  {Vaezi}, \citenamefont {Manchon}, \citenamefont {Kim}, \citenamefont
  {Samarth},\ and\ \citenamefont {Ralph}}]{Mellnik:Nature2014}%
  \BibitemOpen
  \bibfield  {author} {\bibinfo {author} {\bibfnamefont {A.~R.}\ \bibnamefont
  {Mellnik}}, \bibinfo {author} {\bibfnamefont {J.~S.}\ \bibnamefont {Lee}},
  \bibinfo {author} {\bibfnamefont {A.}~\bibnamefont {Richardella}}, \bibinfo
  {author} {\bibfnamefont {J.~L.}\ \bibnamefont {Grab}}, \bibinfo {author}
  {\bibfnamefont {P.~J.}\ \bibnamefont {Mintun}}, \bibinfo {author}
  {\bibfnamefont {M.~H.}\ \bibnamefont {Fischer}}, \bibinfo {author}
  {\bibfnamefont {A.}~\bibnamefont {Vaezi}}, \bibinfo {author} {\bibfnamefont
  {A.}~\bibnamefont {Manchon}}, \bibinfo {author} {\bibfnamefont {E.~A.}\
  \bibnamefont {Kim}}, \bibinfo {author} {\bibfnamefont {N.}~\bibnamefont
  {Samarth}}, \ and\ \bibinfo {author} {\bibfnamefont {D.~C.}\ \bibnamefont
  {Ralph}},\ }\href {\doibase 10.1038/nature13534} {\bibfield  {journal}
  {\bibinfo  {journal} {Nature}\ }\textbf {\bibinfo {volume} {511}},\ \bibinfo
  {pages} {449} (\bibinfo {year} {2014})}\BibitemShut {NoStop}%
\bibitem [{\citenamefont {MacNeill}\ \emph {et~al.}(2017)\citenamefont
  {MacNeill}, \citenamefont {Stiehl}, \citenamefont {Guimaraes}, \citenamefont
  {Buhrman}, \citenamefont {Park},\ and\ \citenamefont
  {Ralph}}]{MacNeill:NatPhys2017}%
  \BibitemOpen
  \bibfield  {author} {\bibinfo {author} {\bibfnamefont {D.}~\bibnamefont
  {MacNeill}}, \bibinfo {author} {\bibfnamefont {G.~M.}\ \bibnamefont
  {Stiehl}}, \bibinfo {author} {\bibfnamefont {M.~H.~D.}\ \bibnamefont
  {Guimaraes}}, \bibinfo {author} {\bibfnamefont {R.~A.}\ \bibnamefont
  {Buhrman}}, \bibinfo {author} {\bibfnamefont {J.}~\bibnamefont {Park}}, \
  and\ \bibinfo {author} {\bibfnamefont {D.~C.}\ \bibnamefont {Ralph}},\ }\href
  {\doibase 10.1038/nphys3933
  http://www.nature.com/nphys/journal/v13/n3/abs/nphys3933.html#supplementary-information}
  {\bibfield  {journal} {\bibinfo  {journal} {Nature Physics}\ }\textbf
  {\bibinfo {volume} {13}},\ \bibinfo {pages} {300} (\bibinfo {year}
  {2017})}\BibitemShut {NoStop}%
\bibitem [{\citenamefont {Zhu}\ \emph {et~al.}(2019)\citenamefont {Zhu},
  \citenamefont {Ralph},\ and\ \citenamefont {Buhrman}}]{Zhu:PRL2019}%
  \BibitemOpen
  \bibfield  {author} {\bibinfo {author} {\bibfnamefont {L.}~\bibnamefont
  {Zhu}}, \bibinfo {author} {\bibfnamefont {D.~C.}\ \bibnamefont {Ralph}}, \
  and\ \bibinfo {author} {\bibfnamefont {R.~A.}\ \bibnamefont {Buhrman}},\
  }\href {\doibase 10.1103/PhysRevLett.122.077201} {\bibfield  {journal}
  {\bibinfo  {journal} {Physical Review Letters}\ }\textbf {\bibinfo {volume}
  {122}},\ \bibinfo {pages} {077201} (\bibinfo {year} {2019})}\BibitemShut
  {NoStop}%
\bibitem [{\citenamefont {Sinova}\ \emph {et~al.}(2015)\citenamefont {Sinova},
  \citenamefont {Valenzuela}, \citenamefont {Wunderlich}, \citenamefont
  {Back},\ and\ \citenamefont {Jungwirth}}]{Sinova:RMP2015}%
  \BibitemOpen
  \bibfield  {author} {\bibinfo {author} {\bibfnamefont {J.}~\bibnamefont
  {Sinova}}, \bibinfo {author} {\bibfnamefont {S.~O.}\ \bibnamefont
  {Valenzuela}}, \bibinfo {author} {\bibfnamefont {J.}~\bibnamefont
  {Wunderlich}}, \bibinfo {author} {\bibfnamefont {C.~H.}\ \bibnamefont
  {Back}}, \ and\ \bibinfo {author} {\bibfnamefont {T.}~\bibnamefont
  {Jungwirth}},\ }\href {https://link.aps.org/doi/10.1103/RevModPhys.87.1213}
  {\bibfield  {journal} {\bibinfo  {journal} {Reviews of Modern Physics}\
  }\textbf {\bibinfo {volume} {87}},\ \bibinfo {pages} {1213} (\bibinfo {year}
  {2015})}\BibitemShut {NoStop}%
\bibitem [{\citenamefont {Brataas}\ \emph {et~al.}(2002)\citenamefont
  {Brataas}, \citenamefont {Tserkovnyak}, \citenamefont {Bauer},\ and\
  \citenamefont {Halperin}}]{Brataas:PRB2002}%
  \BibitemOpen
  \bibfield  {author} {\bibinfo {author} {\bibfnamefont {A.}~\bibnamefont
  {Brataas}}, \bibinfo {author} {\bibfnamefont {Y.}~\bibnamefont
  {Tserkovnyak}}, \bibinfo {author} {\bibfnamefont {G.~E.~W.}\ \bibnamefont
  {Bauer}}, \ and\ \bibinfo {author} {\bibfnamefont {B.~I.}\ \bibnamefont
  {Halperin}},\ }\href {http://link.aps.org/doi/10.1103/PhysRevB.66.060404}
  {\bibfield  {journal} {\bibinfo  {journal} {Physical Review B}\ }\textbf
  {\bibinfo {volume} {66}},\ \bibinfo {pages} {060404} (\bibinfo {year}
  {2002})}\BibitemShut {NoStop}%
\bibitem [{\citenamefont {Brataas}\ \emph
  {et~al.}(2006{\natexlab{a}})\citenamefont {Brataas}, \citenamefont {Bauer},\
  and\ \citenamefont {Kelly}}]{Brataas:PhysRep2006}%
  \BibitemOpen
  \bibfield  {author} {\bibinfo {author} {\bibfnamefont {A.}~\bibnamefont
  {Brataas}}, \bibinfo {author} {\bibfnamefont {G.~E.~W.}\ \bibnamefont
  {Bauer}}, \ and\ \bibinfo {author} {\bibfnamefont {P.~J.}\ \bibnamefont
  {Kelly}},\ }\href
  {http://www.sciencedirect.com/science/article/B6TVP-4JF97B4-1/2/499d943549f8b52e7a2fdde6e9fdc472}
  {\bibfield  {journal} {\bibinfo  {journal} {Physics Reports}\ }\textbf
  {\bibinfo {volume} {427}},\ \bibinfo {pages} {157} (\bibinfo {year}
  {2006}{\natexlab{a}})}\BibitemShut {NoStop}%
\bibitem [{\citenamefont {Foros}\ \emph {et~al.}(2007)\citenamefont {Foros},
  \citenamefont {Brataas}, \citenamefont {Bauer},\ and\ \citenamefont
  {Tserkovnyak}}]{Foros:PRB2007}%
  \BibitemOpen
  \bibfield  {author} {\bibinfo {author} {\bibfnamefont {J.}~\bibnamefont
  {Foros}}, \bibinfo {author} {\bibfnamefont {A.}~\bibnamefont {Brataas}},
  \bibinfo {author} {\bibfnamefont {G.~E.~W.}\ \bibnamefont {Bauer}}, \ and\
  \bibinfo {author} {\bibfnamefont {Y.}~\bibnamefont {Tserkovnyak}},\ }\href
  {http://link.aps.org/abstract/PRB/v75/e092405} {\bibfield  {journal}
  {\bibinfo  {journal} {Physical Review B}\ }\textbf {\bibinfo {volume} {75}},\
  \bibinfo {pages} {092405} (\bibinfo {year} {2007})}\BibitemShut {NoStop}%
\bibitem [{\citenamefont {Brataas}\ \emph
  {et~al.}(2006{\natexlab{b}})\citenamefont {Brataas}, \citenamefont {Bauer},\
  and\ \citenamefont {Kelly}}]{Brataas:PRep2006}%
  \BibitemOpen
  \bibfield  {author} {\bibinfo {author} {\bibfnamefont {A.}~\bibnamefont
  {Brataas}}, \bibinfo {author} {\bibfnamefont {G.~E.~W.}\ \bibnamefont
  {Bauer}}, \ and\ \bibinfo {author} {\bibfnamefont {P.~J.}\ \bibnamefont
  {Kelly}},\ }\href
  {http://www.sciencedirect.com/science/article/B6TVP-4JF97B4-1/2/499d943549f8b52e7a2fdde6e9fdc472}
  {\bibfield  {journal} {\bibinfo  {journal} {Physics Reports}\ }\textbf
  {\bibinfo {volume} {427}},\ \bibinfo {pages} {157} (\bibinfo {year}
  {2006}{\natexlab{b}})}\BibitemShut {NoStop}%
\bibitem [{\citenamefont {Brataas}\ \emph {et~al.}(2011)\citenamefont
  {Brataas}, \citenamefont {Tserkovnyak}, \citenamefont {Bauer},\ and\
  \citenamefont {Kelly}}]{Brataas:arXiv2011}%
  \BibitemOpen
  \bibfield  {author} {\bibinfo {author} {\bibfnamefont {A.}~\bibnamefont
  {Brataas}}, \bibinfo {author} {\bibfnamefont {Y.}~\bibnamefont
  {Tserkovnyak}}, \bibinfo {author} {\bibfnamefont {G.~E.~W.}\ \bibnamefont
  {Bauer}}, \ and\ \bibinfo {author} {\bibfnamefont {P.~J.}\ \bibnamefont
  {Kelly}},\ }\href@noop {} {\bibfield  {journal} {\bibinfo  {journal}
  {arXiv:1108.0385}\ } (\bibinfo {year} {2011})}\BibitemShut {NoStop}%
\bibitem [{\citenamefont {Kamenev}(2011)}]{Kamenev:2011}%
  \BibitemOpen
  \bibfield  {author} {\bibinfo {author} {\bibfnamefont {A.}~\bibnamefont
  {Kamenev}},\ }\href {\doibase DOI: 10.1017/CBO9781139003667} {\emph {\bibinfo
  {title} {Field Theory of Non-Equilibrium Systems}}}\ (\bibinfo  {publisher}
  {Cambridge University Press},\ \bibinfo {address} {Cambridge},\ \bibinfo
  {year} {2011})\BibitemShut {NoStop}%
\bibitem [{Note1()}]{Note1}%
  \BibitemOpen
  \bibinfo {note} {We generalize to a complex spin pumping coefficient that is
  consistent with our finding for the spin transfer torque.}\BibitemShut
  {Stop}%
\bibitem [{\citenamefont {Zwierzycki}\ \emph {et~al.}(2005)\citenamefont
  {Zwierzycki}, \citenamefont {Tserkovnyak}, \citenamefont {Kelly},
  \citenamefont {Brataas},\ and\ \citenamefont {Bauer}}]{Zwierzycki:PRB2005}%
  \BibitemOpen
  \bibfield  {author} {\bibinfo {author} {\bibfnamefont {M.}~\bibnamefont
  {Zwierzycki}}, \bibinfo {author} {\bibfnamefont {Y.}~\bibnamefont
  {Tserkovnyak}}, \bibinfo {author} {\bibfnamefont {P.~J.}\ \bibnamefont
  {Kelly}}, \bibinfo {author} {\bibfnamefont {A.}~\bibnamefont {Brataas}}, \
  and\ \bibinfo {author} {\bibfnamefont {G.~E.~W.}\ \bibnamefont {Bauer}},\
  }\href {http://link.aps.org/doi/10.1103/PhysRevB.71.064420} {\bibfield
  {journal} {\bibinfo  {journal} {Physical Review B}\ }\textbf {\bibinfo
  {volume} {71}},\ \bibinfo {pages} {064420} (\bibinfo {year}
  {2005})}\BibitemShut {NoStop}%
\end{thebibliography}
\end{document}